\documentclass[12pt, a4paper]{article}
 \usepackage[font=small,format=plain,labelfont=bf,up,textfont=normal,up,justification=justified,singlelinecheck=false]{caption}
\usepackage{subcaption}

\usepackage[a4paper, left=2cm,right=2cm]{geometry}
\usepackage[colorlinks=true,linkcolor=black,citecolor=teal,urlcolor=MidnightBlue,filecolor=black]{hyperref}
\usepackage{amsfonts}
\usepackage{amsmath,amssymb}
\usepackage{setspace}
\usepackage{slashed}
\usepackage{braket}

\usepackage[dvipsnames]{xcolor}
\definecolor{SchoolColor}{rgb}{0.6471, 0.1098, 0.1882} 
\usepackage{subfloat}

\usepackage[utf8,applemac]{inputenc}
\usepackage{tensor}
\usepackage{cite}
\usepackage{tikz}
\usetikzlibrary{calc}
\usetikzlibrary{patterns}
\usetikzlibrary{arrows.meta}

\usepackage{graphicx}
\usepackage{bm} 

\usepackage[utf8,applemac]{inputenc}
\usepackage{tensor}
\usepackage{cite}
\usepackage{tikz}
\usepackage{graphicx}
\graphicspath{{figure/}}
\usepackage{array}
\usepackage{booktabs}

\usepackage{multirow}

\usepackage{dcolumn}
\usepackage{bm}

\usepackage{verbatim}

\usepackage{textcomp} 
\usepackage{graphicx} 

\numberwithin{equation}{section}
\newcommand{\bea}{\begin{eqnarray}}
\newcommand{\eea}{\end{eqnarray}}
\newcommand{\be}{\begin{equation}}
\newcommand{\ee}{\end{equation}}
\newcommand{\bs}{\begin{subequations}}
\newcommand{\es}{\end{subequations}}
\def\nn{\nonumber}

\newcommand{\beqs}{\begin{eqnarray}}
\newcommand{\eeqs}{\end{eqnarray}}
\numberwithin{equation}{section}

\newcommand{\Rmnum}[1]{\uppercase\expandafter{\romannumeral #1\relax}}

\newcommand{\yxq}{\color{green}}

\begin{document}
\begin{titlepage}

\begin{flushright}\vspace{-3cm}
{\small
\today }\end{flushright}
\vspace{0.5cm}
\begin{center}
	{{ \LARGE{\bf{
Feynman rules and loop structure of	Carrollian\vspace{8pt}\\ amplitudes }}}}\vspace{5mm}

	\centerline{\large{\bf Wen-Bin Liu\footnote{liuwenbin0036@hust.edu.cn}, Jiang Long\footnote{
				longjiang@hust.edu.cn} and Xiao-Quan Ye\footnote{u202010191@hust.edu.cn}}}
	\vspace{2mm}
	\normalsize
	\bigskip\medskip

	\textit{School of Physics, Huazhong University of Science and Technology, \\ Luoyu Road 1037, Wuhan, Hubei 430074, China
	}
	
	\vspace{25mm}
	
	\begin{abstract}
		\noindent
		{
		In this paper, we derive the Carrollian amplitude in the framework of bulk reduction. The Carrollian amplitude is shown to relate to the scattering amplitude by a Fourier transform in this method. We propose Feynman rules to calculate the Carrollian amplitude where the Fourier transforms emerge as the integral representation of the external lines in the Carrollian space. Then we study the four-point Carrollian amplitude at  loop level in massless $\Phi^4$ theory. As a consequence of Poincar\'e invariance, the four-point Carrollian amplitude can be transformed to the amplitude that only depends on the cross ratio $z$ of the celestial sphere and a variable $\chi$ invariant under translation.  The four-point Carrollian amplitude is a polynomial of the two-point Carrollian amplitude whose argument is replaced with $\chi$. The coefficients of the polynomial have branch cuts in the complex $z$ plane. We also show that the renormalized Carrollian amplitude obeys the Callan-Symanzik equation. Moreover, we initiate a generalized $\Phi^4$ theory by designing the Feynman rules for more general Carrollian amplitudes.  }\end{abstract}
	

\end{center}

\end{titlepage}
\tableofcontents

\section{Introduction}
Flat holography has attracted much attention due to its potential connection with realistic physical processes. Up to now, there are various approaches to handle this intriguing problem.
The first approach is  asymptotic symmetry analysis which may trace back to the famous BMS group \cite{1962RSPSA.269...21B,1962RSPSA.270..103S,1962PhRv..128.2851S}. In this approach, the boundary theory obeys the asymptotic symmetries in the bulk \cite{Arcioni:2003td,Barnich:2006av,Barnich:2010eb} and one may construct various  theories based on these symmetries.  In the second method, the BMS groups are identified as geometric global symmetries of a Carrollian manifold \cite{Duval:2014uva,Duval:2014lpa} on which field theories can be obtained by taking the speed of light to zero \cite{Une, Gupta1966OnAA} in the usual QFTs. The third approach is based on scattering amplitude on the  celestial sphere \cite{Pasterski:2016qvg}. This method is motivated by the deep connections between BMS groups and soft theorems \cite{Strominger:2013jfa} in the IR region. Basically, it maps the usual $S$ matrix by a Mellin transform and the resulting amplitude is called celestial amplitude \cite{Pasterski:2017kqt,Pasterski:2017ylz}. Along this line, soft theorems \cite{2017arXiv170305448S}, MHV celestial gluon amplitudes \cite{Schreiber:2017jsr}, four-point celestial amplitudes \cite{Arkani-Hamed:2020gyp,Puhm:2019zbl},  and loop corrections \cite{Banerjee:2017jeg,Albayrak:2020saa,Gonzalez:2020tpi} have been studied. Interestingly, the scattering amplitude can also be mapped to the  so-called Carrollian amplitude by a simple Fourier transform  \cite{Donnay:2022aba, Donnay:2022wvx}. Therefore, there is a beautiful triangle among scattering amplitude, celestial amplitude and Carrollian amplitude. The two-point Carrollian amplitudes have been derived in \cite{Liu:2022mne, Donnay:2022wvx} and the three-point Carrollian amplitudes can also be found in \cite{Salzer:2023jqv,Nguyen:2023miw}. The four-point scalar Carrollian amplitudes have been studied by modified Mellin transformation \cite{Banerjee:2019prz} and have been extended to gluons and gravitons systematically in recent article \cite{Mason:2023mti}. It is also \cite{Mason:2023mti} where the name of ``Carrollian amplitudes" was first suggested, and the authors also considered various properties of Carrollian amplitudes, such as the  collinear limit, UV and IR behaviours,  connections with twistors and so on.

In addition, the authors in  \cite{Liu:2022mne,Liu:2024nkc} have proposed studying flat holography through bulk reduction from well-known QFTs in asymptotically flat spacetime. The framework is to reduce the bulk theory to the null boundary by asymptotic expansion. There is always a fundamental field $F$ which represents the radiative  degree of freedom at the leading order of the fluctuation. The canonical quantization of the bulk field $\tt{f}$ implies the canonical quantization of the fundamental field with which one may define the quantum flux operators  that form an infinite dimensional Lie algebra \cite{Liu:2022mne,Liu:2023qtr,Liu:2023gwa,Li:2023xrr,Liu:2023jnc} with central extension whose classical version is the infinitesimal (intertwined) Carrollian diffeomorphism. These impressive results indicate that the framework is correct and fruitful. Actually, it has been noted in \cite{Liu:2022mne} that  the asymptotic state in scattering amplitude is related to the fundamental field  acting on the free vacuum through a Fourier transform. It is  pointed out explicitly that there may be a  correspondence between the asymptotic state $|\bm p\rangle$ in the momentum space and the boundary fundamental operator $F(u,\Omega)$ in the Carrollian space  
\be 
|\bm p\rangle\quad\leftrightarrow\quad F(u,\Omega).
\ee As a consequence, the scattering amplitude is mapped  by a Fourier transform to the correlator of the fundamental fields  which are inserted at the null boundary.  In this paper, we will work out this point explicitly and derive the relation between the scattering amplitude and  Carrollian amplitude from bulk reduction. We will  derive the perturbative Feynman rules to obtain Carrollian amplitude in the Carrollian space. The four-point Carrollian amplitude has been studied up to two loops for massless $\Phi^4$ theory. Amazingly, their forms are as simple as the scattering amplitudes in momentum space. We will also find a Callan-Symanzik equation for Carrollian amplitude in our method. 

This paper is organized as follows. In section \ref{pre}, we review the framework of bulk reduction in massless scalar theory and the asymptotically free states have been created by inserting the fundamental field at the null boundary. We work out the antipodal map and derive the Carrollian amplitude from the scattering amplitude in this framework. In section \ref{poin}, we will derive the transformation law of the Carrollian amplitude under Poincar\'e transformation. In the following section, we propose the Feynman rules to calculate the Carrollian amplitude in Carrollian space. In section \ref{4ptscalar}, we study the four-point Carrollian amplitude for massless $\Phi^4$  theory up to  two-loop level. In the following section, we investigate the Carrollian amplitude in a more general $\Phi^4$ theory. We will conclude in section \ref{conc}. Technical details are  relegated to several appendices.

\section{Preliminaries}\label{pre}
In this work, we will study the Carrollian amplitude in four-dimensional $\Phi^4$ theory which  is described by the action 
\bea 
S[\Phi]=\int d^4 x [-\frac{1}{2}(\partial_\mu\Phi)^2-\frac{\lambda}{4!}\Phi^4]
\eea  in Minkowski spacetime. The signature of the metric is $(-,+,+,+)$ and we will use $x^\mu$ with  indices in the Greek  alphabet $\mu,\nu,\cdots$ to denote Cartesian coordinates. By imposing the fall-off condition 
\bea \label{asympexp}
\Phi(t,\bm x)=\left\{\begin{array}{cc}\frac{\Sigma(u,\Omega)}{r}+\mathcal{O}(r^{-2}),& \text{near} \ \mathcal{I}^+\\
\frac{\Xi(v,\Omega)}{r}+\mathcal{O}(r^{-2}),&\text{near} \ \mathcal{I}^-\end{array}\right.
\eea the theory has been reduced to the boundary \cite{Liu:2022mne} at future/past null infinity ($\mathcal{I}^{+}/\mathcal{I}^-$). Here the coordinates $u=t-r$ and $v=t+r$ are the retarded and advanced time and $r$ is the spatial distance in the bulk. We have also used $\Omega=\theta^A=(\theta,\phi)$ to denote the angular direction in spherical coordinates. In some cases, we will also use the stereographic coordinates $(z,\bar z)$ for convenience. 

The fundamental field $\Sigma(u,\Omega)/\Xi(v,\Omega)$ encodes the propagating degree of freedom of the bulk theory. However, we should emphasize that the fundamental field is not a dynamical scalar at the boundary since there is no constraint equation for this field. This radiative data is distinguished from the Carrollian field which obeys nontrivial equation of motion on the boundary. Although the single field $\Sigma(u,\Omega)/\Xi(v,\Omega)$ has no dynamics, the overlap between the radiative modes at $\mathcal{I}^-$ and $\mathcal{I}^+$  actually defines the Carrollian amplitudes which reflect the interactions in the bulk.

\paragraph{Canonical quantization of the fundamental field.} In the canonical quantization, the bulk field $\Phi(t,\bm x)$ may be expanded asymptotically as plane waves 
\bea 
\Phi(t,\bm x)&=&\int \frac{d^3\bm p}{(2\pi)^3}\frac{1}{\sqrt{2\omega_{\bm{p}}}}(e^{-i\omega t+i\bm{p}\cdot\bm{x}}b_{\bm{p}}+e^{i\omega t-i\bm{p}\cdot\bm{x}}b_{\bm{p}}^\dagger)\label{expanPhi}
\eea where $b_{\bm p}$ and $b_{\bm p}^\dagger$ are annihilation and creation operators that satisfy the standard commutation relations 
\bea 
\ [b_{\bm p},b_{\bm p'}^\dagger]=(2\pi)^3\delta(\bm p-\bm p'),\quad [b_{\bm p},b_{\bm p'}]=0,\quad [b_{\bm p}^\dagger,b_{\bm p'}^\dagger]=0.\label{anncom}
\eea 
It has been shown that the field $\Phi(t,\bm x)$ may be expanded as spherical waves such that \cite{Liu:2022mne}
\bea 
\Sigma(u,\Omega)&=&\int_0^\infty \frac{d\omega}{\sqrt{4\pi\omega}}\sum_{\ell m}[a_{\omega,\ell,m}e^{-i\omega u}Y_{\ell,m}(\Omega)+a^\dagger_{\omega,\ell,m}e^{i\omega u}Y^*_{\ell,m}(\Omega)]
\eea at $\mathcal{I}^+$ where $Y_{\ell,m}(\Omega)$ are spherical harmonic functions on $S^2$. The boundary annihilation and creation operators $a_{\omega,\ell,m}$ and $a^\dagger_{\omega,\ell,m}$ are related to $b_{\bm p}$ and $b_{\bm p}^\dagger$ through
\bs\label{aadager}\begin{align}
a_{\omega,\ell,m}&=\frac{\omega}{2\sqrt{2}\pi^{3/2} i}\int d\Omega b_{\bm p}Y_{\ell,m}^*(\Omega),\label{annihi}\\
a_{\omega,\ell,m}^\dagger&=\frac{\omega i}{2\sqrt{2}\pi^{3/2}}\int d\Omega b^\dagger_{\bm p}Y_{\ell,m}(\Omega)\label{create}
\end{align}\es
 and they are labeled by three quantum numbers where $\omega$ is the frequency of the corresponding particle and $\ell,m$ are quantum numbers of angular momentum. The momentum $\bm p$ may also be written in spherical coordinates 
 \bea 
 \bm p=(\omega,\Omega).
 \eea 
 The inverse transformation of \eqref{aadager} is 
 \bs\begin{align}
b_{\bm p}&=\frac{2\sqrt{2}\pi^{3/2}i}{\omega}\sum_{\ell,m}a_{\omega,\ell,m}Y_{\ell,m}(\Omega),\\
b_{\bm p}^\dagger&= -\frac{2\sqrt{2}\pi^{3/2}i}{\omega}\sum_{\ell,m}a^\dagger_{\omega,\ell,m}Y^*_{\ell,m}(\Omega).
\end{align}\es 

We can also evaluate the commutator between the boundary fields $\Sigma(u,\Omega)$ which is of the same form as the one in \cite{Ashtekar:1981hw,Ashtekar:1981sf,Ashtekar:1987tt} where the author developed the method of asymptotic quantization, and derived commutation relations from the asymptotic symplectic form.

\paragraph{Asymptotic states.} Since $a_{\omega,\ell,m}$ is a linear superposition of $b_{\bm p}$, we can define the free vacuum $|0\rangle$ by 
\bea 
b_{\bm p}|0\rangle=0\quad\Leftrightarrow\quad a_{\omega,\ell,m}|0\rangle=0.
\eea 
The equation \eqref{create} shows that $a_{\omega,\ell,m}^\dagger$ creates a state  \bea 
|\omega,\ell,m\rangle= a^\dagger_{\omega,\ell,m}|0\rangle
\eea from vacuum. This is also a  superposition of states $|\bm p\rangle$ 
\bea 
|\omega,\ell,m\rangle=\frac{\sqrt{\omega} i}{4\pi^{3/2}}\int d\Omega Y_{\ell,m}(\Omega)|\bm p\rangle,\quad |\bm p\rangle=\sqrt{2\omega_{\bm p}}b_{\bm p}^\dagger |0\rangle.
\eea We can also create a state located at $(u,\Omega)$ 
\bea 
|\Sigma(u,\Omega)\rangle= \Sigma(u,\Omega)|0\rangle=\frac{i}{8\pi^2}\int_0^\infty d\omega e^{i\omega u}|\bm p\rangle\label{outstate}
\eea which is an asymptotic state that represents a superposition of outgoing particles at $\mathcal{I}^+$. The state $|\bm p\rangle$ can  be obtained by an inverse Fourier transform 
\bea 
|\bm p\rangle=-4\pi i\int_{-\infty}^\infty du e^{-i\omega u}|\Sigma(u,\Omega)\rangle.
\eea For one-particle state, the completeness relation 
\bea 
1=\int\frac{d^3\bm p}{(2\pi)^32\omega_{\bm p}}|\bm p\rangle\langle\bm p|
\eea is transformed to 
\bea 
1=i\int du d\Omega\left( |\Sigma(u,\Omega)\rangle\langle\dot{\Sigma}(u,\Omega)|-|\dot{\Sigma}(u,\Omega)\rangle\langle{\Sigma}(u,\Omega)|\right).\label{complete}
\eea Integrating by parts and ignoring the boundary term, the completeness relation becomes 
\bea 
1=2i\int du d\Omega |\Sigma(u,\Omega)\rangle\langle\dot{\Sigma}(u,\Omega)|=-2i\int du d\Omega |\dot{\Sigma}(u,\Omega)\rangle\langle{\Sigma}(u,\Omega)|.
\eea 
The Fock space is constructed by acting the creation operators repeatedly on the vacuum state. For example, an $n$-particle outgoing state with definite momentum $|\bm p_1\bm p_2\cdots\bm p_{n}\rangle$ is 
\bea 
|\bm p_1\bm p_2\cdots\bm p_{n}\rangle=\prod_{j=1}^n \sqrt{2\omega_{\bm  p_j}} b_{\bm p_j}^\dagger |0\rangle.
\eea Switching to the Carrollian space, we find 
\bea 
|\bm p_1\bm p_2\cdots\bm p_{n}\rangle&=&: \prod_{j=1}^n(-4\pi i)\int du_j  e^{-i\omega_ju_j}\Sigma(u_j,\Omega_j):|0\rangle\nn\\&=&\int d\mu_{1,2,\cdots,n} |\prod_{k=1}^n\Sigma(u_k,\Omega_k)\rangle.\label{fourierout}
\eea where the integration measure is defined as
\bea 
d\mu_{1,2,\cdots,n}=\prod_{j=1}^n (-4\pi i) du_j  e^{-i\omega_ju_j}.
\eea 
Note that we have inserted the normal ordering operator $:\cdots:$ in the first line of \eqref{fourierout}, otherwise there will be  nonvanishing functions from  exchanging the positions of fundamental fields. Taking into account the multi-particle states, the completeness relation \eqref{complete} becomes 
\bea 
1&=&\sum_{n}\prod_{j=1}^n i\int du_j d\Omega_j \left( |\Sigma(u_j,\Omega_j)\rangle\langle\dot{\Sigma}(u_j,\Omega_j)|-|\dot{\Sigma}_(u_j,\Omega_j)\rangle\langle{\Sigma}(u_j,\Omega_j)|\right)\nn\\&=&\sum_{n}\prod_{j=1}^n 2i\int du_j d\Omega_j  |\Sigma(u_j,\Omega_j)\rangle\langle\dot{\Sigma}(u_j,\Omega_j)|\nn\\&=&\sum_{n}\prod_{j=1}^n \left(-2i\right)\int du_j d\Omega_j  |\dot{\Sigma}(u_j,\Omega_j)\rangle\langle{\Sigma}(u_j,\Omega_j)|\label{comp}
\eea where the summation is over all possible multi-particle states labeled by $n$.

 Similarly, one may also expand $\Xi(v,\Omega)$ at $\mathcal{I}^-$ as 
\bea
\Xi(v,\Omega)&=&\int_0^\infty \frac{d\omega}{\sqrt{4\pi\omega}}\sum_{\ell m}[\bar{a}_{\omega,\ell,m}e^{-i\omega v}Y_{\ell,m}(\Omega)+\bar{a}^\dagger_{\omega,\ell,m}e^{i\omega v}Y^*_{\ell,m}(\Omega)]
\nn\\&=&\int_0^\infty \frac{d\omega}{\sqrt{4\pi\omega}}\sum_{\ell,m}[(-1)^{\ell+1}e^{-i\omega v}Y_{\ell,m}(\Omega)a_{\omega,\ell,m}+(-1)^{\ell+1}e^{i\omega v}Y^*_{\ell,m}(\Omega)a^\dagger_{\omega,\ell,m}].\label{sigmaexpscim}
\eea 
Therefore, we may define an incoming state at $\mathcal{I}^-$ by 
\bea 
|\Xi(v,\Omega)\rangle\equiv\Xi(v,\Omega)|0\rangle=-\frac{i}{8\pi^2}\int_0^\infty d\omega e^{i\omega v}|\bm p^P\rangle, \label{invstate}
\eea where the momentum $\bm p^P$ is defined by 
\bea 
\bm p^P=(\omega,\Omega^P)
\eea in the spherical coordinates with $\Omega^P$ the antipodal point of $\Omega=(\theta,\phi)$
\bea 
\Omega^P=(\pi-\theta,\pi+\phi).
\eea Therefore, we may transform the state $|v,\Omega\rangle$ to its corresponding antipodal state via
\bea 
|\Xi(v,\Omega^P)\rangle=\Xi(v,\Omega^P)|0\rangle=-\frac{i}{8\pi^2}\int_0^\infty d\omega e^{i\omega v}|\bm p\rangle.
\eea Hence, an incoming state with definite  momentum $\bm p$ at $\mathcal{I}^-$
can be written as 
\bea 
|\bm p\rangle=4\pi i\int_{-\infty}^\infty dv e^{-i\omega v}|\Xi(v,\Omega^P)\rangle.
\eea An $n$-particle incoming state $|\bm p_1\bm p_2\cdots\bm p_n\rangle$ is 
\bea 
|\bm p_1\bm p_2\cdots\bm p_n\rangle&=&:\prod_{j=1}^n(4\pi i) \int dv_j e^{-i\omega_j v_j}\Xi(v_j,\Omega_j^P):|0\rangle\nn\\&=&\int d\nu_{1,2,\cdots,n}|\prod_{j=1}^n\Xi(v_j,\Omega^P_j)\rangle
\eea where 
\be 
d\nu_{1,2,\cdots,n}=\prod_{j=1}^n(4\pi i)  dv_j e^{-i\omega_j v_j}.
\ee 

\paragraph{Carrollian amplitude.} Now we consider the $m\to n$ scattering process 
\bea 
{}_{\text{out}}\langle \bm p_{m+1}\bm p_{m+2}\cdots\bm p_{m+n}|\bm p_{1}\bm p_{2}\cdots\bm p_m\rangle_{\text{in}}=\langle \bm p_{m+1}\bm p_{m+2}\cdots\bm p_{m+n}|S|\bm p_{1}\bm p_{2}\cdots\bm p_m\rangle. \label{fourierin}
\eea 
The $S$ matrix can also be transformed to Carrollian space using \eqref{fourierout} and \eqref{fourierin}
\bea 
&& \langle \bm p_{m+1}\bm p_{m+2}\cdots\bm p_{m+n}|S|\bm p_1\bm p_2\cdots\bm p_m\rangle\nn\\&=&\int d\mu^*_{m+1,\cdots,m+n}d\nu_{1,\cdots,m}\langle \prod_{k=m+1}^{m+n}\Sigma(u_k,\Omega_k)|S|\prod_{k=1}^m \Xi(v_k,\Omega_k^P)\rangle\nn\\&=&\int d\mu^*_{m+1,\cdots,m+n}d\nu_{1,\cdots,m}\  {}_{\text{out}}\langle \prod_{k=m+1}^{m+n}\Sigma(u_k,\Omega_k)|\prod_{k=1}^m \Xi(v_k,\Omega_k^P)\rangle_{\text{in}}\nn
\eea where we have defined 
\bea 
{}_{\text{out}}\langle \prod_{k=m+1}^{m+n}\Sigma(u_k,\Omega_k)|\prod_{k=1}^m \Xi(v_k,\Omega_k^P)\rangle_{\text{in}}=\langle \prod_{k=m+1}^{m+n}\Sigma(u_k,\Omega_k)|S|\prod_{k=1}^m \Xi(v_k,\Omega_k^P)\rangle.
\eea Inversely, we will find the $m\to n$ Carrollian amplitude as the Fourier transform of the scattering amplitude
\bea 
&&{}_{\text{out}}\langle \prod_{k=m+1}^{m+n}\Sigma(u_k,\Omega_k)|\prod_{k=1}^m \Xi(v_k,\Omega_k^P)\rangle_{\text{in}}\nn\\&=&(\frac{1}{8\pi^2 i})^{m+n}\int d\omega_1\cdots d\omega_{m+n}e^{-i\sum_{j=m+1}^{m+n}\omega_j u_j+i\sum_{j=1}^m \omega_j v_j}\langle \bm p_{m+1}\bm p_{m+2}\cdots\bm p_{m+n}|S|\bm p_1\bm p_2\cdots\bm p_m\rangle.\nn\\
\eea 

On the left-hand side, the Carrollian amplitude may be understood as $(m+n)$-point  correlators with $m$ fields $\Xi(v,\Omega)$ inserted at $(v_1,\Omega_1^P),\cdots, (v_m,\Omega_m^P)$ and $n$ fields $\Sigma(u,\Omega)$  inserted at $(u_{m+1},\Omega_{m+1}),\cdots,(u_{m+n},\Omega_{m+n})$, respectively. On the right-hand side, it is the Fourier transformation of the $m\to n$ scattering matrix. Note that we derive this relation by the standard method of QFT without using any knowledge of flat holography or asymptotic symmetry. One may also transform it to celestial amplitude by Mellin transformation, though we will not elaborate on it in this work. In the expression, we may redefine  
\bea 
u_j=v_j\quad \text{for}\quad j=1,2,\cdots,m.
\eea At the same time, we transform $\Omega_j^P, j=1,2,\cdots,m$ to their antipodal points $\Omega_j$ and relabel 
\bea 
\Sigma(u,\Omega)=\Xi(v,\Omega^P).\label{antipodal}
\eea 
Note that the above identification may be obtained by comparing \eqref{outstate} with \eqref{invstate} up to a constant phase\footnote{In these expressions, the phase is $e^{i\pi}=-1$.}. Since the phase factor is irrelevant in $S$ matrix, we will not care about it later. 
Then the Carrollian amplitude may be written in a more familiar form 
\bea 
&&{}_{\text{out}}\langle \prod_{k=m+1}^{m+n}\Sigma(u_k,\Omega_k)|\prod_{k=1}^m \Sigma(u_k,\Omega_k)\rangle_{\text{in}}\nn\\&=&(\frac{1}{8\pi^2 i})^{m+n}\prod_{j=1}^{m+n} \int d\omega_j e^{-i\sigma_j\omega_j u_j} \langle \bm p_{m+1}\bm p_{m+2}\cdots\bm p_{m+n}|S|\bm p_1\bm p_2\cdots\bm p_m\rangle
\eea with 
\bea 
p^\mu_j=\sigma_j \omega_j n_j^\mu,\quad j=1,2,\cdots,m+n\label{momentumnull}
\eea where $n^\mu_j$ is the null vector associated with $j$-th particle
\bea 
n_j^\mu=(1,\sin\theta_j\cos\phi_j,\sin\theta_j\sin\phi_j,\cos\theta_j).
\eea The symbol $\sigma_j, j=1,2,\cdots,m+n$ is designed to distinguish the outgoing and incoming states through the relation 
\bea 
\sigma_j=\left\{\begin{array}{cl}+1&\text{outgoing state},\\
-1&\text{incoming state}.\end{array}\right.
\eea 

The $S$ matrix may be factorized as 
\bea 
S=1+iT
\eea where the $T$ matrix denotes the connected part and it should respect  the 4-momentum conservation. Therefore, one may extract a Lorentz invariant matrix element $\mathcal{M}$ by \cite{1995iqft.book.....P}
\bea 
\langle \bm p_{m+1}\bm p_{m+2}\cdots\bm p_{m+n}|iT|\bm p_1\bm p_2\cdots\bm p_m\rangle=(2\pi)^4\delta^{(4)}(\sum_{j=1}^{m+n}p_j)i\mathcal{M}(p_1,p_2,\cdots,p_{m+n}).
\eea Therefore, we may reduce the Carrollian amplitude to the $\mathcal{M}$ matrix which is related to the amputated and connected Feynman diagrams 
\bea 
&&{}_{\text{out}}\langle \prod_{k=m+1}^{m+n}\Sigma(u_k,\Omega_k)|\prod_{k=1}^m \Sigma(u_k,\Omega_k)\rangle_{\text{in}}\Big|_{\text{connected and amputated}}\nn\\&=&(\frac{1}{8\pi^2 i})^{m+n}\prod_{j=1}^{m+n} \int d\omega_j e^{-i\sigma_j\omega_j u_j} (2\pi)^4\delta^{(4)}(\sum_{j=1}^{m+n} p_j)i\mathcal{M}(p_1,p_2,\cdots,p_{m+n}).\nn
\eea 

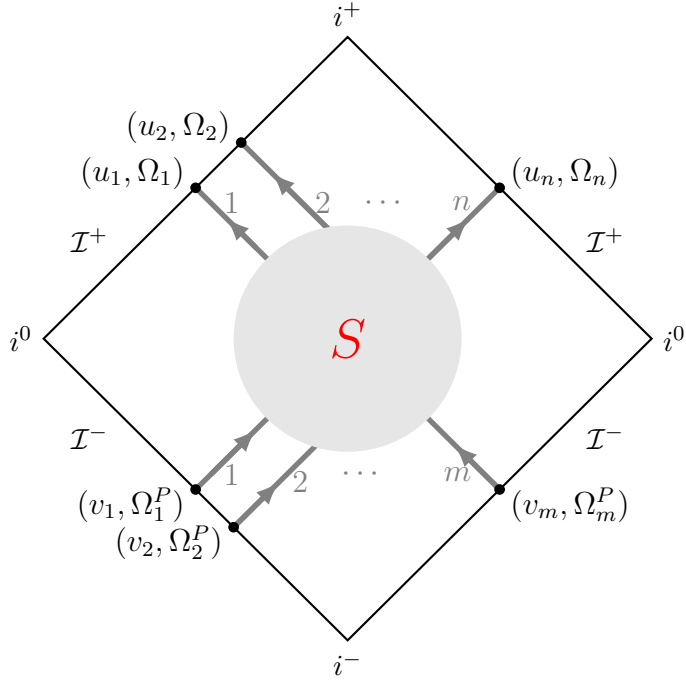
\begin{figure}[h]
  \centering
  \hspace{0.8cm}
  \begin{tikzpicture}
    \filldraw[fill=white,draw,thick] (0,4) node[above]{\small $i^+$} -- (3,1) node[above right]{\small $\mathcal{I}^+$} -- (4,0) node[right]{\small $i^0$}  -- (3,-1) node[below right]{\small $\mathcal{I}^-$} -- (0,-4) node[below]{\small $i^-$} -- (-3,-1) node[below left]{\small $\mathcal{I}^-$} -- (-4,0)node[left]{\small $i^0$} -- (-3,1) node[above left]{\small $\mathcal{I}^+$} -- cycle;
    \draw[-latex,black!50,line width=2pt] (-2,-2) -- (-1.8,-1.8) node[below,right] {$1$} -- (-1.2,-1.2);
    \draw[black!50,line width=2pt] (-1.8,-1.8)  -- (-1.06,-1.06);
    
    \draw[black!50,line width=2pt] (-1.06,-1.06) -- (1.06,1.06);  
    \draw[-latex,black!50,line width=2pt] (1.06,1.06)--(1.6,1.6);
    \draw[black!50,line width=2pt] (1.06,1.06) --(1.8,1.8) node[left] {$n$} -- (2,2);
    
    \draw[-latex,black!50,line width=2pt] (-1.2,1.2)-- (-1.6,1.6); 
    \draw[black!50,line width=2pt] (-2,2)--  (-1.8,1.8) node[right] {$1$} -- (-1.06,1.06); 
    
     \draw[black!50,line width=2pt] (-1.06,1.06) -- (1.06,-1.06);
     \draw[black!50,line width=2pt]  (2,-2)--(1.8,-1.8) node[left] {$m$}--(1.06,-1.06) ;
     \draw[-latex,black!50,line width=2pt]  (2,-2)--(1.4,-1.4)  ;
     
     \draw[black!50,line width=2pt] (-1.5,-2.5) -- (-0.9,-1.9) node[below,right] {$2$} -- (-0.4,-1.4);
     \draw[-latex,black!50,line width=2pt] (-1.5,-2.5) -- (-0.9,-1.9);
     
    \draw[black!50,line width=2pt] (-1.4,2.6) -- (-0.6,1.8) node[right] {$2$} -- (0.4,0.8);
     \draw[-latex,black!50,line width=2pt]  (0.4,0.8) -- (-1,2.2);
    \node[right] at (2.0,-2.2)  {$(v_m,\Omega_m^P)$};
    \node[right] at (2.0,2.2)  {$(u_n,\Omega_n)$};
    \node[left] at (-2.0,-2.2)  {$(v_1,\Omega_1^P)$};
    \node[left] at (-2.0,2.2)  {$(u_1,\Omega_1)$};
    \node[left] at (-1.5,-2.7)  {$(v_2,\Omega_2^P)$};
    \node[left] at (-1.4,2.8)  {$(u_2,\Omega_2)$};
    \fill (-2,-2) circle (2pt);
    \fill (2,2) circle (2pt);
    \fill (-2,2) circle (2pt);
    \fill (2,-2) circle (2pt);
    \fill (-1.5,-2.5) circle (2pt);
    \fill (-1.4,2.6) circle (2pt);
    \node[black!50] at (0.2,-1.8) {$\cdots$};
    \node[black!50] at (0.5,1.8) {$\cdots$};
    \fill[gray!20] (0,0) circle (1.5cm);
    \node at (0,0) {\LARGE{{\color{red} $S$}}};
  \end{tikzpicture}
  \caption{Carrollian amplitude in Penrose diagram. There are $m$ operators inserted at $\mathcal{I}^-$ which correspond to the incoming states in the past and $n$ operators inserted at $\mathcal{I}^+$ which correspond to the outgoing states in the future.  The overlap between the in and out states  is shown in the shaded region, which  corresponds to the $S$ matrix due to bulk interactions. Carrollian amplitude is actually the  scattering amplitude in Carrollian space. The operator $\Xi$ located at $(v,\Omega^P)$ is identified as an operator $\Sigma$  with position $(u,\Omega)$. With the antipodal map, the Carrollian amplitude is written as the correlator at the null boundary. }
  \label{Carramp}
\end{figure}

We will also write this Carrollian amplitude as $(m+n)$-point correlator  for the boundary Carrollian field theory \cite{Donnay:2022wvx,Mason:2023mti}
\bea &&\langle \prod_{j=1}^{m+n}\Sigma_j(u_j,\Omega_j;\sigma_j)\rangle=(\frac{1}{8\pi^2 i})^{m+n}\prod_{j=1}^{m+n} \int d\omega_j e^{-i\sigma_j\omega_j u_j} (2\pi)^4\delta^{(4)}(\sum_{j=1}^{m+n} p_j)i\mathcal{M}(p_1,p_2,\cdots,p_{m+n}).\nn\\\label{Carrollianamp}
\eea 
On the left-hand side, we add a label $\sigma_j$ for each operator to denote the incoming  or outgoing state. The Carrollian amplitude \eqref{Carrollianamp} is a function in the Carrollian space
\bea C(u_1,\Omega_1,\sigma_1;u_2,\Omega_2,\sigma_2;\cdots;u_{m+n},\Omega_{m+n},\sigma_{m+n})\equiv \langle \prod_{j=1}^{m+n}\Sigma_j(u_j,\Omega_j;\sigma_j)\rangle.\eea 
Without causing confusion, we may abbreviate the $m\to n$ Carrollian amplitude as 
\bea 
C(m\to n)=C(u_1,\Omega_1,\sigma_1;u_2,\Omega_2,\sigma_2;\cdots;u_{m+n},\Omega_{m+n},\sigma_{m+n}).
\eea 
In figure \ref{Carramp}, we draw the picture of $m\to n$ Carrollian amplitude. 

\paragraph{Unitarity.} Since the $S$ matrix is unitary, we find 
\bea 
S^\dagger S=1\quad\Rightarrow\quad -i(T-T^\dagger)=T^\dagger T.\label{unitarity}
\eea Usually, we may use momentum representation to express the above identity. Instead, we may consider the scattering from the state $|\prod_{j=1}^m\Sigma(u_j,\Omega_j)\rangle$ to $|\prod_{j=m+1}^{m+n}\Sigma(u_j,\Omega_j)\rangle$. The unitarity condition \eqref{unitarity} becomes 
\bea 
\langle\prod_{j=m+1}^{m+n}\Sigma(u_j,\Omega_j)|(-i)(T-T^\dagger)|\prod_{j=1}^m\Sigma(u_j,\Omega_j)\rangle=\langle \prod_{j=m+1}^{m+n} \Sigma(u_j,\Omega_j)| T^\dagger T |\prod_{j=1}^m \Sigma(u_j,\Omega_j)\rangle.
\eea By inserting the completeness relation \eqref{comp}, we find 
\bea 
C(m\to n)+C^*(n\to m)={ -}\sum_{k}\left(\prod_{j=1}^k 2i\int du_j d\Omega_j\right) C^*(n\to k)\left(\prod_{j=1}^k \frac{\partial}{\partial u_j}\right)C(m\to k).\label{discon}
\eea This is the unitarity of the $S$ matrix in Carrollian space whose momentum space representation may be found in the textbook \cite{1995iqft.book.....P}. Note that the relative sign on the left-hand side is positive since $C$ corresponds to $i\mathcal{M}$ in our notation. 

\section{Poincar\'e transformation}\label{poin}
In this section, we will use two methods to derive the transformation law of the Carrollian amplitude \eqref{Carrollianamp} under Poincar\'e transformations. 
\subsection{Bulk reduction}
In this method, the transformation law of the bulk scalar field is 
\bea 
\Phi'(x')=\Phi(x)
\eea under bulk Poincar\'e transformations\footnote{Actually, this transformation is also valid for general bulk diffeomorphisms.}. The Poincar\'e transformation is the semidirect product of Lorentz transformation and spacetime translation. 
\paragraph{Spacetime translation.} For the spacetime translation 
\bea 
x'^\mu=x^\mu+a^\mu
\eea parameterized by a constant vector $a^\mu=(a^0,a^i)$, the coordinate $r$ becomes
\bea 
r'=\sqrt{\bm x'^2}=\sqrt{(\bm x+\bm a)^2}=r\sqrt{1+2\bm n\cdot{\frac{\bm a}{r}+(\bm n\cdot\frac{\bm a}{r})^2}}
\eea where 
\bea 
\bm n=\frac{\bm x}{r}=(\sin\theta\cos\phi,\sin\theta\sin\phi,\cos\theta)
\eea is the unit normal vector of $S^2$. At large distances, we may find the following expansion of $r'$
\bea 
r'=r+\bm n\cdot \bm a+\mathcal{O}(r^{-1}).
\eea Similarly, the retarded and advanced time transform as 
\bs\begin{align}
u'&=t'-r'=u-a\cdot n+\mathcal{O}(r^{-1}),\\
v'&=t'+r'=v+a\cdot\bar{n}+\mathcal{O}(r^{-1})
\end{align}\es where 
\bea 
n^\mu=(1,n^i),\quad \bar n^\mu=(-1,n^i)
\eea are null vectors and their scalar product is $n\cdot\bar n=2$. The transformation of the spherical angle $\Omega'=(\theta',\phi')$ may be equivalently described by the transformation of the unit vector $\bm n\to \bm n'$
\bea \bm n'=\bm n+\mathcal{O}(r^{-1}).
\eea Therefore, the spherical angle is invariant under spacetime translation in the large $r$ expansion
\bea 
\Omega'=\Omega+\mathcal{O}(r^{-1}).
\eea Combining with fall-off condition \eqref{asympexp}, we obtain the following spacetime  translation of the boundary field
\bs\begin{align}
\Sigma'(u',\Omega')&=\Sigma(u,\Omega),\quad u'=u-a\cdot n,\quad \Omega'=\Omega,\\
\Xi'(v',\Omega')&=\Xi(v,\Omega),\quad v'=v+a\cdot\bar n,\quad \Omega'=\Omega.
\end{align}\es
The spacetime translation can also induce the transformation of the state 
\bs\begin{align}
|\Sigma(u,\Omega)\rangle&\to |\Sigma'(u',\Omega')\rangle=|\Sigma(u,\Omega)\rangle\quad\Leftrightarrow\quad |\Sigma'(u,\Omega)\rangle=|\Sigma(u+a\cdot n,\Omega)\rangle,\\
|\Xi(v,\Omega)\rangle&\to|\Xi'(v',\Omega')\rangle=|\Xi(v,\Omega)\rangle\quad\Leftrightarrow\quad |\Xi'(v,\Omega)\rangle=|\Xi(v-a\cdot\bar{n},\Omega)\rangle.
\end{align}\es 

\paragraph{Lorentz transformation.} Similarly, for the Lorentz transformation 
\bea 
x'^\mu=\Lambda^\mu_{\ \nu}x^\nu, \qquad \Lambda^\mu_{\ \nu}\Lambda^{\rho}_{\ \sigma}\eta_{\mu\rho}=\eta_{\nu\sigma},
\eea we find 
\bs\begin{align}
    r'&=r |\Lambda^i_{\ \mu}n^\mu|+u \frac{\Lambda^i_{\ \mu}n^\mu \Lambda^i_{\ 0}}{|\Lambda^i_{\ \nu}n^\nu|}+\mathcal{O}(r^{-1}),\\
    u'&=u (\Lambda^0_{\ 0}-\frac{\Lambda^i_{\ \mu} n^\mu\Lambda^i_{0}}{|\Lambda^i_\nu n^\nu|})+\mathcal{O}(r^{-1}),\\
    n'^i&=\frac{\Lambda^i_{\ \mu}n^\mu}{|\Lambda^i_{\ \nu}n^\nu|}+\mathcal{O}(r^{-1})
\end{align}\es for the retarded coordinates near $\mathcal{I}^+$ and 
\bs\begin{align}
    r'&=r |\Lambda^i_{\ \mu}\bar n^\mu|+v \frac{\Lambda^i_{\ \mu}\bar n^\mu\Lambda^i_{\ 0}}{|\Lambda^i_{\ \nu}\bar n^\nu|}+\mathcal{O}(r^{-1}),\\
    v'&=v(\Lambda^0_{\ 0}+\frac{\Lambda^i_{\ \mu} \bar n^\mu\Lambda^i_{0}}{|\Lambda^i_\nu \bar n^\nu|})+\mathcal{O}(r^{-1}),\\
    n'^i&=\frac{\Lambda^i_{\ \mu}\bar n^\mu}{|\Lambda^i_{\ \nu}\bar n^\nu|}+\mathcal{O}(r^{-1})
\end{align}\es for the advanced coordinates near $\mathcal{I}^-$.
We have defined the following norms in these expressions
\bea 
|\Lambda^i_{\ \mu}n^{\mu}|=(\Lambda^i_{\ \mu}\Lambda^i_{\ \nu}n^\mu n^\nu)^{1/2},\quad |\Lambda^i_{\ \mu}\bar n^\mu|=(\Lambda^i_{\ \mu}\Lambda^i_{\ \nu}\bar n^\mu\bar n^\nu)^{1/2}.
\eea 
Using the identities 
\bs\begin{align}
&|\Lambda^i_{\ \mu}n^\mu|=\Lambda^0_{\ \mu}n^\mu,\quad  |\Lambda^i_{\ \mu}n^\mu|(\Lambda^0_{\ 0}-\frac{\Lambda^i_{\ \rho} n^\rho\Lambda^i_{0}}{|\Lambda^i_\nu n^\nu|})=1,\\
&|\Lambda^{ i}_{\ \mu}\bar n^\mu|=-\Lambda^0_{\ \mu}\bar n^\mu,\quad |\Lambda^i_{\ \mu}\bar n^\mu|(\Lambda^0_{\ 0}+\frac{\Lambda^i_{\ \rho} \bar n^\rho\Lambda^i_{0}}{|\Lambda^i_\nu \bar n^\nu|})=1,
\end{align}\es the transformations of the coordinates become
\bs\label{lorentzu}\begin{align}
    r'&=\Lambda^0_{\ \mu}n^\mu r+\mathcal{O}(1)\\
    u'&=\frac{1}{\Lambda^0_{\ \mu}n^\mu}u+\mathcal{O}(r^{-1}),\\
    n'^i&=\frac{\Lambda^i_{\ \mu}n^\mu}{|\Lambda^i_{\ \nu}n^\nu|}+\mathcal{O}(r^{-1})
\end{align}\es near $\mathcal{I}^+$ and 
\bs\begin{align}
    r'&=-\Lambda^0_{\ \mu}\bar n^\mu r+\mathcal{O}(1),\\
    v'&=-\frac{1}{\Lambda^0_{\ \mu}\bar n^\mu} v+\mathcal{O}(r^{-1}),\\
    n'^i&=\frac{\Lambda^i_{\ \mu}\bar n^\mu}{|\Lambda^i_{\ \nu}\bar n^\nu|}+\mathcal{O}(r^{-1})
\end{align}\es  near $\mathcal{I}^-$. The Lorentz transformation can be decomposed into spatial rotation and Lorentz boost. We will discuss these two cases explicitly in the following.
\begin{enumerate}
    \item  For a general spatial rotation around an axis \(\bm{\ell}\) $(\bm\ell^2=1)$ with any angle \(\varphi\), one can find the transformation matrix (Rodrigues' rotation formula \cite{linear})
\begin{align}
  \Lambda_{ij}(\bm{\ell}, \varphi) =  \delta_{ij}\cos \varphi + \ell_i\ell_j(1 - \cos \varphi) -  \epsilon_{ijk} \ell^k\sin \varphi.
\end{align}
Besides spatial components, there are other components related to time direction
\begin{align}
  \Lambda^0{}_{j}(\bm{\ell}, \varphi)=\Lambda^j{}_{0}(\bm{\ell}, \varphi)=0,\qquad \Lambda^0{}_{0}(\bm{\ell}, \varphi)=1.
\end{align}
Under such a rotation, we find
\begin{align}
  &u'=u,\qquad r'=r,\\ 
  &n'^i=n^i\cos \varphi+\bm \ell\cdot\bm n    \ \ell^i(1 - \cos \varphi)+(\bm\ell\times\bm n)^i \sin \varphi.
\end{align} 

One can calculate the factor
\begin{align}
  &\Gamma(\bm{\ell}, \varphi)\equiv \Lambda^0_{\ \mu}(\bm{\ell}, \varphi)n^\mu= 1,
\end{align}
and
\begin{align}
    &\Gamma^i(\bm{\ell}, \varphi)\equiv \Lambda^i_{\ \mu}(\bm{\ell}, \varphi)n^\mu=n^i\cos \varphi+\bm \ell\cdot\bm n\ \ell^i(1 - \cos \varphi)+ (\bm\ell\times\bm n)^i \sin \varphi.
\end{align}
It is easy to find that $\Gamma^i\Gamma_i$ equals $\Gamma^2$. With such a definition, the coordinate transformation becomes
  \begin{align}
    &r'= r+\mathcal{O}(1),\qquad u'=u+\mathcal{O}(r^{-1}),\qquad \bm n'=\bm\Gamma+\mathcal{O}(r^{-1}).
 \end{align}
 
 For advanced coordinates, it is easy to compute the Weyl factor of such a rotation
 \begin{align}
  &\bar{\Gamma}(\bm\ell, \varphi)=-\Lambda^0_{\ \mu}(\bm\ell, \varphi)\bar n^\mu=1,\\ 
  &\bar{\Gamma}^i(\bm\ell, \varphi)=\Lambda^i_{\ \mu}(\bm\ell, \varphi)\bar n^\mu=n^i\cos \varphi+\bm\ell\cdot\bm n \ell^i(1 - \cos \varphi)+(\bm \ell\times\bm n)^i \sin \varphi,
\end{align}
and therefore 
\begin{align}
  v'=v,\qquad r'=r,\qquad n'^i=n^i\cos \varphi+\bm\ell\cdot\bm n \ell^i(1 - \cos \varphi)+(\bm \ell\times\bm n)^i \sin \varphi.
\end{align}

    \item 
For a Lorentz boost which is parameterized by a velocity $\bm \beta$,
\bs\begin{align}
t'&=\gamma(t-\bm \beta\cdot\bm r),\\
\bm r'&=\bm r+(\gamma-1)\frac{\bm \beta\cdot\bm r}{\beta^2}\bm \beta-\gamma {\bm \beta}t,\end{align}\es
the transformation \eqref{lorentzu} is reduced to
\bs\begin{align}
u'&=\frac{u}{\gamma(1-\bm \beta\cdot\bm n)}+\mathcal{O}(r^{-1}),\\
r'&=\gamma(1-\bm \beta\cdot\bm n)r+\mathcal{O}(1),\\
\bm n'&=\frac{\bm n+(\gamma-1)\frac{\bm \beta\cdot\bm n}{\beta^2}\bm \beta-\gamma\bm \beta}{\gamma(1-\bm \beta\cdot\bm n)}+\mathcal{O}(r^{-1})
\end{align}\es which is consistent with \cite{Hollands:2016oma,Compere:2019gft}. Note that  $\gamma$ is the Lorentz factor 
\bea 
\gamma=\frac{1}{\sqrt{1-\beta^2}}
\eea and we may use it to define the redshift factor
\bea 
\Gamma\equiv \gamma(1-\bm\beta\cdot\bm n)
\eea  for a light propagating in the direction $\bm n$ and detected by a moving observer with constant velocity $\bm\beta$. Please find more details in Appendix \ref{doppler}.  
\end{enumerate} 
Based on  the previous discussion, we can define a redshift factor 
\bea 
\Gamma=\Lambda^0_{\ \mu}n^\mu
\eea associated with any Lorentz transformation. We will also use the notation 
\bea 
\Gamma^i=\Lambda^i_{\ \mu}n^\mu
\eea whose norm is the redshift factor 
\bea 
\Gamma=|\Gamma^i|=\sqrt{\bm\Gamma^2}.
\eea 
There is also a similar redshift factor at  $\mathcal{I}^-$
\bea 
\bar{\Gamma}=-\Lambda^0_{\ \mu}\bar n^\mu,\qquad \bar{\Gamma}^i=\Lambda^i_{\ \mu}\bar n^\mu.
\eea 
Recalling the fall-off condition \eqref{asympexp}, we can find the finite transformation of the boundary field under general Lorentz transformations. More explicitly, the fundamental field at $\mathcal{I}^+$ transforms as 
\be
\Sigma'(u',\Omega')=\Gamma\Sigma(u,\Omega)\label{transSigma}
\ee with 
\bea 
u'=\Gamma^{-1}u,\quad \bm n'=\Gamma^{-1}\bm\Gamma.
\eea 
The state $|\Sigma(u,\Omega\rangle)$ transforms to another state $|\Sigma'(u',\Omega')\rangle$
\bea 
|\Sigma'(u',\Omega')\rangle=\Gamma|\Sigma(u,\Omega)\rangle.
\eea 
In a similar way, the fundamental field at $\mathcal{I}^-$ transforms as 
\be
    \Xi'(v',\Omega')=\bar{\Gamma}\Xi(v,\Omega)\label{transXi}
\ee with 
\bea
v'=\bar{\Gamma}^{-1}v,\quad \bm n'=\bar{\Gamma}^{-1}\bar{\bm\Gamma}.
\eea We also find the transformation of the state 
\bea 
|\Xi'(v',\Omega')\rangle=\bar{\Gamma}|\Xi(v,\Omega)\rangle.
\eea 
\subsection{Intrinsic derivation}
The transformation laws \eqref{transSigma} and \eqref{transXi} can also be obtained  in an intrinsic way from boundary Carrollian field theory. A Carrollian field $\Sigma(u,\Omega)$ with  weight $1/2$ will transform as  \cite{Li:2023xrr}
\bea 
-\delta_{f,Y}\Sigma(u,\Omega)=f(u,\Omega)\dot{\Sigma}(u,\Omega)+Y^A(\Omega)\nabla_A\Sigma(u,\Omega)+\frac{1}{2}\nabla_AY^A(\Omega)\Sigma(u,\Omega)\label{carrovar}
\eea 
under infinitesimal Carrollian diffeomorphism which is generated by the vector
\be 
\bm\xi_{f,Y}=f(u,\Omega)\partial_u+Y^A(\Omega)\partial_A.
\ee For a general supertranslation which is generated by 
\be 
\bm\xi_f=f(u,\Omega)\partial_u, 
\ee we can find the finite transformation of the field $\Sigma$ 
\bea 
\Sigma'(u',\Omega')=\Sigma(u,\Omega)
\eea where 
\bea 
u'=\mathcal{F}(u,\Omega),\qquad \Omega'=\Omega.
\eea The function $\mathcal{F}(u,\Omega)$ is generated by the vector $\bm\xi_f$ through the exponential map 
\bea 
\mathcal{F}(u,\Omega)=e^{f(u,\Omega)\partial_u}u
\eea such that the infinitesimal variation of the coordinate $u$ is 
\bea 
\delta u=u'-u=\epsilon f(u,\Omega)+\cdots
\eea where $\epsilon$ is a bookkeeping factor. 

For a special superrotation which is generated by 
\be 
\bm\xi_Y=Y^A(\Omega)\partial_A,
\ee the finite transformation of the field $\Sigma$ has been found in \cite{Li:2023xrr}
\be 
\Sigma(u,\Omega)\to \Sigma'(u',\Omega')=\Big|\frac{\partial \Omega'}{\partial\Omega}\Big|^{-1/2}\left(\frac{\det\gamma(\Omega)}{\det\gamma(\Omega')}\right)^{1/4}\Sigma(u,\Omega)\,\,,\label{weight}
\ee where $|\frac{\partial \Omega'}{\partial\Omega}|$ is the Jacobian under finite special superrotation
\be 
(u,\Omega)\to(u',\Omega')\label{sr}
\ee with
\bea 
u'=u,\quad \Omega'=\Omega'(\Omega).
\eea 
We will also denote the pre-factor as the  Weyl factor
\bea 
W=\Big|\frac{\partial \Omega'}{\partial\Omega}\Big|^{-1/2}\left(\frac{\det\gamma(\Omega)}{\det\gamma(\Omega')}\right)^{1/4}.\label{weyldef}
\eea Under the infinitesimal transformation 
\bea \theta'^A=\theta^A+\epsilon Y^A,
\eea we find 
\bea 
W&=&\Big|\delta^A_B+\epsilon \partial_BY^A\Big|^{-1/2}\left(\frac{\det\gamma}{\det\gamma(1+2\epsilon \Gamma_{CA}^{C}Y^A)}\right)^{1/4}\nn\\&=&1-\frac{1}{2}\epsilon\partial_AY^A-\frac{1}{2}\epsilon\Gamma^C_{CA}Y^A\nn\\&=&1-\frac{1}{2}\epsilon\nabla_AY^A.\label{Weylinf}
\eea 

The bulk Lorentz transformation reduces to the Carrollian diffeomorphism $\bm\xi_{f,Y}$ with 
\bea 
f(u,\Omega)=\frac{1}{2}u \nabla_AY^A(\Omega)
\eea where $Y^A(\Omega)$ is a conformal Killing vector (CKV) of $S^2$ which satisfies the conformal Killing equation 
\bea 
\nabla_AY_B+\nabla_BY_A=\gamma_{AB}\nabla_CY^C.
\eea Therefore, the finite Lorentz transformation of $\Sigma$ is 
\bea 
\Sigma'(u',\Omega')=W\Sigma(u,\Omega)\label{transSigma2}
\eea where 
\bea 
u'=W^{-1} u,\quad \Omega'=\Omega'(\Omega).
\eea To check this point, we calculate the infinitesimal variation of the scalar field 
\bea 
\delta_\epsilon\Sigma(u,\Omega)&=&\Sigma'(u,\Omega)-\Sigma(u,\Omega)\nn\\&\approx&W\Sigma(Wu,\Omega-\epsilon Y)-\Sigma(u,\Omega)\nn\\&\approx&(1-\frac{1}{2}\epsilon\nabla_{A}Y^A)\Sigma(u-\frac{u}{2}{ \epsilon}\nabla_CY^C,\Omega-\epsilon Y)-\Sigma(u,\Omega)\nn\\&\approx&-\frac{1}{2}\epsilon u\nabla_CY^C\dot{\Sigma}-\epsilon Y^A\nabla_A\Sigma-\frac{1}{2}\epsilon\nabla_CY^C \Sigma. \label{3.61}
\eea This is exactly the variation \eqref{carrovar} with $f=\frac{1}{2}u\nabla_AY^A$.

\subsection{Transformation of the Carrollian amplitude}

\paragraph{Spacetime translation.}
As has been shown, the finite spacetime translation is 
\be 
u'=u-a\cdot n,\quad \Omega'=\Omega
\ee in retarded coordinates. Therefore, we can write down the identity for the Carrollian amplitude under finite spacetime translation\footnote{We use the short notation $\Sigma(u,\Omega)=\Sigma(u,\Omega;\sigma)$ unless it is necessary to distinguish between incoming and outgoing states.}
\bea 
\langle\prod_{j=1}^n\Sigma_j(u'_j,\Omega_j)\rangle=\langle\prod_{j=1}^n\Sigma_j(u_j,\Omega_j)\rangle,\quad u_j'=u_j-a\cdot n_j.\label{sinv}
\eea 
To prove this identity, we note that the $\mathcal{M}$ matrix is invariant under spacetime translation, and recall the relation between the $\mathcal{M}$ matrix and Carrollian amplitude \eqref{Carrollianamp}, then
\bea 
\langle \prod_{j=1}^{n}\Sigma(u'_j,\Omega_j)\rangle&=&(\frac{1}{8\pi^2 i})^{n}\prod_{j=1}^n \int d\omega_j e^{-i\sigma_j\omega_ju_j'}(2\pi)^4\delta^{(4)}(\sum_{j=1}^n p_j)i\mathcal{M}(p_1,\cdots,p_n)\nn\\&=&(\frac{1}{8\pi^2 i})^{n}\prod_{j=1}^n \int d\omega_j e^{-i\sigma_j\omega_j(u_j-a\cdot n_j)}(2\pi)^4\delta^{(4)}(\sum_{j=1}^n p_j)i\mathcal{M}(p_1,\cdots,p_n)\nn\\&=&(\frac{1}{8\pi^2 i})^{n}\prod_{j=1}^n \int d\omega_j e^{-i\sigma_j\omega_ju_j+ia\cdot p_j}(2\pi)^4\delta^{(4)}(\sum_{j=1}^n p_j)i\mathcal{M}(p_1,\cdots,p_n)
\eea At the last step, we have used the definition of the null momentum \eqref{momentumnull}. Due to the conservation of the external momentum 
\be 
\sum_{j=1}^n p_j=0, 
\ee the phase factors $e^{ia\cdot p_j}$ are canceled and the Carrollian amplitude is invariant under spacetime translation 
\bea 
\langle \prod_{j=1}^{n}\Sigma(u'_j,\Omega_j)\rangle&=&\langle \prod_{j=1}^n\Sigma(u_j,\Omega_j)\rangle=\langle \prod_{j=1}^n\Sigma'(u'_j,\Omega_j)\rangle.
\eea 

\paragraph{Lorentz transformation.}
Note that the form of the transformation \eqref{transSigma} is exactly the same as \eqref{transSigma2}, provided the following matching condition
\bea 
\Gamma=W.
\eea Since both of the redshift factor and the Weyl factor are generated by Lorentz transformations, we only need to find the correspondence between the infinitesimal transformations. When $Y^A$ is a Killing vector of $S^2$, it will obey the equation $\nabla_AY^A=0$ which means that the Weyl factor is always 1. This matches with the redshift factor for pure spatial rotations. Now we will focus on the Lorentz boost. It is generated by a strictly CKV and the corresponding redshift factor is 
\bea 
\Gamma\approx 1-\bm\beta\cdot\bm n
\eea for infinitesimal $\bm\beta$. This is the same as the Weyl factor \eqref{Weylinf} with the identification 
\bea 
\bm\beta\cdot\bm n=\epsilon\nabla_AY^A. \label{identification}
\eea By choosing $\bm\beta$ along the $i$-th direction 
\bea 
\bm\beta\cdot\bm n=2\epsilon n_i,
\eea the equation \eqref{identification} is satisfied by the identity 
\bea 
2n_i=\nabla_AY^A_i
\eea where $Y^A_i$ is the $i$-th  strictly CKV defined in \cite{Liu:2022mne}. 

\paragraph{Stereographic coordinates.} The redshift (Weyl) factor can be written down explicitly by stereographic coordinates $(z,\bar z)$ which is related to the spherical coordinates by 
\be 
z=\cot\frac{\theta}{2}e^{i\phi},\quad \bar z=\cot\frac{\theta}{2}e^{-i\phi}.
\ee The corresponding metric would be 
\bea 
ds^2_{S^2}=\frac{4}{(1+z\bar z)^2}dz d\bar z.
\eea The Lorentz transformation does not change the metric of $S^2$ at the null boundary, except that the stereographic coordinates become $(z',\bar{z}')$. It is well-known that the Lorentz transformation $SO(1,3)$ induces a M\"{o}bius transformation $SL(2,\mathbb{C})$ on $S^2$ 
\bea 
z'=\frac{az+b}{cz+d},\quad ad-bc=1.
\eea Substituting into the definition of Weyl factor \eqref{weyldef}, we find 
\bea 
\Gamma=W=\frac{|az+b|^2+|cz+d|^2}{1+z\bar z}=\frac{1+|z'|^2}{|a-cz'|^2+|b-dz'|^2}=\frac{(1+|z'|^2)|cz+d|^2}{1+|z|^2}.
\eea 
We will calculate the redshift factor for three special M\"{o}bius transformations below.
\begin{enumerate}
    \item The transformation from $z=0$ to $z'$. The corresponding redshift factor is 
    \bea 
    \Gamma=|d|^2(1+|z'|^2).
    \eea 
    \item The transformation from $z=1$ to $z'$. The corresponding redshift factor is 
    \bea 
    \Gamma=\frac{1}{2}|c+d|^2(1+|z'|^2).
    \eea 
    \item The transformation from $z=\infty$ to $z'$. The corresponding redshift factor is 
    \bea
    \Gamma=|c|^2(1+|z'|^2).
    \eea 
\end{enumerate}

\paragraph{Transformation law of Carrollian amplitudes.} Now we can prove the following identity for the Carrollian amplitude under finite Lorentz transformation
\bea 
\langle\prod_{j=1}^n\Sigma_j(u'_j,\Omega'_j)\rangle=\left(\prod_{j=1}^n \Gamma_j\right)\langle\prod_{j=1}^n\Sigma_j(u_j,\Omega_j)\rangle.\label{carrollianlorentz}
\eea Using stereographic coordinates, this is 
\bea 
\langle\prod_{j=1}^n\Sigma_j(u'_j,z'_j,\bar{z}_j')\rangle=\left(\prod_{j=1}^n \Gamma_j\right)\langle\prod_{j=1}^n\Sigma_j(u_j,z_j,\bar{z}_j)\rangle \label{rinv}
\eea where 
\bea 
u_j'=\Gamma_j^{-1}u_j,\quad z_j'=\frac{az_j+b}{cz_j+d},\quad \Gamma_j=\frac{|az_j+b|^2+|cz_j+d|^2}{1+z_j\bar z_j}.\label{Lorentzrotation}
\eea 
 Note that the momentum $\bm p=(\omega,\Omega)$ is transformed to a new one
 \bea 
 p'^\mu=\Lambda^\mu_{\ \nu}p^\nu
 \eea under Lorentz transformation. This is equivalent to 
 \bea 
\bm p'=(\omega',\Omega'),\quad \text{with}\quad \omega'=\omega\Lambda^0_{\ \mu} n^\mu=\Gamma \omega,\quad \Omega'=\Omega'(\Omega).
 \eea Using \eqref{Carrollianamp}, we find 
 \bea 
 \langle\prod_{j=1}^n\Sigma(u_j',\Omega_j')\rangle&=&(\frac{1}{8\pi^2 i})^{n}\prod_{j=1}^n \int d\omega_j e^{-i\sigma_j\omega_ju_j'}(2\pi)^4\delta^{(4)}(\sum_{j=1}^n \tilde{p}_j)i\mathcal{M}(\tilde{p}_1,\cdots,\tilde{p}_n),\label{sigmadef}
 \eea where $\tilde{p}_j$ is the four-momentum associated with the three momentum
 \bea 
 \tilde{\bm p}_j=(\omega_j,\Omega_j')
 \eea Note that $\tilde{\bm p}_j\not=\bm p_j$ since we have just replaced the coordinates $(u_j,\Omega_j)$ in \eqref{sigmadef}. Now we use the relation $u'_j=\Gamma_j^{-1}u_j$ , change the integration variable $\omega_j$ to $\omega_j'=\Gamma_j\omega_j$, and combine with the invariance of the $\mathcal{M}$ matrix under Lorentz transformations, then
 \bea 
 \langle\prod_{j=1}^n\Sigma(u_j',\Omega_j')\rangle&=&(\frac{1}{8\pi^2 i})^{n}\prod_{j=1}^n \int d\omega'_j e^{-i\sigma_j\omega'_ju_j'}(2\pi)^4\delta^{(4)}(\sum_{j=1}^n {p}'_j)i\mathcal{M}({p}'_1,\cdots,{p}'_n)\nn\\&=&(\frac{1}{8\pi^2 i})^{n}\prod_{j=1}^n \int d\omega_j \Gamma_j e^{-i\sigma_j\omega_ju_j}(2\pi)^4\delta^{(4)}(\sum_{j=1}^n {p}_j)i\mathcal{M}({p}_1,\cdots,{p}_n)\nn\\&=&\left(\prod_{j=1}^n \Gamma_j\right) \langle \prod_{j=1}^n \Sigma(u_j,\Omega_j)\rangle.
 \eea 

\section{Feynman rules}\label{feynman}
In this section, we will derive the Feynman rules in Carrollian space to compute the Carrollian amplitude perturbatively. Given \eqref{Carrollianamp}, this is not necessary since we already know the Feynman rules in momentum space to calculate  the  $S$ matrix even though the additional Fourier transform lacks a diagram interpretation. However, as we will show, the Feynman rule in Carrollian space fits nicely with the Fourier transformation in Carrollian amplitude \eqref{Carrollianamp}.  From a more practical point of view, there are some advantages to evaluating Feynman diagrams in configuration space, e.g., GPXT method \cite{Chetyrkin:1980pr}. 
\subsection{Boundary-to-boundary propagator}
\begin{figure}[h]
  \centering
  \begin{tikzpicture}[scale=0.7]
    \filldraw[fill=white,draw,thick] (0,4) node[above]{\small $i^+$} -- (3,1) node[above right]{\small $\mathcal{I}^+$} -- (4,0) node[right]{\small $i^0$}  -- (3,-1) node[below right]{\small $\mathcal{I}^-$} -- (0,-4) node[below]{\small $i^-$} -- (-3,-1) node[below left]{\small $\mathcal{I}^-$} -- (-4,0)node[left]{\small $i^0$} -- (-3,1) node[above left]{\small $\mathcal{I}^+$} -- cycle;
    \coordinate (in) at (-2,-2);
    \coordinate (out) at (2,2);
    \node [below left] at (in) {\footnotesize $(v_1,\Omega_1)$};
    \node [above right] at (out) {\footnotesize $(u_2,\Omega_2)$};
    \draw[black!50,line width=1.5pt] (in) -- (out);
    \node[above] at (-0.4,-1) {\rotatebox{45}{\footnotesize $C(u_1,\Omega^P_1,u_2,\Omega_2)$}};
    \fill (2,2) circle (2pt);
    \fill (-2,-2) circle (2pt);
  \end{tikzpicture}
  \caption{\centering{Boundary-to-boundary propagator in Penrose diagram: from $\mathcal{I}^-$ to $\mathcal{I}^+$}.}
  \label{bb2pt}
\end{figure}
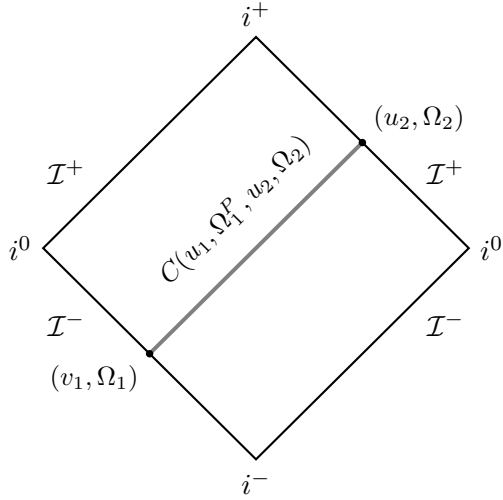
We will start with the two-point  Carrollian amplitude which is shown in figure \ref{bb2pt}. In this diagram, we insert one operator $\Xi(v_1,\Omega_1)$ at $\mathcal{I}^-$ and another operator $\Sigma(u_2,\Omega_2)$ at $\mathcal{I}^+$. Therefore, with the relation \eqref{outstate} and \eqref{invstate}, the two-point Carrollian amplitude would be 
\bea 
{}_{\text{out}}\langle \Sigma(u_2,\Omega_2)|\Xi(v_1,\Omega_1)\rangle_{\text{in}}&=&\left(\frac{1}{8\pi^2 i}\right)^2\int_0^\infty d\omega_2 e^{-i\omega_2u_2}\int_0^\infty d\omega_1 e^{i\omega_1 v_1} \langle \bm p_2|\bm p_1^P\rangle\nn\\&=&-\beta(u_2-v_1)\delta^{}(\Omega^P_1-\Omega_2).\label{bbprp}
\eea In the second line, we used the normalization 
\bea 
\langle\bm p_2|\bm p_1\rangle=(2\pi)^3 2\omega_1 \delta^{(3)}(\bm p_1-\bm p_2)\label{delta3}
\eea and defined the  function $\beta(u)$ as 
\bea 
\beta(u)=\frac{1}{4\pi}\int_0^\infty \frac{d\omega}{\omega}e^{-i\omega (u-i\epsilon)}.\label{beta}
\eea Note that \eqref{bbprp} can be rewritten as 
\bea 
{}_{\text{out}}\langle \Sigma(u_2,\Omega_2)|\Xi(v_1,\Omega^P_1)\rangle_{\text{in}}&=&-\beta(u_2-v_1)\delta^{}(\Omega_1-\Omega_2).
\eea After the replacement
\bea 
\Xi(v_1,\Omega^P_1)\to \Sigma(u_1,\Omega_1),
\eea we obtain the following boundary-to-boundary propagator 
\bea 
C(u_1,\Omega_1,-;u_2,\Omega_2,+)\equiv \langle \Sigma(u_2,\Omega_2)|\Sigma(u_1,\Omega_1)\rangle=-\beta(u_2-u_1)\delta(\Omega_1-\Omega_2).
\eea  Still, this is a boundary-to-boundary propagator from $\mathcal{I}^-$ to $\mathcal{I}^+$, although the coordinates are both retarded coordinates. Similarly, there would be another form which is represented by advanced coordinates. All the expressions are related to each other by the antipodal map.

\paragraph{Regularization.} The $\beta(u)$ function is infrared divergent
which comes from the infrared degrees of 
freedom in the definition of the boundary state
\bea 
|\Sigma(u,\Omega)\rangle=\frac{i}{8\pi^2}\int_0^\infty d\omega e^{i\omega u}|\bm p\rangle.
\eea
We may separate the infrared degrees of freedom by introducing an IR cutoff $\omega_0>0$ by redefining 
\bea 
|\Sigma(u,\Omega;\omega_0)\rangle\equiv\frac{i}{8\pi^2}\int_{\omega_0}^\infty d\omega e^{i\omega u}|\bm p\rangle=\frac{i}{8\pi^2}\int_{0}^\infty d\omega e^{i\omega u}\Theta(\omega-\omega_0)|\bm p\rangle,\label{sigmaomega0}
\eea where we have inserted a step function into the integrand. Therefore, the boundary-to-boundary propagator is modified to 
\bea 
C(u_1,\Omega_1,-,u_2,\Omega_2,+;\omega_0)&\equiv& \langle \Sigma(u_2,\Omega_2;\omega_0)|\Sigma(u_1,\Omega_1;\omega_0)\rangle \nn\\&=&\left(\frac{1}{8\pi^2i}\right)^2\int_0^\infty d\omega_2 e^{-i\omega_2u_2}\Theta(\omega_2-\omega_0)\int_0^\infty d\omega_1 e^{i\omega_1u_1}\Theta(\omega_1-\omega_0)\langle \bm p_2|\bm p_1\rangle\nn\\&=&-\frac{1}{4\pi}\int_0^\infty \frac{d\omega}{\omega}\Theta(\omega-\omega_0)e^{-i\omega(u_2-u_1)}{ \delta(\Omega_1-\Omega_2)}\nn\\&=&\frac{1}{4\pi}\Gamma[0,i\omega_0(u_2-u_1-i\epsilon)]\delta(\Omega_1-\Omega_2)
\eea where $\Gamma(0,x)$ is the incomplete Gamma function which is reviewed in Appendix \ref{gamma}. We have inserted a small positive constant $\epsilon\to 0^+$ to guarantee the convergence of the integration. In the limit $\omega_0\to 0^+$, we find 
\bea 
C(u_1,\Omega_1,-;u_2,\Omega_2,+;\omega_0)&=&-\frac{1}{4\pi}\left(\gamma_E+\log i\omega_0(u_2-u_1-i\epsilon)\right){ \delta(\Omega_1-\Omega_2)}\nn\\&\equiv&-\frac{1}{4\pi}I_0(u_2-u_1){ \delta(\Omega_1-\Omega_2)}.
\eea The Euler constant $\gamma_E$ may be absorbed into the IR cutoff $\omega_0$. Note that the IR cutoff is introduced by regularizing the divergent integral \eqref{beta} in \cite{Liu:2022mne} which is a bit ad hoc. However, the treatment here is to modify the $\Sigma(u,\Omega)$ by $\Sigma(u,\Omega;\omega_0)$ such that the infrared modes are automatically discarded in the definition.  Imagine that one inserts an operator at the position $u$ with some uncertainty $\delta u$. Then by  the Heisenberg uncertainty principle, there will be a lower bound $\omega_0\sim\frac{1}{\delta u}$ on the fluctuation of the energy. It is natural to insert a step function $\Theta(\omega-\omega_0)$ in \eqref{sigmaomega0}. Note that we have defined the function 
\be
I_0(u)=\gamma_E+\log i\omega_0(u-i\epsilon)
\ee which is essentially the regularized $\beta(u)$ function.

\paragraph{Poincar\'e invariance.} The two-point Carrollian amplitude is fixed by Poincar\'e invariance. The spacetime translation invariance implies 
\bea 
C(u_1,\Omega_1;u_2,\Omega_2)=\langle\Sigma(u_2,\Omega_2)\Sigma(u_1,\Omega_1)\rangle=\langle\Sigma(u_2-a\cdot n_2,\Omega_2)\Sigma(u_1-a\cdot n_1,\Omega_1)\rangle
\eea where $a^\mu$ is a constant vector. By choosing $a^\mu=(1,0,0,0)$, we find 
\bea 
C(u_1,\Omega_1;u_2,\Omega_2)=\langle\Sigma(u_2+a,\Omega_2)\Sigma(u_1+a,\Omega_1)\rangle=C(u_1+a;\Omega_1,u_2+a,\Omega_2).
\eea This implies that the two-point Carrollian amplitude only depends on the difference $u_2-u_1$ 
\bea 
C(u_1,\Omega_1;u_2,\Omega_2)&=&C(0,\Omega_1;u_2-u_1,\Omega_2)=C(u_1-a\cdot n_1,\Omega_1;u_2-a\cdot n_2,\Omega_2)\nn\\&=&C(0,\Omega_1;u_2-u_1-a\cdot(n_2-n_1),\Omega_2).
\eea Since the Carrollian amplitude is independent of the arbitrary constant $a$, the two-point Carrollian amplitude is nonvanishing only for $\Omega_1=\Omega_2$, otherwise we have
\bea 
C(u_1,\Omega_1;u_2,\Omega_2)=0,\quad \Omega_1\not=\Omega_2.
\eea This implies 
\bea 
C(u_1,\Omega_1;u_2,\Omega_2)=g(u_1-u_2)\delta(\Omega_1-\Omega_2).
\eea The function $g(u)$ should be independent of $\Omega$, otherwise the two-point Carrollian amplitude would depend on the angular direction.  Now the Lorentz invariance of the Carrollian amplitude  \eqref{carrollianlorentz} can be written as
\bea 
\langle \Sigma(u_2,\Omega_2')\Sigma(u_1,\Omega_1')\rangle=\Gamma_1\Gamma_2\langle\Sigma(\Gamma_2u_2,\Omega_2)\Sigma(\Gamma_1u_1,\Omega_1)\rangle.
\eea This is equivalent to 
\bea 
g(u_1-u_2)\delta(\Omega_1'-\Omega_2')=\Gamma_1^2\ g(\Gamma_1(u_1-u_2))\delta(\Omega_1-\Omega_2)
\eea 
By definition, the Lorentz transformation does not change the metric,
\bea 
\gamma'_{AB}(\Omega)=\gamma_{AB}(\Omega).\label{invgamma}
\eea Therefore, the Lorentz transformation of the Dirac delta function is 
\bea 
\delta(\Omega_1'-\Omega_2')&=&\frac{1}{\sqrt{\det\gamma'(\Omega')}}\delta(\theta'_1-\theta_2')\delta(\phi'_1-\phi_2')=\frac{1}{\sqrt{\det\gamma(\Omega')}\Big|\frac{\partial\Omega'}{\partial\Omega}\Big|}\delta(\theta_1-\theta_2)\delta(\phi_1-\phi_2)\nn\eea which implies \bea \delta(\Omega_1'-\Omega_2')=\Gamma_1^2\delta(\Omega_1-\Omega_2).
\eea The function $g(u)$ should be invariant under Lorentz transformation 
\bea 
g(u)=g(\Gamma u).
\eea This is only possible for 
\bea 
g(u)=\text{const.}
\eea without introducing any energy scale. By introducing an IR cutoff $\omega_0$, the function $g(u)$ may be written as 
\bea 
g(u)= N\times\log\omega_0u\label{functiong}
\eea since the redshift factor $\Gamma$ may be absorbed into the cutoff $\omega_0$, leaving the function $g$ invariant. Here the normalization is denoted as $N$. The result is consistent with the perturbative calculation. 

\paragraph{Primary scalar operators.} We will discuss more on this two-point Carrollian amplitude. Considering a boundary primary scalar operator $V(u,\Omega)$ at the boundary with conformal weight $h$ under $SL(2,\mathbb{C})$, the infinitesimal transformation of the field $V(u,\Omega)$ under Lorentz transformation is 
\bea 
-\delta_Y V(u,\Omega)=\frac{1}{2}u \nabla_CY^C\dot{V}(u,\Omega)+Y^A\nabla_AV(u,\Omega)+h\nabla_CY^C V(u,\Omega)
\eea  whose finite transformation is \cite{Li:2023xrr}
\bea 
V'(u',\Omega')=\Gamma^{2h}V(u,\Omega).
\eea The Carrollian amplitude should satisfy the conditions
\bea 
\langle \prod_{j=1}^n V_j(u_j',\Omega_j)\rangle=\langle \prod_{j=1}^n V_j(u_j,\Omega_j)\rangle
\eea for spacetime translation $u'=u-a\cdot n$ and\bea \langle
\prod_{j=1}^n V_j(u_j',\Omega'_j)\rangle=\left(\prod_{j=1}^n \Gamma_j^{2h_j}\right) \langle \prod_{j=1}^n V_j(u_j,\Omega_j)\rangle\label{lorentzinvariance}
\eea for Lorentz rotation \eqref{Lorentzrotation}. 
In these expressions, the conformal weight of the primary field $V_i$  is $h_i$ under $SL(2,\mathbb{C})$. Similar to the previous discussion, the two-point Carrollian amplitude is fixed to 
\bea 
\langle V_1(u_1,\Omega_1)V_2(u_2,\Omega_2)\rangle=\frac{C_{h_1,h_2}}{(u_1-u_2)^{2(h_1+h_2)-2}}\delta(\Omega_1-\Omega_2)\label{twoptcarroll}
\eea where $C_{h_1,h_2}$ is the normalization constant. Note that $C_{h_1,h_2}$ is not necessarily proportional to $\delta_{h_1,h_2}$. As an example, we consider the previous massless scalar field and define the $u$-descendants
\be V_n(u,\Omega)=\left(\frac{\partial}{\partial u}\right)^n\Sigma(u,\Omega).
\ee It is easy to find its conformal weight 
\be 
h=\frac{1+n}{2}.
\ee Obviously, the correlator 
\bea 
\langle V_n(u,\Omega)V_{n'}(u',\Omega')\rangle
\eea is nonvanishing even for $n\not=n'$.

Note that the two-point Carrollian amplitude would increase for $h_1+h_2<1$ when $|u_1-u_2|$ increases. We may rule out this case since it indicates a strong correlation between two operators which are ``far away'' from each other. For $h_1+h_2=1$, the two-point Carrollian amplitude is also divergent logarithmically. However, we will allow the logarithmic divergence since it appears in previous examples. Therefore, we find a lower bound for the conformal weights 
\be 
h_1+h_2\ge 1
\ee which leads to 
\bea 
h\ge \frac{1}{2}\label{boundh}
\eea when $h_1=h_2=h$. 

Before we close this section, we will emphasize the invariance of the metric \eqref{invgamma}. Note that the invariance  comes from the Lorentz invariance of the Minkowski spacetime in four dimensions,  which contrasts with the intrinsic M\"{o}bius transformation of $S^2$. To see this point, the intrinsic M\"{o}bius transformation of the metric is \bea 
\gamma'_{AB}(\Omega')\frac{\partial\theta'^A}{\partial\theta^C}\frac{\partial\theta'^B}{\partial\theta^D}=\gamma_{CD}(\Omega).\label{Mobiustrans}
\eea The metric $\gamma_{AB}$ is invariant under pure spatial rotations. On the other hand, it is only invariant under conformal transformations up to a scaling factor. This confirms that the transformation law of the metric of the Carrollian manifold is different from the usual coordinate transformation. Actually, the invariance of the metric \eqref{invgamma} has been imposed as an indispensable ingredient to define covariant variation for the theory with nonvanishing helicity \cite{Liu:2023qtr,Liu:2023gwa, Liu:2023jnc}. 
Here we emphasize that the invariance of the metric has already appeared implicitly for scalar theory.

\subsection{External lines}
\begin{figure}[h]
  \centering
  \hspace{0.8cm}
  \begin{tikzpicture}
    \filldraw[fill=white,draw,thick] (0,4) node[above]{\small $i^+$} -- (3,1) node[above right]{\small $\mathcal{I}^+$} -- (4,0) node[right]{\small $i^0$}  -- (2,-2) node[below right]{\small $\mathcal{I}^-$} -- (0,-4) node[below]{\small $i^-$} -- (-2,-2) node[below left]{\small $\mathcal{I}^-$} -- (-4,0)node[left]{\small $i^0$} -- (-2,2) node[above left]{\small $\mathcal{I}^+$} -- cycle;
    \draw[black!50,line width=2pt] (0.5,0.5)  -- (1.8,2.2);
    \node[below left] at (0.5,0.5) {$x$};
    \node[above right] at (1.8,2.2) {$(u,\Omega)$};
    \node[left] at (1.2,1.5) {$D_+(u,\Omega;x)$};
    \fill (0.5,0.5) circle (2pt);
    \fill (1.8,2.2) circle (2pt);
  \end{tikzpicture}
  \caption{External line in Penrose diagram: from bulk to $\mathcal{I}^+$. There is a similar external line from bulk to $\mathcal{I}^-$.}
  \label{extleg}
\end{figure}
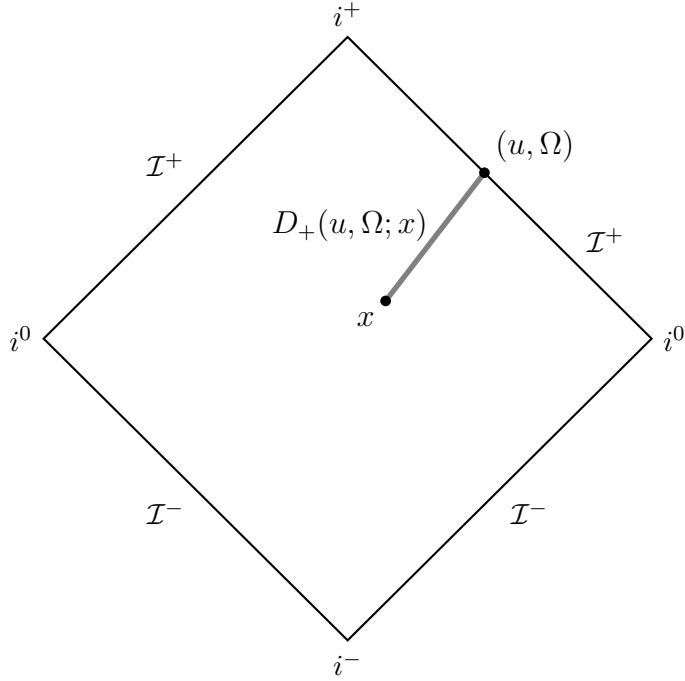
The external line from bulk to $\mathcal{I}^+$ is shown in figure \ref{extleg}. We insert one operator $\Sigma$ at $\mathcal{I}^+$ with position $(u,\Omega)$ and draw a  line to connect it with a bulk field $\Phi$ located at $x$. Then the external line may be defined as 
\bea 
D_+(u,\Omega;x)=\langle0|\Sigma(u,\Omega)\Phi(x)|0\rangle.
\eea With the mode expansion \eqref{expanPhi} and \eqref{outstate}, we can easily obtain 
\bea 
D_+(u,\Omega;x)=-\frac{1}{8\pi^2(u+n\cdot x-i\epsilon)}.\label{el1}
\eea Here the null vector $n^\mu$ is determined by the angular coordinate as before
\be 
n^\mu=(1,\sin\theta\cos\phi,\sin\theta\sin\phi,\cos\theta).
\ee Interestingly, the external line is invariant under spacetime translation
\bea 
u\to u-a\cdot n,\quad x\to x+a.
\eea We may also insert an operator $\Xi$ at $\mathcal{I}^-$ with position $(v,\Omega)$ and connect it with a bulk field $\Phi$ whose position is $x$. Then we define the external line from  $\mathcal{I}^-$ to bulk  
\bea 
D_-(x;v,\Omega)=\langle 0|\Phi(x)\Xi(v,\Omega)|0\rangle=\frac{1}{8\pi^2(v-\bar{n}\cdot x+i\epsilon)}.\label{el2}
\eea This external line is also invariant under spacetime translation
\bea 
v\to v+a\cdot\bar{n},\quad x\to x+a.
\eea 
We still replace $(v,\Omega)$ to $(u,\Omega^P)$ and the external line \eqref{el2} becomes 
\bea 
D_-(x;u,\Omega)=\frac{1}{8\pi^2(u+{n}\cdot x+i\epsilon)}.
\eea 
We find the unified formula 
\bea 
D_{\sigma}(u,\Omega;x)=-\frac{\sigma}{8\pi^2(u+n\cdot x-i\sigma\epsilon)}
\eea where $\sigma$ is $+1$ for outgoing particles and $-1$ for incoming particles. 
The two external lines are related by the complex conjugate
\bea 
D_+(u,\Omega;x)=-\left(D_-(x;u,\Omega)\right)^*.
\eea 

We may also separate the infrared modes as \eqref{sigmaomega0} and define the modified external line
\bea 
D_+(u,\Omega;x;\omega_0)=\langle 0| \Sigma(u,\Omega;\omega_0)\Phi(x)|0\rangle=-\frac{e^{-i\omega_0(u+n\cdot x)}}{8\pi^2(u+n\cdot x-i\epsilon)}.
\eea After inserting the signature $\sigma$ to distinguish outgoing and incoming states, we find 
\bea 
D_\sigma(u,\Omega;x;\omega_0)=-\frac{\sigma e^{-i\sigma\omega_0(u+n\cdot x)}}{8\pi^2(u+n\cdot x-i
\sigma\epsilon)}.
\eea In the following, we may abbreviate $D_{\sigma}$ to $D$ when it causes no confusion. 

\begin{figure}
    \centering
    \begin{tikzpicture}[scale=1.5]
    \draw [-latex] (-2.2, 0) -- (2.2, 0);
    \draw [-latex] (0, -1) -- (0, 2.2);
    \draw (2,1.6)--(1.6,1.6)--(1.6,2);
    \node at(1.8,1.8){$u$};
    \node  at (-0.6, 0.2) {\footnotesize $\times$};
       \node  at (-0.6, 0.4) {\footnotesize $-n\cdot x+i\epsilon$};
     \draw[thick, blue!60] (1.6,0) arc (0:180:1.6);
    \draw[-{Stealth[length=2mm,width=1.5mm]},thick, blue!60] (0.4,0) -- (0.6,0);
    \draw [thick, blue!60] (-1.6, 0)-- (1.6, 0);
    \draw[-{Stealth[length=2mm,width=1.5mm]},thick, blue!60] (1.6,0) arc (0:120:1.6);
    \node[above] at (-0.8,1.5) {$\mathcal{C}$};
  \end{tikzpicture}
    \caption{\centering{The contour $\mathcal{C}$ is the combination of the real $u$ axis and a half circle with radius $R\to\infty$ in the upper complex $u$ plane.}}
    \label{countourux}
\end{figure}
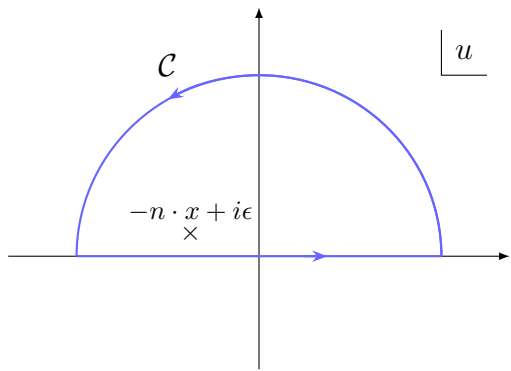

\begin{figure}[h]
  \centering
  \hspace{0.8cm}
  \begin{tikzpicture}
    \filldraw[fill=white,draw,thick] (0,4) node[above]{\small $i^+$} -- (3,1) node[above right]{\small $\mathcal{I}^+$} -- (4,0) node[right]{\small $i^0$}  -- (2,-2) node[below right]{\small $\mathcal{I}^-$} -- (0,-4) node[below]{\small $i^-$} -- (-2,-2) node[below left]{\small $\mathcal{I}^-$} -- (-4,0)node[left]{\small $i^0$} -- (-2,2) node[above left]{\small $\mathcal{I}^+$} -- cycle;
    \draw[dashed,black!50,line width=2pt] (0.5,0.5)  -- (1.8,2.2);
    \node[below left] at (0.5,0.5) {$x$};
    \node[above right] at (1.8,2.2) {$(u,\Omega)$};
    \node[left] at (1.2,1.5) {$K_+(u,\Omega;x)$};
    \fill (0.5,0.5) circle (2pt);
    \fill (1.8,2.2) circle (2pt);
  \end{tikzpicture}
  \caption{Bulk-to-boundary propagator in Penrose diagram: from bulk to $\mathcal{I}^+$. There is a similar bulk-to-boundary propagator from bulk to $\mathcal{I}^-$.}
  \label{bbpK}
\end{figure}
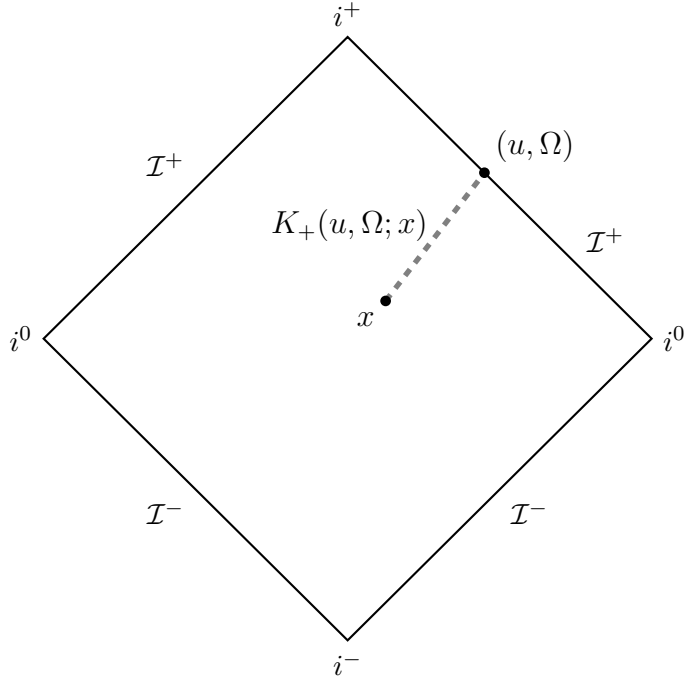

\paragraph{Bulk-to-boundary propagator.} The external line may be used to reconstruct the bulk field. To see this point, we define the state 
\be 
|\Phi(x)\rangle=\Phi(x)|0\rangle.
\ee Then by inserting the completeness relation \eqref{comp} and ignoring all the multi-particle states, we find 
\bea 
|\Phi(x)\rangle&=&-2i\int du d\Omega |\dot{\Sigma}(u,\Omega)\rangle \langle {\Sigma}(u,\Omega)|\Phi(x)\rangle\nn\\&=&-2i\int du d\Omega |\dot\Sigma(u,\Omega)\rangle D(u,\Omega;x).\label{btobulk}
\eea  
For $\sigma=+$, the outgoing state \eqref{outstate} is a superposition of modes 
$ e^{i\omega u}
$ with $\omega>0$. Therefore, the integrand of  \eqref{btobulk} decays exponentially in the upper complex $u$ plane. We may choose a contour $\mathcal{C}$ (seeing  figure \ref{countourux}) to include the pole $u=-n\cdot x+i\epsilon$ and use the residue theorem to evaluate the $u$ integration and thus \bea 
|\Phi(x)\rangle=-\frac{1}{2\pi}\int d\Omega |\dot\Sigma(u=-n\cdot x,\Omega)\rangle.\label{Kirch}
\eea 
This is the Kirchhoff-d'Adh\'emar formula \cite{1986ssv..book.....P} which has also been derived in \cite{Donnay:2022wvx}.  For $\sigma=-$, we find the same formula using the antipodal map and the convention $p^\mu=-\omega n^\mu$ for incoming states. It is obvious that there could be more contributions since we only include the one particle state in the derivation. We will not discuss the corrections in this work. Integration by parts in \eqref{btobulk}, we obtain 
\bea 
|\Phi(x)\rangle=2i\int du d\Omega \partial_u D(u,\Omega;x)|\Sigma(u,\Omega)\rangle.
\eea We define the bulk-to-boundary propagator $K_{\sigma}(u,\Omega;x)$ as
\bea 
 K_{\sigma}(u,\Omega;x)&=&\begin{tikzpicture}[baseline=(current bounding box.center)]
      \fill (0,0) circle (1.5pt);
      \fill (1.5,0) circle (1.5pt);
      \draw[dashed] (0,0) node[below] {$(u,\Omega)$} -- (1.5,0);
      \node at (1.5,-0.1) [below] {$x$};
    \end{tikzpicture}=2i\langle \dot{\Sigma}(u,\Omega)\Phi(x)\rangle=2i\partial_u D(u,\Omega;x)\nn\\
    &=&\frac{i\sigma}{4\pi^2(u+n\cdot x-i\sigma\epsilon)^2}\label{btobsigma}
\eea 
which is shown in figure \ref{bbpK}, and thus we have
\bea 
|\Phi(x)\rangle=\int dud\Omega K_\sigma(u,\Omega;x)|\Sigma(u,\Omega)\rangle.
\eea 

The  bulk-to-boundary propagator $K_{\sigma}(u,\Omega;x)$ is slightly different from the one defined in \cite{Donnay:2022wvx} (denoted by $P(u,\Omega;x)$). This is fine since only the part with creation operators  contributes to the state $|\Phi(x)\rangle$ by definition. On the other hand, the field $\Phi(x)$ in \cite{Donnay:2022wvx} has more contributions from the annihilation operators. However, one should note that the bulk-to-boundary propagator \eqref{btobsigma} leads to the same Kirchhoff-d'Adh\'emar formula \eqref{Kirch}. Actually, the bulk-to-boundary propagator $P(u,\Omega;x)=\frac{1}{2\pi}\partial_u\delta(u+n\cdot x)$ is related to ours by  
\bea 
P(u,\Omega;x)=K_+(u,\Omega;x)+K_-(u,\Omega;x).
\eea To check this point, we use the formula 
\be \frac{1}{A\pm i\epsilon}=\mathcal{P}(\frac{1}{A})\mp i\pi \delta(A)
\ee and rewrite the right-hand side as 
\bea 
\text{RHS}&=&\frac{i}{4\pi^2}[\frac{1}{(u+n\cdot x-i\epsilon)^2}-\frac{1}{(u+n\cdot x+i\epsilon)^2}]\nn\\
&=&-\frac{i}{4\pi^2}\partial_u[\frac{1}{u+n\cdot x-i\epsilon}-\frac{1}{u+n\cdot x+i\epsilon}]\nn\\&=&\frac{1}{2\pi}\partial_u\delta(u+n\cdot x).
\eea 

\subsection{Carrollian amplitudes from Feynman rules}
The Feynman rules in the bulk are the same as those for the bulk theory. These include the Feynman propagators and vertices, as well as symmetry factors. Combining with the previous Feynman rules which connect the bulk and boundary, 
we will summarize the Feynman rules to compute the Carrollian amplitudes. We will omit the null boundary in the Feynman diagrams.
\begin{itemize}
\item For each state, we define a signature $\sigma$ 
\bea 
\sigma=\left\{\begin{array}{cl}1&\text{outgoing state}\\ -1&\text{incoming state}\end{array}\right.
\eea 
    \item For each boundary-to-boundary propagator, 
    \bea 
    \begin{tikzpicture}[baseline=(current bounding box.center)]
      \fill (0,0) circle (1.5pt);
      \fill (2,0) circle (1.5pt);
      \draw (0,0) node[below] {$(u_1,\Omega_1)$} -- (2,0) node[below] {$(u_2,\Omega_2)$};
    \end{tikzpicture}=C(u_1,\Omega_1;u_2,\Omega_2)=-\beta(u_2-u_1)\delta(\Omega_1-\Omega_2)
    \eea or 
    \bea 
    C(u_1,\Omega_1;u_2,\Omega_2;\omega_0)=\frac{1}{4\pi}\Gamma[0,i\omega_0(u_2-u_1-i\epsilon)]{ \delta(\Omega_1-\Omega_2)}
    \eea by discarding the infrared modes.
    \item For each external line,
    \begin{align}
        \begin{tikzpicture}[baseline=(current bounding box.center)]
      \fill (0,0) circle (1.5pt);
      \fill (1.5,0) circle (1.5pt);
      \draw (0,0) node[below] {$(u,\Omega)$} -- (1.5,0);
      \node at (1.5,-0.1) [below] {$x$};
    \end{tikzpicture}=D(u,\Omega;x)=-\frac{\sigma}{8\pi^2(u+n\cdot x-i\sigma\epsilon)}
    \end{align}
    or 
    \bea 
    D(u,\Omega;x;\omega_0)=-\frac{\sigma e^{-i\sigma\omega_0(u+n\cdot x)}}{8\pi^2(u+n\cdot x-i
\sigma\epsilon)}
    \eea after discarding the infrared modes.
    \item For each Feynman propagator,
    \begin{align}
        \begin{tikzpicture}[baseline=(current bounding box.center)]
      \fill (0,0) circle (1.5pt);
      \fill (1,0) circle (1.5pt);
      \draw (0,0) node[below] {$x$} -- (1,0) node[below] {$y$};
    \end{tikzpicture} \
    =G_F(x,y)=\int \frac{d^4p}{(2\pi)^4}G_F(p) e^{ip\cdot(x-y)}=\frac{1}{4\pi^2\left((x-y)^2+i\epsilon\right)}\label{Feynmanmassless}
    \end{align}
    where the momentum space Feynman  propagator is 
    \bea 
    G_F(p)=\frac{i}{-p^2+i\epsilon}.
    \eea 
    The $i\epsilon$ prescription of Feynman propagator in position space can be found in \cite{Birrell:1982ix}.
    We should also mention that the external line may be obtained by asymptotic expansion of the Feynman propagator. To see this point, we may expand $\Phi(y)$ in the Feynman propagator near $\mathcal{I}^{+}$ and read out the leading coefficient of $r^{-1}$. The result is the same as the external line. Inversely, we may use the external line to derive the Feynman propagator. To simplify derivation, we assume $x^0>y^0$, then 
    \bea 
    G_F(x,y)&=&\langle \Phi(x)\Phi(y)\rangle\nn\\&=&2i\int du d\Omega \langle \Phi(x)\Sigma(u,\Omega)\rangle\langle \dot\Sigma(u,\Omega)\Phi(y)\rangle\nn\\&=&2i\int du d\Omega D^*(u,\Omega;x)\partial_u D(u,\Omega;y)
    \nn\\&=&\frac{1}{4\pi^2[(\bm x-\bm y)^2-(x^0-y^0-i\epsilon)^2]}
    \eea 
    which is consistent with \eqref{Feynmanmassless}. In the derivation, the $i\epsilon$ description is crucial and interested readers may find more details in Appendix \ref{iep}. Taking into account the possibility $y^0>x^0$, we find the following  formula \bea 
    G_F(x,y)&=&2i\int du d\Omega \left( \theta(x^0-y^0)D^*(u,\Omega;x)\partial_u D(u,\Omega;y)+\theta(y^0-x^0)D^*(u,\Omega;y)\partial_u D(u,\Omega;x)\right)\nn\\&=&\int du d\Omega \left(\theta(x^0-y^0)D^*(u,\Omega;x)K(u,\Omega;y)+\theta(y^0-x^0)D^*(u,\Omega;y)K(u,\Omega;x)\right).\label{split}
    \eea The bulk-to-bulk propagator (Feynman propagator) is expressed as the product of the external line and bulk-to-boundary propagator in the  Carrollian space. This is shown in figure \ref{splitrep}. 
    \begin{figure}
  \centering
  \hspace{0.8cm}
  \begin{tikzpicture}
    \filldraw[fill=white,draw,thick] (0,4) node[above]{\small $i^+$} -- (3,1) node[above right]{\small $\mathcal{I}^+$} -- (4,0) node[right]{\small $i^0$}  -- (2,-2) node[below right]{\small $\mathcal{I}^-$} -- (0,-4) node[below]{\small $i^-$} -- (-2,-2) node[below left]{\small $\mathcal{I}^-$} -- (-4,0)node[left]{\small $i^0$} -- (-2,2) node[above left]{\small $\mathcal{I}^+$} -- cycle;
    \draw[black!50,line width=2pt] (-0.5,0.5)  -- (1.8,2.2);
    \node[below left] at (-0.5,0.5) {$x$};
    \node[above right] at (1.8,2.2) {$(u,\Omega)$};
    \draw[dashed,black!50,line width=2pt] (1.5,0)  -- (1.8,2.2);
    \draw[black!50,line width=2pt] (-0.5,0.5)--(1.5,0);
    \node[below] at (0.25,0) {$G_F(x,y)$};
    \node[left] at (1,1.8) {$D^*(u,\Omega;x)$};
    \node[left] at (3.6,0.4) {$K(u,\Omega;y)$};
    \fill (-0.5,0.5) circle (2pt);
     \fill (1.5,0) node[below] {$y$} circle (2pt);
    \fill (1.8,2.2) circle (2pt);
  \end{tikzpicture}
  \caption{\centering{Split representation of Feynman propagator for $x^0>y^0$.}}
  \label{splitrep}
\end{figure}
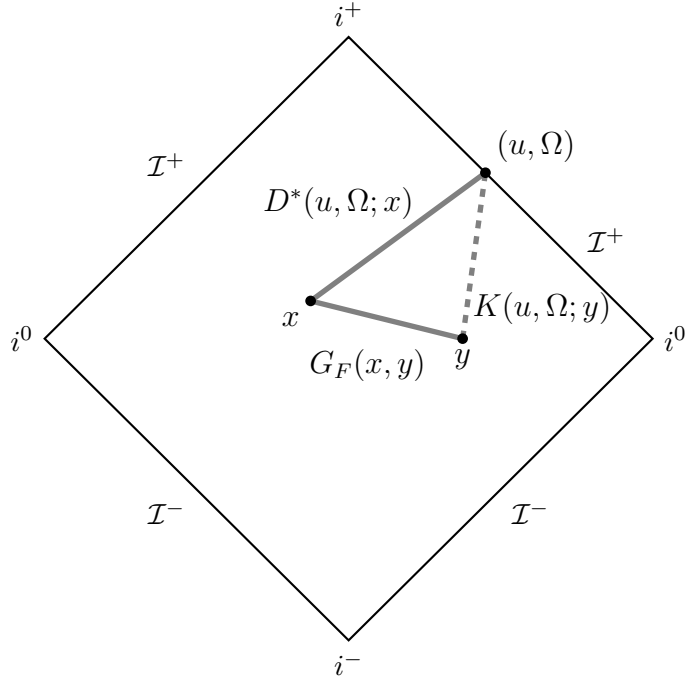

    The Feynman propagator is a superposition of the Wightman functions 
    \bea 
    G_F(x,y)=\theta(x^0-y^0)W^+(x,y)+\theta(y^0-x^0)W^-(x,y)
    \eea where 
    \bea 
    W^{+}(x,y)=\langle \Phi(x)\Phi(y)\rangle,\qquad W^-=\langle \Phi(y)\Phi(x)\rangle.
    \eea 
    Therefore, the Wightman functions can be written as a split representation 
    \bs\begin{align}
        W^+(x,y)&=\int du d\Omega D^*(u,\Omega;x)K(u,\Omega;y),\\
        W^-(x,y)&=\int du d\Omega D^*(u,\Omega;y)K(u,\Omega;x).
    \end{align}\es
    In AdS/CFT, there is a split representation of bulk-to-bulk propagator which is based on the product of bulk-to-boundary propagators \cite{Costa:2014kfa}. It would be interesting to study this similarity in the future.
    \item For each vertex,
    \bea 
    \begin{tikzpicture}[scale=0.7,baseline=(current bounding box.center)]
      \fill (0,0) circle (2pt);
      \draw (0,0) node[below] {$x$} -- (1,1);
      \draw (0,0) -- (1,-1);
      \draw (0,0) -- (-1,1);
      \draw (0,0) -- (-1,-1);
    \end{tikzpicture} \ \ =-i\lambda\int d^4x
    \eea 
    \item Divide by the symmetry factor.
\end{itemize} 
Now we will claim that the Carrollian amplitude \eqref{Carrollianamp} can be obtained by the previous Feynman rules. We will illustrate it in  two examples to show the general idea. In the first example, we will consider the one-loop correction for the two-point Carrollian amplitude. The one-loop Carrollian amplitude may be obtained by 
\bea 
&& C^{\text{1-loop}}(u_1,\Omega_1;u_2,\Omega_2)\nn\\
&=&\begin{tikzpicture}[baseline=(current bounding box.center)]
    \draw (0, 0) node [left] {$(u_1,\Omega_1)$} -- (1.2,0) node[below]{$x$} -- (2.4, 0) node [right] {$(u_2,\Omega_2)$};
    \fill (0,0) circle (1.5pt);
    \fill (1.2,0) circle (1.5pt);
    \fill (2.4,0) circle (1.5pt);
    \draw (1.2, 0.5) circle [radius=0.5];
  \end{tikzpicture}\nn\\
  &=&(-i\lambda) \int d^4x D(u_1,\Omega_1;x)G_F(x;x)D(u_2,\Omega_2;x)\nn\\
  &=&-i\lambda\sigma_1\sigma_2\left(\frac{1}{8\pi^2}\right)^2 \int d^4x \frac{1}{(u_1+n_1\cdot x-i\sigma_1\epsilon)(u_2+n_2\cdot x-i\sigma_2\epsilon)}\int \frac{d^4p}{(2\pi)^4}\frac{-i}{p^2}\nn\\&=&-i\lambda\left(\frac{1}{8\pi^2i}\right)^2\int d^4x\int_0^\infty d\omega_1 e^{-i\sigma_1\omega_1(u_1+n_1\cdot x-i\sigma_1\epsilon)}\int_0^\infty d\omega_2 e^{-i\sigma_2\omega_2(u_2+n_2\cdot x-i\sigma_2\epsilon)}\int \frac{d^4p}{(2\pi)^4}\frac{-i}{p^2}\nn\\&=&-i\lambda \left(\frac{1}{8\pi^2i}\right)^2\int_0^\infty d\omega_1 e^{-i\sigma_1\omega_1u_1}\int_0^\infty d\omega_2e^{-i\sigma_2\omega_2u_2}(2\pi)^4\delta^{(4)}(\sigma_1\omega_1 n_1+\sigma_2\omega_2n_2)\int \frac{d^4p}{(2\pi)^4}\frac{-i}{p^2}\nn\\&=&\left(\frac{1}{8\pi^2i}\right)^2\int_0^\infty d\omega_1 e^{-i\sigma_1\omega_1u_1}\int_0^\infty d\omega_2 e^{-i\sigma_2\omega_2 u_2}(2\pi)^4\delta^{(4)}(p_1+p_2)i\mathcal{M}^{\text{1-loop}}(p_1,p_2).
\eea In the third line, we used the integral representation of the external line
\bea 
D(u,\Omega;x)=\frac{1}{8\pi^2i} \int_0^\infty d\omega e^{-i\sigma\omega(u+n\cdot x-i\sigma\epsilon)}.
\eea In the last line, we rewrote the $\sigma_j\omega_jn_j$ as the null momentum $p_j$ and used the  one-loop $\mathcal{M}$  matrix of $\Phi^4$ theory
\bea 
i\mathcal{M}^{\text{1-loop}}(p_1,p_2)=-i\lambda \times\int \frac{d^4p}{(2\pi)^4} \frac{-i}{p^2}
\eea The result matches with \eqref{Carrollianamp}. In the second example, we  consider the tree-level four-point Carrollian amplitude. Using the Feynman rules, we find 
\bea&& C^{\text{tree}}(u_1,\Omega_1;u_2,\Omega_2;u_3,\Omega_3;u_4,\Omega_4)\nn\\
    &=&\begin{tikzpicture}[scale=0.6,baseline=(current bounding box.center)]\vspace{1cm}
      \fill (0,0) circle (2pt);
      \draw (0,0) node[below]{$x$} -- (1.5,1.5) node[right]{$(u_1,\Omega_1)$};
      \draw (0,0) -- (-1.5,1.5) node[left]{$(u_2,\Omega_2)$};
      \draw (0,0) -- (1.5,-1.5) node[right]{$(u_4,\Omega_4)$};
      \draw (0,0) -- (-1.5,-1.5) node[left]{$(u_3,\Omega_3)$};
    \end{tikzpicture}\nn\\
&=&-i\lambda \int d^4x D(u_1,\Omega_1;x)D(u_2,\Omega_2;x)D(u_3,\Omega_3;x)D(u_4,\Omega_4;x)\nn\\&=&-i\lambda\sigma_1\sigma_2\sigma_3\sigma_4 \left(\frac{1}{8\pi^2}\right)^4\int d^4 x \frac{1}{\prod_{j=1}^4 (u_j+n_j\cdot x-i\sigma_i\epsilon)}\nn\\&=&-i\lambda\left(\frac{1}{8\pi^2i}\right)^4\int d^4 x \prod_{j=1}^4 \int_0^\infty d\omega_j e^{-i\sigma_j\omega_j(u_j+n_j\cdot x-i\sigma_j\epsilon)}\nn\\&=&-i\lambda\left(\frac{1}{8\pi^2i}\right)^4\left(\prod_{j=1}^4\int_0^\infty d\omega_j e^{-i\sigma_j\omega_j u_j}\right)(2\pi)^4\delta^{(4)}(\sum_{j=1}^4\sigma_j\omega_jn_j)\nn\\&=&\left(\frac{1}{8\pi^2i}\right)^4 \left(\prod_{j=1}^4\int_0^\infty d\omega_j e^{-i\sigma_j\omega_j u_j}\right)(2\pi)^4\delta^{(4)}(p_1+p_2+p_3+p_4)i\mathcal{M}^{\text{tree}}(p_1,p_2,p_3,p_4)\nn\\\label{treelevel4pt}
\eea Again, the result matches with the formula \eqref{Carrollianamp}. In the following, we will prove that the Carrollian amplitudes can be obtained from the Feynman rules presented in the previous subsections.

\paragraph{Proof.} It is well-known that the Fourier transform of the time ordered correlation function is related to the $S$ matrix through the LSZ reduction formula \cite{1995iqft.book.....P}.
	In massless $ \Phi^4 $ theory, with the notation $ p_j = \sigma_j \omega_j n_j $, the LSZ reduction formula can be simplified to be:
	\begin{equation}
		\begin{aligned}
	      (\prod\limits_{j=1}^{m+n}\int d^4 y_j e^{-ip_j\cdot y_j}) 
		 \langle \Omega|T \{\Phi(y_1)\cdots\Phi(y_{m})\Phi(y_{m+1})\cdots\Phi(y_{m+n})\}|\Omega\rangle\\
		 \mathop{\sim} \limits_{p^0_j\rightarrow \sigma_j \omega_j}(\prod\limits_{j=1}^{m+n} \frac{-i\sqrt{Z}}{p_j^2-i\epsilon})\langle p_{m+1}\cdots  p_{m+n}|S| p_{1} \cdots  p_{m}\rangle
		\end{aligned}
	\end{equation} where the quantity $Z$ is the renormalization factor which could be different whenever the fields are not the same. The Feynman rules for correlation functions in $ \Phi^4 $ theory have already been proved in standard QFT textbook\cite{1995iqft.book.....P}:
	\begin{itemize}
		\item For each Feynman propagator,
		\begin{align}
			\begin{tikzpicture}[baseline=(current bounding box.center)]
				\fill (0,0) circle (1.5pt);
				\fill (1,0) circle (1.5pt);
				\draw (0,0) node[below] {$x$} -- (1,0) node[below] {$y$};
			\end{tikzpicture} \
			=G_F(x,y)=\int \frac{d^4p}{(2\pi)^4}\frac{i}{-p^2+i\epsilon} e^{ip\cdot(x-y)}.
		\end{align}
	\item For each vertex,
	    \begin{eqnarray}	     
	    \begin{tikzpicture}[scale=0.7,baseline=(current bounding box.center)]
	    	\fill (0,0) circle (2pt);
	    	\draw (0,0) node[below] {$x$} -- (1,1);
	    	\draw (0,0) -- (1,-1);
	    	\draw (0,0) -- (-1,1);
	    	\draw (0,0) -- (-1,-1);
	    \end{tikzpicture} \ \ =-i\lambda\int d^4x.
	\end{eqnarray}
        \item For each external point,
		\begin{align}
	      \begin{tikzpicture}[baseline=(current bounding box.center)]
		           \fill (0,0) circle (1.5pt);
		           \draw (0,0) node[below] {$x$} --(1,0);
   	         \end{tikzpicture} \
	       = 1.
        \end{align}        
         
         \item Divide by the symmetry factor.
	\end{itemize}
	
	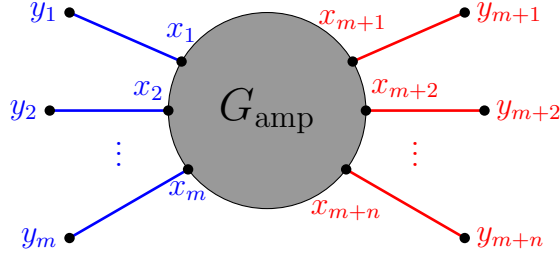
\begin{figure}
	    \centering
	    \begin{tikzpicture}[scale=1.3]
	    \draw[blue,line width=1pt] (-0.866,0.5) node[above,yshift=0.11cm]{$x_1$}  -- (-2,1) node[left]{$y_1$};
	    \draw[blue,line width=1pt] (-1,0)  -- (-2.2,0) node[left] {$y_2$};
	    \node[above,blue] at (-1.2,0) {$x_2$};
	     \draw[blue,line width=1pt] (-0.8,-0.6) node[below]{$x_m$}  -- (-2,-1.3) node[left]{$y_m$};
	     \node[blue] at (-1.5,-0.36) {$\vdots$};
	     
	     \draw[red,line width=1pt] (0.866,0.5) node[above,yshift=0.25cm]{$x_{m+1}$}  -- (2,1) node[right]{$y_{m+1}$};
	    \draw[red,line width=1pt] (1,0)  -- (2.2,0) node[right] {$y_{m+2}$};
	    \node[above,red] at (1.4,0) {$x_{m+2}$};
	     \draw[red,line width=1pt] (0.8,-0.6) node[below,yshift=-0.25cm]{$x_{m+n}$}  -- (2,-1.3) node[right]{$y_{m+n}$};
	     \node[red] at (1.5,-0.36) {$\vdots$};
	   
	      \fill[gray!80] (0,0) circle (1);
	      \draw (0,0) circle (1);
	      \fill (-0.866,0.5) circle (1.5pt);
	       \fill (-0.8,-0.6) circle (1.5pt);
	    \fill (-2,1) circle (1.5pt);
	    \fill (-2.2,0) circle (1.5pt);
	    \fill (-1,0) circle (1.5pt);
	    \fill (-2,-1.3) circle (1.5pt);
	    
	     \fill (0.866,0.5) circle (1.5pt);
	       \fill (0.8,-0.6) circle (1.5pt);
	    \fill (2,1) circle (1.5pt);
	    \fill (2.2,0) circle (1.5pt);
	    \fill (1,0) circle (1.5pt);
	    \fill (2,-1.3) circle (1.5pt);
	    \node at (0,0) {\Large $ G_{\text{amp}}$};
	    \end{tikzpicture}
	    \caption{Time ordered correlation function. The blue and red lines or  points are connected to the incoming and outgoing states, respectively. The external points are $y_j,\ j=1,2,\cdots,m+n$ and the vertices $x_j,\ j=1,2,\cdots,m+n$ are connected to $y_j$ through Feynman propagators which should be integrated out. The shaded part  is the amputated correlation function $G_{\text{amp}}$ which could be constructed by Feynman rules in the position space.}
	    \label{figamp}
	\end{figure}
	
	With these Feynman rules, we  will split the Feynman diagrams of correlation function into external points and amputated diagrams, which is shown in figure \ref{figamp}.  
	The amputated diagrams of the correlation function are described by $ G_{\text{amp}}(x_1,x_2, \cdots,x_{m+n}) $ with $x_j, j = 1,\cdots, m+n$ labeling the internal vertices connected to the external points $y_j, j=1,2,\cdots, m+n$. The $(n+m)$-point correlation function will be modified as:

    \begin{equation}
		\begin{aligned}
			&\langle \Omega|T \{\Phi(y_1)\cdots\Phi(y_{m})\Phi(y_{m+1})\cdots\Phi(y_{m+n})\}|\Omega\rangle \\
			=&\left(\prod_{j=1}^{m+n}\int d^4x_j\right)\left(\prod_{j=1}^{m+n} G_F(x_j-y_j)\right) G_{\text{amp}}(x_1,x_2,\cdots,x_{m+n}).
		\end{aligned}\label{correlation}
	\end{equation}
Note that there will be a vertex for each point $x_j$, so we have the integration $\int d^4 x_j$ in this expression. The factors $-i\lambda$ in the Feynman rule have been absorbed into the amputated correlation function.
     Inserting (\ref{correlation}) into the LSZ reduction formula, we get 
    \begin{equation}
    	\begin{aligned}
    \text{LHS}	&=\left(\prod_{j=1}^{m+n}\int d^4 x_j \right)\left(\prod_{j=1}^{m+n} G_F(p_j)e^{-i p_j \cdot x_j}\right)G_{\text{amp}}(x_1,x_2,\cdots, x_{m+n}).
    	\end{aligned}
    \end{equation}
	We can redefine the field  to absorb the renormalization factor $ Z $. Thus the scattering amplitude can be deduced as:
	 \begin{equation}
	 	\langle \bm p_{m+1}\cdots \bm p_{m+n}|S|\bm p_{1} \cdots \bm p_{m}\rangle =\left(\prod_{j=1}^{m+n}\int d^4x_j e^{-i p_j\cdot x_j}\right)G_{\text{amp}}(x_1,x_2,\cdots,x_{m+n})
	 \end{equation} The Fourier transform factor $e^{-ip\cdot x}$ is interpreted as the external lines in the momentum space \cite{1995iqft.book.....P}.
	 Note that both sides of the  previous equation contain the disconnected Feynman diagrams. We will only focus on  the connected part. On the left-hand side, the $S$ matrix is reduced to the $\mathcal{M}$ matrix up to a Dirac delta function which is associated with the conservation of energy and momentum. On the right-hand side, the amputated correlation function becomes the connected and amputated correlation function. Hence, the Carrollian amplitude can be reduced to
		\bea 
			&&\langle\prod_{j=1}^{m+n}\Sigma_j(u_j,\Omega_j)\rangle\nn\\
			&=&(\frac{1}{8\pi^2 i})^{m+n}\prod_{j=1}^{m+n} \int d\omega_j e^{-i\sigma_j\omega_j u_j} (2\pi)^4\delta^{(4)}(\sum_{j=1}^{m+n}p_j)i\mathcal{M}(p_1,p_2,\cdots,p_{m+n})\nn\\
			&=&(\frac{1}{8\pi^2 i})^{m+n}\left(\prod_{j=1}^{m+n} \int d\omega_j e^{-i\sigma_j\omega_j u_j}\right)\left(\prod_{j=1}^{m+n}\int d^4x_j e^{-ip_j\cdot x_j}\right) G_{\text{connected and amputated }}(x_1,x_2,\cdots,x_{m+n})\nn\\
			&=&(\frac{1}{8\pi^2 i})^{m+n}\prod_{j=1}^{m+n}\int d^4x_j  G_{\text{connected and amputated}}(x_1,x_2 \cdots,x_{m+n}) \prod_{j=1}^{m+n} \int d\omega_j e^{-i\sigma_j\omega_j (u_j+ n_j \cdot x_j)}\nn \\
			&=&(\frac{1}{8\pi^2 i})^{m+n} \prod_{j=1}^{m+n}\int d^4x_j  G_{\text{connected and amputated}}(x_1,x_2 \cdots,x_{m+n}) \prod_{j=1}^{m+n} \frac{-i \sigma_j}{(u_j +n_j x_j -i\sigma_j\epsilon)}\nn \\
			&=&\prod_{j=1}^{m+{ n}}\int d^4x_j  \prod_{j=1}^{m+n} D(u_j,\Omega_j;x_j) G_{\text{connected and amputated}}(x_1,x_2 \cdots,x_{m+n}) \label{amputatedcon}
		\eea 
    \color{black}
These are the Feynman rules for the Carrollian amplitude in Carrollian space. From the derivation, it is understood that  the function  $D(u,\Omega;x)$ is the external line, similar to the factor $e^{-ip\cdot x}$ in momentum space. 
 
\section{Four-point Carrollian amplitude}\label{4ptscalar}
In this section, we will evaluate the four-point Carrollian amplitude in $\Phi^4$ theory. The four boundary operators are inserted at $(u_j,\Omega_j),\ j=1,2,3,4$ respectively. We will adopt the stereographic coordinates 
\bea 
\Omega_j=(z_j,\bar z_j)
\eea in the following. Since the Carrollian amplitude is invariant under Poincar\'e transformation, we may use Lorentz transformation to fix three of the coordinates to $0,1,\infty$. The Lorentz transformations induce  M\"{o}bius transformations which form the conformal group on $S^2$. We can define the cross ratio
\be z=\frac{z_{12}z_{34}}{z_{13}z_{24}}\ee which is invariant under M\"{o}bius transformations. More explicitly, we will assume 
\bea 
z_1=0,\quad z_2=z,\quad z_3=1,\quad z_4=\infty
\eea to simplify discussion. The cross ratio $z$ and its complex conjugate $\bar z$ are related to the normal vectors $\bm n_j,\ j=1,2,3,4$ (which correspond to the angular directions of the inserting points) through the relations 
\bea 
\frac{(\bm n_1\cdot\bm n_2)(\bm n_3\cdot\bm n_4)}{(\bm n_1\cdot\bm n_3)(\bm n_2\cdot\bm n_4)}=z\bar z,\quad \frac{(\bm n_1\cdot\bm n_4)(\bm n_2\cdot\bm n_3)}{(\bm n_1\cdot\bm n_3)(\bm n_2\cdot\bm n_4)}=(1-z)(1-\bar z).
\eea 
\subsection{Tree level}
At the tree level, the four-point Carrollian amplitude is \eqref{treelevel4pt}. A massless particle has a null momentum which may be written as 
\bea 
p^\mu=\sigma\omega n^\mu=\sigma\omega(1,\frac{z+\bar z}{1+z\bar z},-i\frac{z-\bar z}{1+z\bar z},-\frac{1-z\bar z}{1+z\bar z})\label{momentumpara}
\eea where we have parameterized the null vector $n^\mu$ by stereographic coordinates. Therefore, the  momenta associated with $z_j,\ j=1,2,3,4$ are 
\bs\label{parap}\begin{align}
&p_1=\sigma_1\omega_1(1,0,0,-1),\\ &p_2=\sigma_2\omega_2(1,\frac{z+\bar z}{1+z\bar z},-i\frac{z-\bar z}{1+z\bar z},-\frac{1-z\bar z}{1+z\bar z}),\\ &p_3=\sigma_3\omega_3(1,1,0,0),\\
&p_4=\sigma_4\omega_4(1,0,0,1).
\end{align}\es It is easy to compute the Dirac delta function 
\bea
\hspace{-8pt}\delta^{(4)}(p_1+p_2+p_3+p_4)=\frac{1+z^2}{2\omega_4}\sigma_1\sigma_2\sigma_3\sigma_4\delta(\omega_1+\frac{\sigma_4\omega_4}{z \sigma_1})\delta(\omega_2-\frac{1+z^2}{z(1-z)}\frac{\sigma_4\omega_4}{\sigma_2})\delta(\omega_3+\frac{2}{1-z}\frac{\sigma_4\omega_4}{\sigma_3})\delta(\bar z-z).\nn\\ \label{delta4}
\eea Therefore, the tree-level Carrollian amplitude is 
\bea 
&& C^{\text{tree}}(u_1,\Omega_1,\sigma_1;u_2,\Omega_2,\sigma_2;u_3,\Omega_3,\sigma_3;u_4,\Omega_4,\sigma_4)\nn\\&=&-i\lambda\frac{1}{(8\pi^2)^4}\prod_{j=1}^4 \int_0^\infty d\omega_j e^{-i\sigma_j\omega_j u_j}(2\pi)^4\nn\\&&\times \frac{1+z^2}{2\omega_4}\sigma_1\sigma_2\sigma_3\sigma_4\delta(\omega_1+\frac{\sigma_4\omega_4}{z \sigma_1})\delta(\omega_2-\frac{1+z^2}{z(1-z)}\frac{\sigma_4\omega_4}{\sigma_2})\delta(\omega_3+\frac{2}{1-z}\frac{\sigma_4\omega_4}{\sigma_3})\delta(\bar z-z)\nn\\&=&-i\lambda\frac{1+z^2}{2(4\pi)^4}\sigma_1\sigma_2\sigma_3\sigma_4\delta(\bar z-z)\nn\\&&\times\int_0^\infty \frac{d\omega_4}{\omega_4}e^{-i\sigma_4\omega_4\chi(u_1,u_2,u_3,u_4;z)}\Theta(-\frac{\sigma_4\omega_4}{z\sigma_1})\Theta(\frac{1+z^2}{z(1-z)}\frac{\sigma_4\omega_4}{\sigma_2})\Theta(-\frac{2}{1-z}\frac{\sigma_4\omega_4}{\sigma_3})
\eea where 
\bea 
\chi(u_1,u_2,u_3,u_4;z)=\frac{-z(1-z)u_4+(1-z)u_1-(1+z^2)u_2+2zu_3}{z(z-1)}.\label{chicomplete}
\eea The Dirac delta function $\delta(\bar z-z)$ constrains the particle $2$ to propagate in the plane with $\phi=0$ or $\phi=\pi$ in spherical coordinates such that the cross ratio is real.
The appearance  of the step function is from  the conservation of energy and momentum. In the integration domain $\omega_4>0$, they lead to nonvanishing results only for 
\bea 
\sigma_1\sigma_2\sigma_3\sigma_4>0.
\eea 
There are only two cases. 
\begin{enumerate}
    \item All particles are incoming or outgoing. Without losing generality, we may set
    \bea 
    \sigma_1=\sigma_2=\sigma_3=\sigma_4=1.
    \eea In this case, the product of the step function is always zero
    \bea 
    \Theta(-\frac{\sigma_4\omega_4}{z\sigma_1})\Theta(\frac{1+z^2}{z(1-z)}\frac{\sigma_4\omega_4}{\sigma_2})\Theta(-\frac{2}{1-z}\frac{\sigma_4\omega_4}{\sigma_3})=\Theta(-z)\Theta(z(1-z))\Theta(z-1)\equiv 0.
    \eea This is trivial since all the particles are created from the vacuum, which violates the conservation of energy obviously.
    \item Two particles are incoming and the other two particles are outgoing. Without losing generality, we may set 
    \bea 
    \sigma_1=\sigma_2=-1,\quad \sigma_3=\sigma_4=1.
    \eea Then the product of the step function becomes 
    \bea 
     \Theta(-\frac{\sigma_4\omega_4}{z\sigma_1})\Theta(\frac{1+z^2}{z(1-z)}\frac{\sigma_4\omega_4}{\sigma_2})\Theta(-\frac{2}{1-z}\frac{\sigma_4\omega_4}{\sigma_3})=\Theta(z)\Theta(z(z-1))\Theta(z-1)=\Theta(z-1).\nn\\
    \eea This is nonvanishing only for\footnote{For the other choices of $\sigma_j,\ j=1,2,3,4$, the domains of $z$ with nonvanishing amplitude are different. Readers can find more details in Appendix \ref{zvalue}.}
    \bea 
    z>1.\label{zlargerthan1}
    \eea 
    
    \begin{figure}
        \centering
        \begin{tikzpicture}
        \fill[ball color=white!10] (0,0) circle (2);
        \draw[-latex] (0,0)--(3,0) node[right]{$y$};
        \draw[-latex] (0,0)--(0,3) node[above]{$z$};
        \draw[-latex] (0,0)--(-1.8,-2.7) node[below,left]{$x$};
        \draw[blue,-latex] (0,0)--(0,0.5) node[right]{4};
        \draw[red,-latex] (0,0)--(0,0.8)node[left]{1} ;
        \draw[blue,-latex] (0,0)--(-0.4,-0.6) node[above,left]{3};
        \draw[dashed] (0.6,3) -- (-1,0.6) node[above,left]{$x$-$z$ plane} -- (-1,-2.8) -- (0.6,-0.4) -- (0.6,3);
         \draw[red,-latex] (0,0) -- (0.4,-0.7)node[right]{2};
        \end{tikzpicture}
        \caption{Kinematic constraints from energy and momentum conservation for $2\to 2$ scattering of massless scalar particles. The particles $1$ and $2$ with red color are incoming and the particles $3$ and $4$ with blue color are outgoing. The four particles are constrained in the $x$-$z$ plane. In this figure, the arrows represent the momenta of the corresponding particles in the real space. }
        \label{fig4z}
    \end{figure}
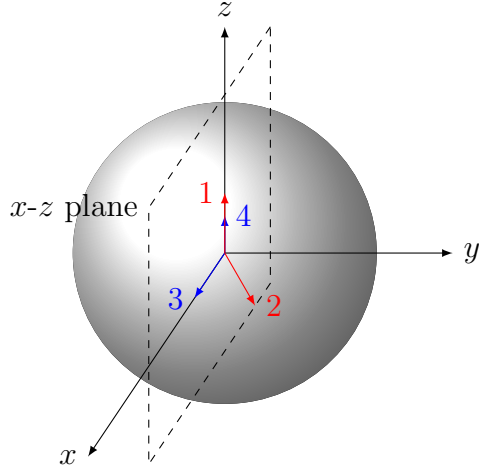
    
    The physical interpretation is shown in figure \ref{fig4z}. The momentum of the incoming particle 1  and outgoing particle 4 point to the north pole  of the celestial sphere $S^2$. At the same time, the outgoing particle $3$ propagates towards the positive $x$ axis. The inequality \eqref{zlargerthan1} leads to 
    \bea 
    0\le\theta<\frac{\pi}{2},\quad\phi=0.
    \eea Therefore, the second particle should direct to the third quadrant of $x$-$z$ plane. This is clear in the figure since the total momentum should be zero. In this case, the tree-level four-point Carrollian amplitude becomes
    \bea 
    && C^{\text{tree}}(u_1,\Omega_1,-;u_2,\Omega_2,-;u_3,\Omega_3,+;u_4,\Omega_4,+)\nn\\&=&-\lambda F(z)\int_0^\infty \frac{d\omega}{\omega}e^{-i\omega\chi},
    \eea  where
\be 
F(z)=\frac{i}{(4\pi)^4}\frac{1+z^2}{2}\delta(\bar z-z)\Theta(z-1).
\ee
In other words, we have 
\bea 
&& C^{\text{tree}}(u_1,\Omega_1,-;u_2,\Omega_2,-;u_3,\Omega_3,+;u_4,\Omega_4,+)=-4\pi \lambda F(z) \beta(\chi).
\eea  Note that the beta function is still divergent in the IR. 
One may regularize it by inserting a step function in the integrand for $\omega_4$. This is equivalent to compute the following four-point Carrollian amplitude 
\bea 
&& {C}^{\text{tree}}(u_1,\Omega_1,-;u_2,\Omega_2,-;u_3,\Omega_3,+;u_4,\Omega_4,+;\omega_0)\nn\\&\equiv&\langle \Sigma(u_1,\Omega_1)\Sigma(u_2,\Omega_2)\Sigma(u_3,\Omega_3)\Sigma(u_4,\Omega_4;\omega_0)\rangle\nn\\&=&-\lambda F(z)\int_0^\infty \frac{d\omega}{\omega}e^{-i\omega\chi}\Theta(\omega-\omega_0)\nn\\&=&-\lambda F(z)\Gamma(0,i\omega_0\chi).\label{tildeC}
\eea  In the IR limit, we find  
\bea 
&& \lim_{\omega_0\to0}{C}^{\text{tree}}(u_1,\Omega_1,-;u_2,\Omega_2,-;u_3,\Omega_3,+;u_4,\Omega_4,+;\omega_0)=\lambda F(z)I_0.
\eea 
One should have consider the four-point Carrollian amplitude by discarding the IR modes in each operator $\Sigma(u_j,\Omega_j)$ which is much more symmetric. This defines the following four-point Carrollian amplitude
\bea 
&&\tilde{C}^{\text{tree}}(u_1,\Omega_1,\sigma_1;u_2,\Omega_2,\sigma_2;u_3,\Omega_3,\sigma_3;u_4,\Omega_4,\sigma_4;\omega_0)\nn\\&\equiv&\langle \Sigma(u_1,\Omega_1;\omega_0)\Sigma(u_2,\Omega_2;\omega_0)\Sigma(u_3,\Omega_3;\omega_0)\Sigma(u_4,\Omega_4;\omega_0)\rangle\nn\\&=&-i\lambda\frac{1}{(8\pi^2)^4}\prod_{j=1}^4 \int_0^\infty d\omega_j e^{-i\sigma_j\omega_j u_j}\Theta(\omega_j-\omega_0)(2\pi)^4\nn\\&&\times \frac{1+z^2}{2\omega_4}\sigma_1\sigma_2\sigma_3\sigma_4\delta(\omega_1+\frac{\sigma_4\omega_4}{z \sigma_1})\delta(\omega_2-\frac{1+z^2}{z(1-z)}\frac{\sigma_4\omega_4}{\sigma_2})\delta(\omega_3+\frac{2}{1-z}\frac{\sigma_4\omega_4}{\sigma_3})\delta(\bar z-z)\nn\\&=&-i\lambda\frac{1+z^2}{2(4\pi)^4}\sigma_1\sigma_2\sigma_3\sigma_4\delta(\bar z-z)\int_0^\infty \frac{d\omega_4}{\omega_4}e^{-i\sigma_4\omega_4\chi(u_1,u_2,u_3,u_4;z)}\nn\\&&\times\Theta(-\frac{\sigma_4\omega_4}{z\sigma_1}-\omega_0)\Theta(\frac{1+z^2}{z(1-z)}\frac{\sigma_4\omega_4}{\sigma_2}-\omega_0)\Theta(-\frac{2}{1-z}\frac{\sigma_4\omega_4}{\sigma_3}-\omega_0)\Theta(\omega_4-\omega_0).
\eea We still consider  $\sigma_1=\sigma_2=-1,\quad \sigma_3=\sigma_4=1$. In this case, the product of the step function is still nonvanishing only for $z>1$
\bea 
&&\Theta(-\frac{\sigma_4\omega_4}{z\sigma_1}-\omega_0)\Theta(\frac{1+z^2}{z(1-z)}\frac{\sigma_4\omega_4}{\sigma_2}-\omega_0)\Theta(-\frac{2}{1-z}\frac{\sigma_4\omega_4}{\sigma_3}-\omega_0)\Theta(\omega_4-\omega_0)\nn\\&=&\Theta(z-1)\Theta(\omega_4-z\omega_0).
\eea Note that the integral domain for $\omega_4$ is modified to 
\bea 
\omega_4>z\omega_0.
\eea Therefore, 
\bea 
&&\tilde{C}^{\text{tree}}(u_1,\Omega_1,-;u_2,\Omega_2,-;u_3,\Omega_3,+;u_4,\Omega_4,+;\omega_0)=-\lambda F(z)\Gamma(0,iz\omega_0\chi).
\eea 
    The result is slightly different from \eqref{tildeC}. This is fine since they correspond to different correlators, depending on the way to discard the IR modes. Note that we can find a finite result by their difference
    \bea 
    &&\hspace{-1cm}\tilde{C}^{\text{tree}}(u_1,\Omega_1,-;u_2,\Omega_2,-;u_3,\Omega_3,+;u_4,\Omega_4,+;\omega_0)-{C}^{\text{tree}}(u_1,\Omega_1,-;u_2,\Omega_2,-;u_3,\Omega_3,+;u_4,\Omega_4,+;\omega_0)\nn\\&=& \lambda F(z)\log z.
    \eea 
\end{enumerate}

\paragraph{General four-point Carrollian amplitudes.} Now we will use Lorentz transformation to transform the previous result to  general four-point Carrollian amplitudes. By a  M\"{o}bius transformation which is parameterized by 
\bea 
a=\pm \sqrt{\frac{z_{31}}{z_{14}z_{34}}}z_4,\quad b=\mp\sqrt{\frac{z_{34}}{z_{31}z_{14}}}z_1,\quad c=\pm \sqrt{\frac{z_{31}}{z_{14}z_{34}}},\quad d=\mp\sqrt{\frac{z_{34}}{z_{31}z_{14}}},
\eea we may transform the points $0,z, 1,\infty$ to $z_1,z_2,z_3,z_4$, respectively. With the transformation property \eqref{rinv}, we find 
\bea 
&&\langle \Sigma(u_1,z_1,\bar{z}_1)\Sigma(u_2,z_2,\bar{z}_2)\Sigma(u_3,z_3,\bar{z}_3)\Sigma(u_4,z_4,\bar{z}_4)\rangle\nn\\&=&\Gamma_1\Gamma_2\Gamma_3\Gamma_4\langle \Sigma(\Gamma_1u_1,0)\Sigma(\Gamma_2u_2,z)\Sigma(\Gamma_3u_3,1)\Sigma(\Gamma_4u_4,\infty)\rangle,
\eea where the redshift factors are 
\bs\begin{align}
    \Gamma_1&=|d|^2(1+|z_1|^2)=|\frac{z_{34}}{z_{31}z_{14}}|(1+|z_1|^2),\\
    \Gamma_2&=\frac{1+|z_2|^2}{1+|z|^2}|cz+d|^2=\frac{|z_{34} z_{14}|}{|z_{24}|^2|z_{31}|}\frac{1+|z_2|^2}{1+|z|^2},\\
    \Gamma_3&=\frac{1}{2}|c+d|^2(1+|z_3|^2)=\frac{1}{2}|\frac{z_{14}}{z_{31}z_{34}}|(1+|z_3|^2),\\
    \Gamma_4&=|c|^2(1+|z_4|^2)=|\frac{z_{31}}{z_{14}z_{34}}|(1+|z_4|^2).
\end{align}
\es Therefore, at the tree level, we find
\bea 
&&\langle \Sigma(u_1,z_1,\bar{z}_1)\Sigma(u_2,z_2,\bar{z}_2)\Sigma(u_3,z_3,\bar{z}_3)\Sigma(u_4,z_4,\bar{z}_4)\rangle\nn\\&=&\frac{1}{2|z_{24}|^2|z_{13}|^2(1+|z|^2)}\prod_{j=1}^4(1+|z_j|^2)\langle \Sigma(\Gamma_1u_1,0)\Sigma(\Gamma_2u_2,z)\Sigma(\Gamma_3u_3,1)\Sigma(\Gamma_4u_4,\infty)\rangle\nn\\&=&-i\lambda \frac{\sigma_1\sigma_2\sigma_3\sigma_4}{4(4\pi)^4|z_{24}|^2|z_{13}|^2}\delta(\bar z-z)\prod_{j=1}^4(1+|z_j|^2)\Theta(-\frac{\sigma_4}{z\sigma_1})\Theta(\frac{1+z^2}{z(1-z)}\frac{\sigma_4}{\sigma_2})\Theta(-\frac{2}{1-z}\frac{\sigma_4}{\sigma_3})\nn\\&&\times\int_0^\infty \frac{d\omega_4}{\omega_4}e^{-i\sigma_4\omega_4\chi(\Gamma_1 u_1,\Gamma_2 u_2,\Gamma_3 u_3,\Gamma_4 u_4;z)},\label{conf4pt}
\eea 
where the function 
\bea 
&&\chi(\Gamma_1 u_1,\Gamma_2 u_2,\Gamma_3 u_3,\Gamma_4 u_4;z)\nn\\&=&\frac{-z(1-z)\Gamma_4u_4+(1-z)\Gamma_1 u_1-(1+z^2)\Gamma_2u_2+2z\Gamma_3 u_3}{z(z-1)}\nn\\&=&\Gamma_4[u_4-z^{-1}\Gamma_4^{-1}\Gamma_1u_1-z^{-1}(z-1)^{-1}(1+z^2)\Gamma_4^{-1}\Gamma_2u_2+2(z-1)^{-1}\Gamma_4^{-1}\Gamma_3 u_3]\nn\\&=&\Gamma_4[u_4-z\frac{1+|z_1|^2}{1+|z_4|^2}|\frac{z_{24}}{z_{12}}|^2 u_1+\frac{1-z}{z}\frac{1+|z_2|^2}{1+|z_4|^2}|\frac{z_{34}}{z_{23}}|^2u_2-\frac{1}{1-z}\frac{1+|z_3|^2}{1+|z_4|^2}|\frac{z_{14}}{z_{13}}|^2 u_3].
\eea The result is 
the same as \cite{Mason:2023mti} after taking into account the fact that the authors in that paper used a different parameterization of the null momentum\footnote{Note that we also flip the sign the fourth component of the momentum. However, this doesn't change the conclusion.}
\bea 
p^\mu=\sigma\omega (1+|z|^2)n^\mu,
\eea which is equivalent to our parameterization \eqref{momentumpara} up to a factor $(1+|z|^2)$. 

Before we close this subsection, we will comment on the dependence of  function $\chi$ in the four-point Carrollian amplitude. This function actually reflects the fact that the Carrollian amplitude is invariant under spacetime  translation. From \eqref{sinv}, we find the following identity for four-point Carrollian amplitude 
\bea 
&&\langle \Sigma(u_1-a\cdot n_1,0)\Sigma(u_2-a\cdot n_2,z)\Sigma(u_3-a\cdot n_3,1)\Sigma(u_4-a\cdot n_4,\infty)\rangle\nn\\&=&\langle \Sigma(u_1,0)\Sigma(u_2,z)\Sigma(u_3,1)\Sigma(u_4,\infty)\rangle.
\eea This implies that the Carrollian amplitude depends only on the function which is invariant under spacetime translation. The invariant may be written as 
\bea 
\hat{\chi}=\alpha_1 u_1+\alpha_2 u_2+\alpha_3u_3+\alpha_4 u_4.
\eea The spacetime translation invariance implies 
\bea 
\alpha_1 n^\mu_1+\alpha_2 n^\mu_2+\alpha_3 n^\mu_3+\alpha_4 n^\mu_4=0.
\eea With the parameterization \eqref{parap}, we find 
\bea 
\alpha_1=-\frac{1}{z}\alpha_4,\quad \alpha_2=\frac{1+z^2}{z(1-z)}\alpha_4,\quad \alpha_3=-\frac{1}{1-z}\alpha_4.
\eea Therefore, the spacetime translation invariant function $\hat\chi$ is proportional to $\chi$ up to a time-independent constant
\bea \hat{\chi}=\alpha_4[-\frac{1}{z}u_1+\frac{1+z^2}{z(1-z)}u_2-\frac{2}{1-z}u_3+u_4]=\alpha_4 \chi(u_1,u_2,u_3,u_4;z).
\eea 

\subsection{1-loop corrections}
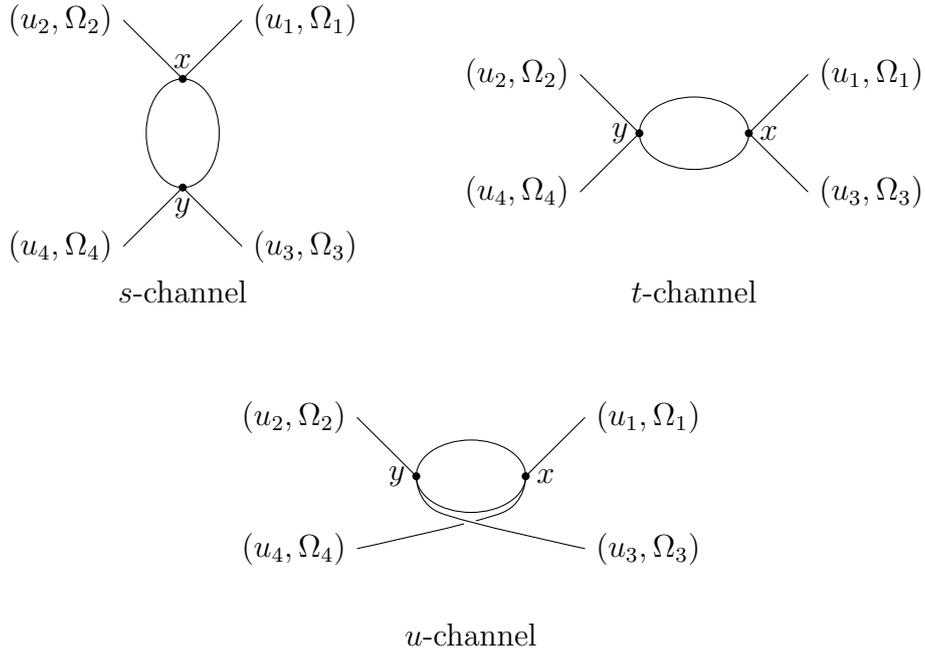
\begin{figure}
    \centering
    \begin{tikzpicture}[scale=0.6]
      \draw (0,1.2) node[above]{$x$} -- (1.3,2.5) node[right]{$(u_1,\Omega_1)$};
      \draw (0,1.2) -- (-1.3,2.5) node[left]{$(u_2,\Omega_2)$};
      \draw (0,-1.2) node[below]{$y$} -- (1.3,-2.5) node[right]{$(u_3,\Omega_3)$};
      \draw (0,0) ellipse (0.8 and 1.2);
      \draw (0,-1.2) -- (-1.3,-2.5) node[left]{$(u_4,\Omega_4)$};
      \fill (0,1.2) circle (2.5pt);
      \fill (0,-1.2) circle (2.5pt);
      \node at (0,-3.5) {$s$-channel};
    \end{tikzpicture}\hspace{1cm}
    \begin{tikzpicture}[scale=0.6]
      \draw (1.2,0) node[right]{$x$} --(2.5,-1.3) node[right]{$(u_3,\Omega_3)$} ;
      \draw (-1.2,0) node[left]{$y$} -- (-2.5,1.3) node[left]{$(u_2,\Omega_2)$};
      \draw (1.2,0)  -- (2.5,1.3) node[right]{$(u_1,\Omega_1)$};
      \draw (0,0) ellipse (1.2 and 0.8);
      \draw (-1.2,0) -- (-2.5,-1.3) node[left]{$(u_4,\Omega_4)$};
      \fill (1.2,0) circle (2.5pt);
      \fill (-1.2,0) circle (2.5pt);
      \node at (0,-3.5) {$t$-channel};
    \end{tikzpicture} \vspace{1cm}
    
    \begin{tikzpicture}[scale=0.6]
      \draw (-1.2,0) node[left]{$y$} -- (-2.5,1.3) node[left]{$(u_2,\Omega_2)$};
      \draw (1.2,0) node[right]{$x$} -- (2.5,1.3) node[right]{$(u_1,\Omega_1)$};
      \draw (0,0) ellipse (1.2 and 0.8);
      \draw plot[smooth,tension=0.5] coordinates{(1.2,0)(0.7,-0.8) (-2.5,-1.6)}; 
      \node[left] at (-2.5,-1.6) {$(u_4,\Omega_4)$};
     \fill[white] (-0.015,-1.05) circle (4pt);
    \draw plot[smooth,tension=0.5] coordinates {(-1.2,0) (-0.7,-0.8)(2.5,-1.6)}; 
     \node[right] at (2.5,-1.6) {$(u_3,\Omega_3)$};
      \fill (1.2,0) circle (2.5pt);
      \fill (-1.2,0) circle (2.5pt);
      \node at (0,-3.5) {$u$-channel};
    \end{tikzpicture}
    \caption{Four-point Carrollian amplitudes at 1-loop level in $s,t,u$ channels}
    \label{stu}
\end{figure}
There are three channels at the 1-loop level which are shown in  figure \ref{stu}. In the $s$ channel, the Carrollian amplitude is 
\bea 
&&C_s^{\text{1-loop}}(u_1,\Omega_1,\sigma_1;u_2,\Omega_2,\sigma_2;u_3,\Omega_3,\sigma_3;u_4,\Omega_4,\sigma_4)\nn\\&=&\frac{(-i\lambda)^2}{2}\int d^4 x \int d^4 y D(u_1,\Omega_1;x)D(u_2,\Omega_2;x)D(u_3,\Omega_3;y)D(u_4,\Omega_4;y)\left(G_F(x-y)\right)^2\nn\\&=&\frac{(-i\lambda)^2}{2}\left(\frac{1}{8\pi^2 i}\right)^{4}\int d^4x\int d^4y\prod_{j=1}^2\int_0^\infty d\omega_j e^{-i\sigma_j \omega_j(u_j+n_j\cdot x-i\sigma_j\epsilon)}\prod_{k=1}^2\int_0^\infty d\omega_k e^{-i\sigma_k\omega_k(u_k+n_k\cdot y-i\sigma_k\epsilon)}\nn\\&&\times \int \frac{d^4p}{(2\pi)^4}\frac{-i}{p^2}e^{ip\cdot(x-y)}\int\frac{d^4p'}{(2\pi)^4}\frac{-i}{p'^2}e^{ip'\cdot(x-y)}\nn\\&=&\frac{(-i\lambda)^2}{2}\left(\frac{1}{8\pi^2 i}\right)^{4}\prod_{j=1}^4 \int_0^\infty d\omega_j e^{-i\sigma_j\omega_ju_j}\nn\\&&\times\int d^4 p\int d^4 p'\frac{(-i)^2}{(p^2)(p'^2)}\delta^{(4)}(p+p'-p_1-p_2)\delta^{(4)}(p+p'+p_3+p_4)\nn\\&=&\left(\frac{1}{8\pi^2 i}\right)^{4}\prod_{j=1}^4 \int_0^\infty d\omega_j e^{-i\sigma_j\omega_ju_j}(2\pi)^4\delta^{(4)}(p_1+p_2+p_3+p_4)i\mathcal{M}^{\text{1-loop}}_s(p_1,p_2,p_3,p_4)
\eea where $\mathcal{M}^{\text{1-loop}}_s(p_1,p_2,p_3,p_4)$ is  $2\to 2$ scattering $\mathcal{M}$ matrix in the $s$ channel at the 1-loop level
\bea 
i\mathcal{M}_s^{\text{1-loop}}(p_1,p_2,p_3,p_4)=\frac{(-i\lambda)^2}{2}\int\frac{d^4p}{(2\pi)^4}\frac{(-i)^2}{p^2(p-p_1-p_2)^2}
\eea whose value could be calculated via dimensional regularization \cite{tHooft:1972tcz, 1995iqft.book.....P}
\bea 
i\mathcal{M}_s^{\text{1-loop}}(p_1,p_2,p_3,p_4)=\frac{i\lambda^2\bar{M}^{\varepsilon}}{16\pi^2}\left(\frac{1}{\varepsilon}-\frac{1}{2}\log\frac{s}{4M^2}\right)
\eea where 
\bea 
\varepsilon=4-d
\eea is a small positive quantity and $s$ is the Mandelstam variable 
\bea 
s=(p_1+p_2)^2.
\eea We have also inserted an energy scale $M$ into the result to balance the dimensions. We use  $\overline{\mathrm{MS}}$ scheme by absorbing the constant into the energy scale\footnote{Strictly speaking, the $\overline{\mathrm{MS}}$ scheme in \cite{1995iqft.book.....P} is slightly different from ours up to a constant factor.}
\bea 
\bar{M}=M e^{-\frac{1}{2}\log 4\pi-\frac{1}{2}\psi(3/2)},
\eea 
 where $\psi(q)$ is Digamma function, namely the derivative of the logarithm of the Gamma function, seeing \eqref{Digamma}.
By taking into account the other two channels, we find 
\bea 
i\mathcal{M}^{\text{1-loop}}(p_1,p_2,p_3,p_4)=i\lambda^2\bar{M}^{\varepsilon}(\frac{3}{16\pi^2\varepsilon}-\frac{1}{32\pi^2}\log \frac{stu}{64M^6}
)
\eea with\footnote{Note that here $u$ is a Mandelstam variable in the $u$ channel, which should be distinguishable from the retarded time. Similarly, $t$ is the Mandelstam variable in the $t$ channel, which is not the coordinate time in Cartesian coordinates.}
\bea 
t=(p_1+p_3)^2,\quad u=(p_1+p_4)^2.
\eea The momentum conservation \eqref{delta4} fixes 
\bea 
\omega_1=\frac{\omega_4}{z},\quad \omega_2=-\frac{1+z^2}{z(1-z)}\omega_4,\quad \omega_3=-\frac{2}{1-z}\omega_4
\eea and thus, 
\bea s=-\frac{4\omega_4^2}{z-1},\quad
    t=\frac{4\omega_4^2}{z(z-1)},\quad
    u=\frac{4\omega_4^2}{z}.
\eea Substituting them into the 1-loop amplitude, and taking into account the counterterm 
\begin{align}
    \begin{tikzpicture}[scale=0.7,baseline=(current bounding box.center)]
      \draw (0,0) circle (0.15);
      \draw (0,0) node[below] {$x$} -- (1,1);
      \draw (0,0) -- (1,-1);
      \draw (0,0) -- (-1,1);
      \draw (0,0) -- (-1,-1);
    \end{tikzpicture} \ \ =-i\delta_\lambda=-i\frac{3\lambda^2}{16\pi^2\varepsilon}+\mathcal{O}(\lambda^3),
\end{align}
we find 
\bea 
i\mathcal{M}^{\text{1-loop}}(p_1,p_2,p_3,p_4)=i(a_0(\lambda)+a_1(\lambda)\log\frac{\omega_4}{ M})\label{M1loop}
\eea where we have expanded the coefficients $a_0(\lambda), a_1(\lambda)$ up to 1-loop level
\bea 
a_0(\lambda)=-\lambda+\frac{\lambda^2}{32\pi^2}[\log z(z-1)+\log z+\log(1-z)],\quad a_1(\lambda)=-\frac{3\lambda^2}{16\pi^2}.
\eea
Note that there are three branch points at $z=0,1$ and $\infty$  for $a_0$ which lead to discontinuity of the $\mathcal{M}$ matrix on the two sides of the branch cuts. The discontinuity of the imaginary part of  $\mathcal{M}$ is related to the unitarity of $S$ matrix.  In the following, the coefficients $a_j(\lambda),j=0,1,\cdots$ will be abbreviated to $a_j, j=0,1,\cdots$, respectively.
The 1-loop Carrollian amplitude becomes
\bea 
&&C^{\text{1-loop}}(u_1,\Omega_1,-;u_2,\Omega_2,-;u_3,\Omega_3,+;u_4,\Omega_4,+)\nn\\&=&F(z)\int_0^\infty \frac{d\omega}{\omega}e^{-i\omega\chi}(a_0+a_1\log\frac{\omega}{M})
\eea The integration suffers IR divergent. We may cure this problem by inserting  a step function $\Theta(\omega-\omega_0)$ as \eqref{tildeC}. By defining a definite integral 
\bea 
J(q,\chi,\omega_0)=\int_0^\infty \frac{d\omega}{\omega^{1-q}}e^{-i\omega\chi}\Theta(\omega-\omega_0),\quad \text{Re}(q)>0
\eea we find 
\bea 
&&C^{\text{1-loop}}(u_1,\Omega_1,-;u_2,\Omega_2,-;u_3,\Omega_3,+;u_4,\Omega_4,+)\nn\\&=&F(z)\lim_{q\to 0}(a_0-a_1\log M+a_1\frac{d}{dq})J(q,\chi,\omega_0)\nn\\
&=&F(z)[(a_0-a_1\log M)J_0(\chi,\omega_0)+a_1J_1(\chi,\omega_0)]\nn\\&=&F(z)[(a_0-a_1\log M)(-I_0)+a_1( \frac{I_0^2}{2}-I_0\log\omega_0+\frac{\pi^2}{12})]\nn\\&=&F(z)[\frac{\pi^2}{12}a_1+(-a_0-a_1\log\frac{\omega_0}{M})I_0+\frac{a_1}{2}I_0^2],
\eea 
 where $J_0,J_1$ are defined in \eqref{Indef} and we have used \eqref{c10b}. 

Now we will discuss the discontinuity of the $\mathcal{M}$ matrix and corresponding Carrollian amplitude. Physically, the discontinuity is associated with the case that the intermediate virtual particle becomes on shell. In our convention, this is only possible in the $s$ channel for $s<0$. Assuming $\omega_4$ is always real, the discontinuity appears only for 
\be 
z>1
\ee which is exactly the requirement of nonvanishing $F(z)$. From \eqref{M1loop}, the discontinuity of the $\mathcal{M}^{\text{1-loop}}$  matrix is 
\be 
\text{Disc}\ \mathcal{M}^{\text{1-loop}}=i\frac{\lambda^2}{16\pi}.
\ee Usually, this discontinuity is associated with the Optical theorem which could be found in the textbook. Here we notice that there is a similar discontinuity for the Carrollian amplitude  
\be 
\text{Disc}\ C^{\text{1-loop}}=-\frac{i\lambda^2}{16\pi}F(z)I_0
\ee which may be related to \eqref{discon}.

\subsection{2-loop corrections}
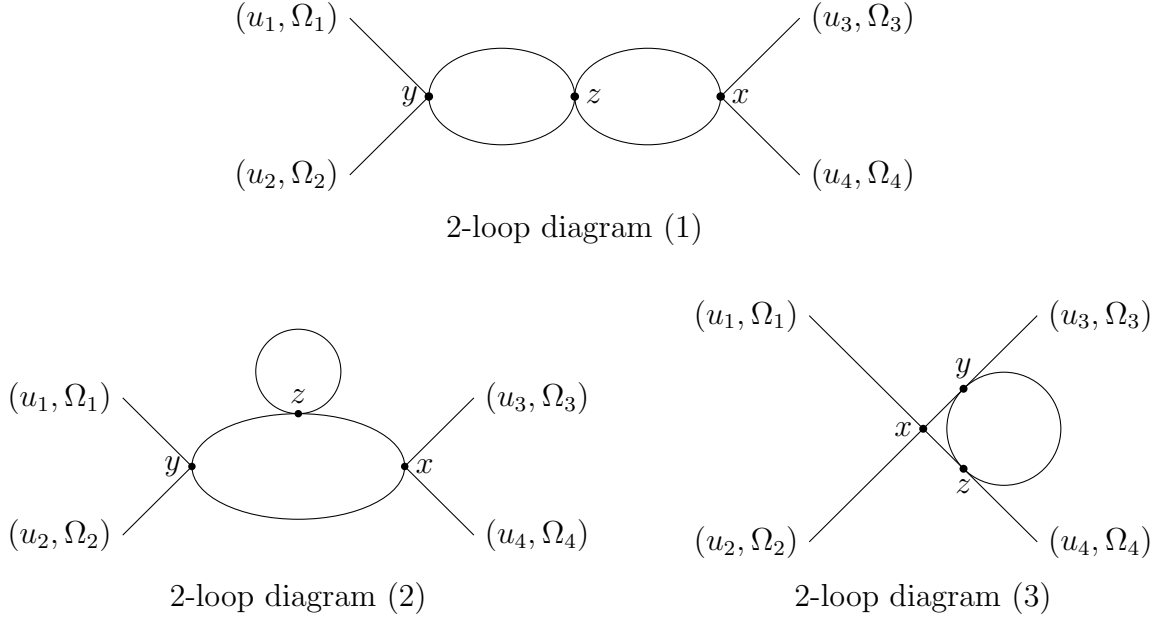
\begin{figure}
    \centering
      \begin{tikzpicture}[scale=0.8]
      \draw (3.6,0) --(4.9,-1.3) node[right]{$(u_4,\Omega_4)$} ;
      \draw (-1.2,0) node[left]{$y$} -- (-2.5,1.3) node[left]{$(u_1,\Omega_1)$};
      \draw (3.6,0) node[right]{$x$}  -- (4.9,1.3) node[right]{$(u_3,\Omega_3)$};
      \draw (0,0) ellipse (1.2 and 0.8);
      \draw (2.4,0) ellipse (1.2 and 0.8);
      \draw (-1.2,0) -- (-2.5,-1.3) node[left]{$(u_2,\Omega_2)$};
      \fill (1.2,0) node[right]{$z$} circle (2pt);
      \fill (3.6,0) circle (2pt);
      \fill (-1.2,0) circle (2pt);
       \node at (1.2,-2.2) {2-loop diagram (1)};
    \end{tikzpicture} \vspace{1em}
    
     \begin{tikzpicture}[scale=0.7]
      \draw (2,0) node[right]{$x$} --(3.3,-1.3) node[right]{$(u_4,\Omega_4)$} ;
      \draw (-2,0) node[left]{$y$} -- (-3.3,1.3) node[left]{$(u_1,\Omega_1)$};
      \draw (2,0)  -- (3.3,1.3) node[right]{$(u_3,\Omega_3)$};
      \draw (0,0) ellipse (2 and 1);
      \draw (-2,0) -- (-3.3,-1.3) node[left]{$(u_2,\Omega_2)$};
      \draw (0,1.8) circle (0.8);
      \fill (2,0) circle (2pt);
      \fill (-2,0) circle (2pt);
      \fill (0,1) node[above]{$z$} circle (2pt);
      \node at (0,-2.5) {2-loop diagram (2)};
    \end{tikzpicture}
    \hspace{2em}
      \begin{tikzpicture}[scale=0.75]
      \draw (-2,-2)  node[left]{$(u_2,\Omega_2)$} -- (2,2) node[right]{$(u_3,\Omega_3)$};
      \draw (-2,2)  node[left]{$(u_1,\Omega_1)$} -- (2,-2) node[right]{$(u_4,\Omega_4)$};
      \fill (0,0) circle (2pt);
      \node[left] at (0,0) {$x$};
      \draw ({sqrt(2)},0) circle (1);
      \fill ({1/sqrt(2)},{1/sqrt(2)}) node[above]{$y$} circle (2pt);
      \fill ({1/sqrt(2)},{-1/sqrt(2)}) node[below]{$z$} circle (2pt);
       \node at (0,-3) {2-loop diagram (3)};
    \end{tikzpicture} 
    \caption{The Feynman diagrams for massless $\Phi^4$ theory at two loops. Here we only show the $s$-channel.}\label{2loop}
\end{figure}
The 2-loop Feynman diagrams with $2\to 2$ scattering may be found in \cite{2021qftc.book.....Z}. In figure \ref{2loop}, we have shown the $s$ channel part. There are three diagrams in the $s$-channel. For the first diagram of 
figure \ref{2loop}, the $\mathcal{M}$ matrix is 
\bea 
&&i\mathcal{M}_{(1)}^{\text{2-loop}}(p_1,p_2,p_3,p_4)\nn\\&=&-i\frac{\lambda^3}{(4\pi)^4}\bar{M}^{\varepsilon}\left(\frac{1}{\varepsilon^2}-\frac{1}{\varepsilon}\log\frac{s}{4M^2}+\frac{1}{2}\log^2\frac{s}{4M^2}-\frac{\pi^2}{24}+1\right)+\text{$t,u$ channel}.
\eea 
For the second diagram of figure \ref{2loop}, the  $\mathcal{M}$ matrix is 
\bea 
&& i\mathcal{M}_{(2)}^{\text{2-loop}}(p_1,p_2,p_3,p_4)\nn\\&=&\frac{(-i\lambda)^3}{4}\int \frac{d^4p}{(2\pi)^4}\left(\frac{i}{-p^2}\right)^2\frac{i}{-(p_1+p_2-p)^2}\int \frac{d^4p'}{(2\pi)^4}\frac{i}{-p'^2}+\text{$t,u$ channel}.
\eea The integral for $p'$ may be evaluated by inserting the small mass term $m_0^2$. In dimensional regularization, 
\bea 
\int\frac{d^dp'}{(2\pi)^d}\frac{1}{-p^2-m_0^2}=-\frac{i}{(4\pi)^{d/2}}\Gamma(1-\frac{d}{2})(m_0^2)^{\frac{d}{2}-1}.
\eea The result may be set to 0 since it is proportional to the positive power of $m_0$. Note that this integral also appears in the 1-loop correction of two-point Carrollian amplitude. It is zero such that the two-point Carrollian amplitude is not affected by 1-loop correction\footnote{Actually, similar integrals can be regularized to 0\cite{Leibbrandt:1975dj, 2022arXiv220103593W}
\bea 
\int \frac{d^dp}{(2\pi)^d}(-p^2)^a=0
\eea for any complex number $a+\frac{d}{2}\not=0$.}.

For the third diagram of figure \ref{2loop}, the $\mathcal{M}$ matrix is 
\bea \label{2-loop}
&& i\mathcal{M}_{(3)}^{\text{2-loop}}(p_1,p_2,p_3,p_4)\nn\\&=&2\times\frac{(-i\lambda)^3\bar M^{3\varepsilon}}{2}\int \frac{d^dp}{(2\pi)^d}\int \frac{d^dp'}{(2\pi)^d}\frac{i}{-p^2}\frac{i}{-p'^2}\frac{i}{-(p_3+p-p')^2}\frac{i}{-(p_1+p_2-p)^2}+\text{$t,u$ channel}.\nn\\\eea
Note that we have shifted $\lambda$ to $\lambda \bar M^{\varepsilon}$ in dimensional regularization.
The integration over $p'$ can be obtained as 1-loop calculation. The integration over $p$ can be obtained by the Feynman's formula 
\bea 
&&\frac{1}{A_1^{a_1}A_2^{a_2}\cdots A_n^{a_n}}\nn\\
&=&\frac{\Gamma(a_1+a_2+\cdots+a_n)}{\Gamma(a_1)\cdots\Gamma(a_n)}\int_0^1 dx_1\cdots\int_0^1 dx_n \frac{\delta(x_1+\cdots+x_n-1)x_1^{a_1-1}\cdots x_n^{a_n-1}}{(x_1A_1+\cdots+x_nA_n)^{a_1+\cdots+a_n}}.
\eea We summarize the result in the following 
\bea 
&& i\mathcal{M}_{(3)}^{\text{2-loop}}(p_1,p_2,p_3,p_4)\nn\\&=&-i\frac{\lambda^3}{(4\pi)^4}\bar M^{\varepsilon}[\frac{2}{\varepsilon^2}+2\frac{\frac{1}{2}-\log \frac{s}{4M^2}}{\varepsilon}+\frac{1}{12} \left(12 (\log \frac{s}{4M^2}-1) \log \frac{s}{4M^2}+\pi ^2+42\right)]+\text{$t,u$ channel}.\nn
\eea 

\begin{figure}
    \centering
      \begin{tikzpicture}[scale=0.8]
      \draw (1.2,0) --(2.5,-1.3) node[right]{$(u_4,\Omega_4)$} ;
      \draw (-1.2,0)  -- (-2.5,1.3) node[left]{$(u_1,\Omega_1)$};
      \draw (1.2,0)  -- (2.5,1.3) node[right]{$(u_3,\Omega_3)$};
      \draw (0,0) ellipse (1.2 and 0.8);
      \draw (-1.2,0) -- (-2.5,-1.3) node[left]{$(u_2,\Omega_2)$};
    \node at (1.2,0) {$\otimes$};
      \fill (-1.2,0) circle (2pt);
       \node at (0,-2.2) {2-loop counterterm (1)};
    \end{tikzpicture}
    \hspace{2em} 
    \begin{tikzpicture}[scale=1.2]
      \draw (0,0) circle (0.1);
      \draw (0,0) node[below,yshift=-0.2cm]{$x$} -- (1,1) node[right]{$(u_3,\Omega_3)$};
      \draw (0,0) -- (1,-1) node[right]{$(u_4,\Omega_4)$};
      \draw (0,0) -- (-1,1) node[left]{$(u_1,\Omega_1)$};
      \draw (0,0) -- (-1,-1) node[left]{$(u_2,\Omega_2)$};
      \node at (0,-1.5) {2-loop counterterm (2)};
    \end{tikzpicture}
    \caption{Counterterms at 2-loop level. There are also diagrams in the $t,u$ channel for the left diagram which are not shown in the figure. }\label{2loopcount}
\end{figure}
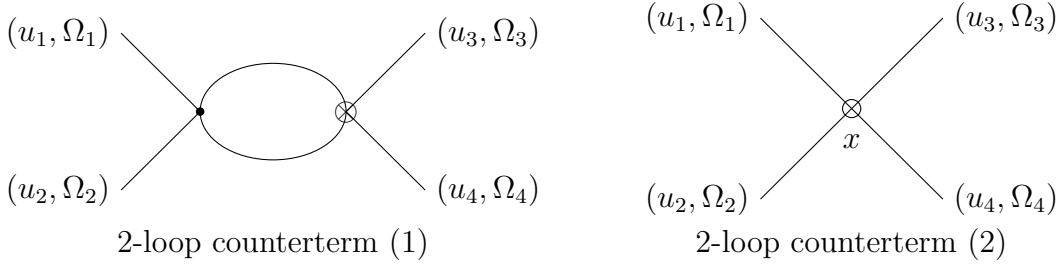

We should also consider the contributions from the counterterms which are shown in the first diagram of figure \ref{2loopcount} 
\bea 
i\mathcal{M}_{(4)}^{\text{2-loop}}(p_1,p_2,p_3,p_4)=2\times (-i\lambda)(-i\delta_{\lambda})\bar{M}^{2\varepsilon}\frac{1}{2}\int \frac{d^dp}{(2\pi)^d}\frac{-i}{p^2}\frac{-i}{(p-p_1-p_2)^2}+\text{$t,u$ channel}.\nn\\
\eea The factor $2$ counts the position of the counterterm and $\frac{1}{2}$ is the symmetry factor.
Up to order $\lambda^3$, this is 
\bea 
i\mathcal{M}_{(4)}^{\text{2-loop}}(p_1,p_2,p_3,p_4)&=&i\frac{\lambda^3}{(4\pi)^4}\bar M^{\varepsilon}[\frac{6}{\varepsilon^2}-\frac{3\log \frac{s}{4M^2}}{\varepsilon}+3\left(\frac{\log ^2\frac{s}{4M^2}}{4}-\frac{\pi ^2}{24}+1\right)]
+\text{$t,u$ channel}.\nn\\
\eea 

Adding the 2-loop results, we find 
\bea 
\sum_{j=1}^4 i\mathcal{M}_{(j)}^{\text{2-loop}}(p_1,p_2,p_3,p_4)&=&i\frac{\lambda^3}{(4\pi)^4}\bar M^{\varepsilon}[\frac{3}{\varepsilon^2}-\frac{1}{\varepsilon}-\frac{3}{4}  \log ^2\frac{s}{4M^2}+\log \frac{s}{4M^2}-\frac{3}{2}-\frac{\pi ^2}{6}]+\text{$t,u$ channel}.\nn\\
\eea 
The divergences should be canceled by the vertex counterterm at order  $\lambda^3$ (seeing the second diagram of figure \ref{2loopcount})
\bea 
-i\delta_\lambda=-i\frac{3\lambda^2}{16\pi^2\varepsilon}-i\frac{3\lambda^3}{(4\pi)^4}(\frac{3}{\varepsilon^2}-\frac{1}{\varepsilon})+\mathcal{O}(\lambda^4).
\eea 
Including all the four-point amplitudes up to 2-loop, we can write it as 
\bea 
i\mathcal{M}(p_1,p_2,p_3,p_4)=i(a_0+a_1\log\frac{\omega_4}{M}+a_2\log^2\frac{\omega_4}{M})
\eea where the coefficients $a_0,a_1,a_2$ may be parameterized as 
\be 
a_j=\sum_{k=j+1}^\infty a_j^{(k)}\lambda^k.
\ee Up to two loop, the expansion constants are
\bs\label{as}\begin{align}
    a_0^{(1)}&=-1,\\
        a_0^{(2)}&=\frac{ \log (1-z)+\log (z)+\log ((z-1) z)}{32 \pi ^2},\\
    a_0^{(3)}&=-\frac{1}{1024 \pi ^4}\Big[3 \log ^2(1-z)+3 \log ^2(z)+3 \log ^2((z-1) z)+4 \log (1-z)\nn\\
    &\quad +4 \log (z)+4 \log ((z-1) z)+2 \pi ^2+18\Big],\\
    a_1^{(2)}&=-\frac{3 }{16 \pi ^2},\\
    a_1^{(3)}&=\frac{3  (\log (1-z)+\log (z)+\log ((z-1) z)+2)}{256 \pi ^4},\\
    a_2^{(3)}&=-\frac{9}{256 \pi ^4}.
\end{align}\es 
The 2-loop Carrollian amplitude becomes
\bea 
&&C^{\text{2-loop}}(u_1,\Omega_1,-;u_2,\Omega_2,-;u_3,\Omega_3,+;u_4,\Omega_4,+)\nn\\&=&F(z)\int_0^\infty \frac{d\omega}{\omega}e^{-i\omega\chi}(a_0+a_1\log\frac{\omega}{M}+a_2\log^2\frac{\omega}{M})\nn\\&=&F(z)\lim_{q\to 0}(a_0-a_1\log M+a_2\log^2M+(a_1-2a_2\log M)\frac{d}{dq}+a_2\frac{d^2}{dq^2})J(q,\chi,\omega_0)\nn\\
&=&F(z)[\frac{1}{12} \left(\pi ^2 a_1+2 \pi ^2 a_2 \log\frac{\omega_0}{M}-8 a_2 \zeta (3)\right)\nn\\&&-\left(a_0+a_1 \log\frac{\omega_0}{M}+a_2 \log ^2\frac{\omega_0}{M}+\frac{\pi ^2 a_2}{6}\right)I_0+\left(\frac{a_1}{2}+a_2 \log \frac{\omega_0}{M}\right)I_0^2-\frac{a_2}{3}I_0^3].\label{2loopcarr}
\eea

\subsection{Callan-Symanzik equation}
To check the consistency of our result, we will show that the four-point Carrollian amplitude obeys the Callan-Symanzik equation up to 2-loop. The four-point Carrollian amplitude is abbreviated to $C_{(4)}$ below
\be 
C_{(4)}=\langle \Sigma(u_1,\Omega_1)\Sigma(u_2,\Omega_2)\Sigma(u_3,\Omega_3)\Sigma(u_4,\Omega_4)\rangle.
\ee According to our previous results, the Carrollian amplitude depends on an arbitrary energy scale $M$ and the coupling constant $\lambda$ at the same scale. The coupling constant is the renormalized one and the Carrollian amplitude $C_{(4)}$ is different from  the bare Carrollian amplitude $C_{(4)}^{(0)}$
\bea 
C_{(4)}^{(0)}=\langle \Sigma_0(u_1,\Omega_1)\Sigma_0(u_2,\Omega_2)\Sigma_0(u_3,\Omega_3)\Sigma_0(u_4,\Omega_4)\rangle
\eea where $\Sigma_0(u,\Omega)$ is the leading order coefficient in the asymptotic expansion of the bare field $\Phi_0(t,\bm x)$
\bea 
\Phi_0(t,\bm x)=\frac{\Sigma_0(u,\Omega)}{r}+\cdots.
\eea On the other hand, the field $\Sigma(u,\Omega)$ corresponds to the leading order coefficient of the renormalized field $\Phi(t,\bm x)$
\be 
\Phi(t,\bm x)=\frac{\Sigma(u,\Omega)}{r}+\cdots.
\ee The renormalized field $\Phi$ and the bare field $\Phi_0$ may be related by a rescaling factor $Z$
\be 
\Phi(t,\bm x)=Z^{-1/2}\Phi_0(t,\bm x).
\ee Therefore, the bare field $\Sigma_0$ and the renormalized field $\Sigma$ are related by the same factor 
\be 
\Sigma(u,\Omega)=Z^{-1/2}\Sigma_0(u,\Omega).
\ee Naively, one may expect the following relations 
\bea 
C_{(4)}=Z^{-2}C_{(4)}^{(0)}.
\eea 
 However, from \eqref{amputatedcon}, the Carrollian amplitude is actually related to the connected and amputated  correlation function, which implies a different behaviour and we will discuss it in detail. It is obvious that the relation between the correlation function of renormalized fields  $ G(x_1,\cdots,x_{m+n})$ 
and that of the bare fields $ G_{0} (x_1,x_2,\cdots,x_{m+n})$ are expressed as
\begin{equation}
    G (x_1,x_2,\cdots,x_{m+n}) = Z^{-\frac{m+n}{2}} G_{0} (x_1,x_2,\cdots,x_{m+n}).
\end{equation}
In our derivation of Feynman rules, the correlation function is split into external Feynman propagators and the internal amputated correlation function. The propagator of the  renormalized field  is related to that of bare field by
\begin{equation}
    \langle 0 | T \{\Phi(y_1)\Phi(y_2)\}| 0 \rangle = Z^{-1} \langle 0 | T \{\Phi_0(y_1)\Phi_0(y_2)\}| 0 \rangle.
\end{equation}
Hence the amputated correlation function for renormalized field $ G_{\text{amputated} }(x_1,x_2\cdots,x_{m+n}) $
and bare field $ G_{0,\text{amputated}}(x_1,x_2,\cdots,x_{m+n}) $ have the relation
\begin{equation}
    G_{\text{amputated}}(x_1,x_2,\cdots,x_{m+n}) = Z^{\frac{m+n}{2}} G_{0,\text{amputated}}(x_1,x_2,\cdots,x_{m+n}).
\end{equation}
Therefore we can use \eqref{amputatedcon} to derive the relation between Carrollian amplitude $C_{(m+n)}$ and the bare Carrollian amplitude $C_{(m+n)}^{(0)}$ :
\begin{equation}
	C_{(m+n)}=Z^{+\frac{m+n}{2}}C_{(m+n)}^{(0)}.
\end{equation}
For four-point Carrollian amplitude, we have:
\begin{equation}
	C_{(4)}=Z^{+2}C_{(4)}^{(0)}.
\end{equation}

\begin{figure}
    \centering
    \begin{tikzpicture}[scale=1.5]
      \draw (-1.2,0) -- (1.2,0);
      \draw (0,0) circle (0.5);
    \end{tikzpicture}
    \caption{\centering{Sunset diagram.} }
    \label{figsunset}
\end{figure}
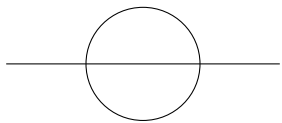

Note that the bare Carrollian amplitude is independent of the arbitrary energy scale $M$
\bea 
M\frac{d}{dM}C_{(4)}^{(0)}=0
\eea which leads to 
\bea 
M\frac{\partial}{\partial M}C_{(4)}+\beta\frac{\partial}{\partial\lambda}C_{(4)}-4\gamma C_{(4)}=0,\label{CSeqn}
\eea where we have defined the $\beta$ and $\gamma$ function as 
\be 
\beta=M\frac{\partial\lambda}{\partial M},\qquad \gamma=\frac{1}{2}M\frac{\partial\log Z}{\partial M}.
\ee The above equation \eqref{CSeqn} is the Callan-Symanzik equation for four-point Carrollian amplitude. It is easy to extend to $n$-point Carrollian amplitude 
\bea 
M\frac{\partial}{\partial M}C_{(n)}+\beta\frac{\partial}{\partial\lambda}C_{(n)}-n\gamma C_{(n)}=0.\label{ncs}
\eea 
The $\beta$ function can be used to define a running coupling constant $\bar\lambda$.
Note that the $\beta,\gamma$ functions have been computed and they are 
\bea 
\beta=\sum_{j=2}^\infty\beta^{(j)}\lambda^j,\quad 
\gamma=\sum_{j=2}^\infty \gamma^{(j)}\lambda^j.
\eea  We only need the result up to 2-loop level whose relevant coefficients are \cite{2001cppt.book.....K}
\bea 
\beta^{(2)}=\frac{3}{16\pi^2},\qquad \beta^{(3)}=-\frac{17}{3\times(4\pi)^4},\qquad \gamma^{(2)}=\frac{1}{12\times(4\pi)^4}\label{betagamma}
\eea 
where the $\gamma^{(2)}$ is obtained by analyzing the sunset diagram which is shown in figure \ref{figsunset}. 

The equation \eqref{CSeqn} should be satisfied order by order. The relevant parts are \bs\begin{align}
    M\frac{\partial}{\partial M}C_{(4)}&=F(z)[-\frac{\pi^2}{6}a_2+(a_1+2a_2\log\frac{\omega_0}{M})I_0-a_2 I_0^2]+\cdots,\\
    \beta \frac{\partial}{\partial\lambda}C_{(4)}&=F(z)\beta[\frac{\pi^2}{12}\partial_\lambda a_1-(\partial_\lambda a_0+\partial_\lambda a_1\log\frac{\omega_0}{M})I_0+\frac{1}{2}\partial_\lambda a_1 I_0^2]+\cdots,\\
    \gamma C_{(4)}&=-F(z)\gamma a_0 I_0.
\end{align}\es Up to order $\lambda^3$, they are equivalent to the following equations
\bs\begin{align}
    a_2^{(3)}&=\beta^{(2)}a_1^{(2)},\\
    a_1^{(2)}&=\beta^{(2)}a_0^{(1)},\\
    a_1^{(3)}&=2\beta^{(2)}a_0^{(2)}+\beta^{(3)}a_0^{(1)}-4\gamma^{(2)}a_0^{(1)}
\end{align}\es whose validity could be checked using the equations \eqref{as} and  \eqref{betagamma}. 

Now we will solve the Callan-Symanzik equation \eqref{ncs} to find the relation between two $n$-point Carrollian amplitudes at two different energy scales $\mu$ and $\mu'$. Note that both of them are related to the bare Carrollian amplitude, therefore we find 
\bea 
C_{(n)}(\mu)=\frac{Z^{n/2}(\mu)}{Z^{n/2}(\mu')}C_{(n)}(\mu')=\exp{[\frac{n}{2}(\log Z(\mu)-\log Z(\mu'))]}C_{(n)}(\mu').
\eea With the definition of the $\gamma$ function, we get
\bea 
C_{(n)}(\bar\lambda(\mu))=\exp\Big[n \int_{\mu'}^\mu \frac{d\mu''}{\mu''}\gamma(\bar\lambda(\mu''))\Big]\ C_{(n)}(\bar\lambda(\mu'))
\eea where we have written out the dependence on the running coupling constant explicitly in the expression. We can obtain the same result by solving \eqref{ncs} directly as in the standard textbook. By expanding the exponential in the solution up to one loop level, i.e., order $\lambda^2$,  it is obvious that the four-point Carrollian amplitude at different scales should obey the equation 
\bea 
C_{(4)}(\bar\lambda(\mu))=C_{(4)}(\bar\lambda(\mu')).\label{1-loopcarr}
\eea To check this equation, we set $\mu'=M$ and $\mu=\omega_0$. Using the fact that the running coupling constant at the scale $\omega_0$ is
\be 
\bar{\lambda}(\omega_0)\approx\frac{\lambda}{1-\frac{3\lambda}{16\pi^2}\log\frac{\omega_0}{M}}\approx \lambda+\frac{3\lambda^2}{16\pi^2}\log\frac{\omega_0}{M}
\ee up to order $\lambda^2$, indeed the four-point Carrollian amplitude at 1-loop level can be rewritten as a function of $\bar\lambda(\omega_0)$ which is abbreviated to $\bar\lambda$
\bea 
 &&C_{(4)}^{\text{1-loop}}=F(z)[\frac{\pi^2}{12}a_1(\bar\lambda)-a_0(\bar\lambda)I_0+\frac{a_1(\bar\lambda)}{2}I_0^2].
\eea At two loops, i.e., up to order $\lambda^3$, the equation \eqref{1-loopcarr} is modified to 
\bea 
C_{(4)}(\bar{\lambda}(\mu))=C_{(4)}(\bar\lambda(\mu'))+4\gamma(\bar{\lambda}(\mu'))\log\frac{\mu}{\mu'}C_{(4)}(\bar\lambda(\mu')).
\eea We still choose $\mu'=M$ and $\mu=\omega_0$, then it is straightforward to check that the four-point Carrollian amplitude \eqref{2loopcarr} at the scale $\omega_0$ is also a function of $\bar\lambda$
\bea 
C_{(4)}^{\text{2-loop}}=F(z)\sum_{j=0} b_j(\bar\lambda)I_0^j
\eea where 
the coefficients $b_j$ are
\bea 
{ b_0(\bar\lambda)=\sum_{k=2}^\infty b_0^{(k)}\bar\lambda^k, \qquad b_j(\bar\lambda)=\sum_{k=j}^\infty b_j^{(k)}\bar\lambda^k,\quad j=1,2,3,\cdots.}
\eea Up to order $\bar\lambda^3$, we have 
\bs\begin{align}
    b_0^{(2)}&=-\frac{1}{64},\\
    b_0^{(3)}&=\frac{\pi ^2 \log (1-z)+\pi ^2 \log (z)+\pi ^2 \log ((z-1) z)+24 \zeta (3)+2 \pi ^2}{1024 \pi ^4},\\
    b_1^{(1)}&=1,\\
    b_1^{(2)}&=-\frac{\log (1-z)+\log (z)+\log ((z-1) z)}{32 \pi ^2},\\
    b_1^{(3)}&=\frac{1}{1024 \pi ^4}\Big[3 \log ^2(1-z)+3 \log ^2(z)+3 \log ^2((z-1) z)+4 \log (1-z)\nn\\
    &\quad +4 \log (z)+4 \log ((z-1) z)+8 \pi ^2+18\Big],\\
    b_2^{(2)}&=-\frac{3 }{32 \pi ^2},\\
    b_2^{(3)}&=\frac{3  (\log (1-z)+\log (z)+\log ((z-1) z)+2)}{512 \pi ^4},\\
    b_3^{(3)}&=\frac{3}{256 \pi ^4}.
\end{align}\es
The relation between the running coupling constant $\bar\lambda$ and $\lambda$ is 
\be 
\lambda=\bar\lambda-\frac{3}{16 \pi ^2}\bar\lambda^2\log\frac{\omega_0}{M}+\frac{9}{256 \pi ^4}\bar\lambda^3\log^2\frac{\omega_0}{M}+\frac{17}{768 \pi ^4}\bar\lambda^3\log\frac{\omega_0}{M}+\cdots.
\ee This is consistent with the solution of the $\beta$ function equation up to $\mathcal{O}(\lambda^3)$:
\bea 
M\frac{\partial}{\partial M}\lambda=\beta(\lambda)\quad\Rightarrow\quad\bar\lambda=\lambda+\frac{3}{16 \pi ^2}\lambda^2\log\frac{\omega_0}{M}+\frac{9}{256 \pi ^4}\lambda^3\log^2\frac{\omega_0}{M}-\frac{17}{768 \pi ^4}\lambda^3\log\frac{\omega_0}{M}+\cdots\nn\\ 
\eea 
In Appendix \ref{cseqn}, we also discussed the Callan-Symanzik equation for two-point Carrollian amplitude.

\section{More Carrollian amplitudes}\label{more}

In the previous sections, we have discussed the four-point Carrollian amplitude in  massless $\Phi^4$ theory. Massless $\Phi^4$ theory is the standard QFT whose properties are studied extensively. In this section, we will reverse the method of bulk reduction and  try to propose a slightly generalized $\Phi^4$ theory in the context of flat holography. Notice that the key ingredients are the Feynman rules which are inherited from bulk in the previous sections. However, it seems that there is no obstruction to designing the Feynman rules for Carrollian amplitude.  We still assume that the boundary Carrollian field theory is Poincar\'e invariant. 
\subsection{Description}
In this generalized $\Phi^4$ theory, we don't have to write out the Lagrangian explicitly. However, the two-point correlator may be parameterized by the K\"all\'en-Lehmann spectral representation
\bea 
G_F(x,y)=\langle \bm 0|T\Phi(x)\Phi(y)|\bm 0\rangle=\int_0^\infty \frac{d\mu^2}{2\pi}\rho(\mu^2)G_F(x-y;\mu^2)
\eea where $G_F(x-y;\mu^2)$ is the Feynman propagator for a particle with mass $\mu$. The bulk field $\Phi$ still encodes a massless degree of freedom. Therefore, there may be a state that is massless in the spectral density $\rho(\mu^2)$
\be 
\rho(\mu^2)\sim \delta(\mu^2).\label{spectral}
\ee However, this spectral density will lead to the Feynman propagator in massless $\Phi^4$ theory. For the generalized $\Phi^4$ theory, we will choose the following spectral density
\bea 
\rho(\mu^2)=2\pi(\mu^2)^{\Delta-2}Z
\eea where $Z$ is a renormalization factor which may be absorbed into the redefinition of $\Phi$. The spectral density appears in the context of low-energy effective field theory which is scale invariant \cite{Georgi:2007ek}. From this spectral density, the two-point correlator in the bulk becomes 
\bea 
G_F(x,y)=\int \frac{d^4p}{(2\pi)^4}G_F(p)e^{ip\cdot(x-y)}\label{feynmanprop}
\eea where 
\be 
G_F(p)=i\left(\frac{1}{-p^2+i\epsilon}\right)^{2-\Delta}.
\ee We have fixed the normalization factor of $\Phi$ in this expression. The position space propagator may be fixed by dimensional analysis
\bea 
G_F(x,y)= \theta(x^0-y^0)W^+(x,y)+\theta(y^0-x^0)W^-(x,y)\label{Feynmandelta}
\eea where the Wightman functions $W^{\pm}$ are 
\bs\begin{align}
W^+(x,y)&=\frac{C_{\Delta}}{\left((x^0-y^0-i\epsilon)^2-(\bm x-\bm y)^2\right)^\Delta},\\
W^-(x,y)&=\frac{C_{\Delta}}{\left((x^0-y^0+i\epsilon)^2-(\bm x-\bm y)^2\right)^\Delta}.
\end{align}\es 
The coefficient $C_{\Delta}$ is an unimportant constant that can be fixed by the integration. We have used the $i\epsilon$ prescription and it may be reduced to 
\bea 
G_F(x,y)=\frac{C_{\Delta}}{\left(-(x-y)^2-i\epsilon\right)^\Delta}
\eea which is consistent with the Feynman propagator \eqref{Feynmanmassless} in the limit $\Delta\to 1$.
Note that the two-point correlator is the same  one in conformal field theory. Since the dimension of $\Phi$ is $\Delta$,  it may be expanded near $\mathcal{I}^+$ as 
\bea 
\Phi(x)=\frac{V(u,\Omega)}{r^\Delta}+\cdots.
\eea It follows that the transformation law of the boundary field $V(u,\Omega)$ is 
\bea 
V'(u',\Omega')=V(u,\Omega),\qquad u'=u-a\cdot n,\quad \Omega'=\Omega
\eea for spacetime translation and 
\bea 
V'(u',\Omega')=\Gamma^\Delta V(u,\Omega),\quad u'=\Gamma^{-1}u,\qquad \bm n'=\Gamma^{-1}\bm\Gamma
\eea for Lorentz transformation. It follows that the conformal weight of $V(u,\Omega)$ is \be 
h=\frac{\Delta}{2}.
\ee The bound \eqref{boundh} on $h$ leads to 
\be 
\Delta\ge 1
\ee which is exactly the lower bound of conformal weight for the primary field with $s=0$ in unitary CFTs of four dimensions \cite{Mack:1975je}. 
Using the relation 
\be 
y^\mu=\frac{u}{2}(n^\mu-\bar{n}^\mu)+r n^\mu
\ee between the Cartesian and retarded coordinates, we find the geodesic distance between $x$ and $y$ 
\be 
(x-y)^2=-2r(u+n\cdot x)+\mathcal{O}(r^0)
\ee in the large $r$ limit. Therefore, the  external line may be found by taking the limit $r\to\infty$ with $u$ fixed
\bea 
D(u,\Omega;x)=\langle V(u,\Omega)\Phi(x)\rangle=\lim\hspace{-0.8mm}_+ r^\Delta G_F(x,y)=\frac{C_\Delta}{2^\Delta(u+n\cdot x-i\epsilon)^\Delta}
\eea where we used the limit 
\bea 
\lim{}\hspace{-0.8mm}_+=\lim_{r\to\infty,\ u \ \text{finite}}\label{lim+}
\eea to push the field to the boundary which has been defined in \cite{Liu:2023gwa}. The $i\epsilon$ prescription is inherited from the Feynman propagator since in this case we have $y^0>x^0$. Similarly, we also have the relation 
\be 
y^\mu=\frac{v}{2}(n^\mu-\bar n^\mu)+r\bar n^\mu
\ee in advanced coordinates. The external line would be 
\be 
D(x;v,\Omega)=\langle \Phi(x)V(v,\Omega)\rangle=\lim{}\hspace{-0.8mm}_- r^\Delta G_F(x,y)=\frac{C_\Delta}{(-2)^\Delta(v-\bar n\cdot x+i\epsilon)^\Delta}
\ee 
where 
\bea 
\lim{}\hspace{-0.8mm}_-=\lim_{r\to\infty,\ v\ \text{finite}}.
\eea  
The antipodal map would be 
\bea 
(v,\Omega)\to (u,\Omega^P),\quad V(v,\Omega)\to V(u,\Omega^P)
\eea and then the external lines may be unified as 
\bea 
D(u,\Omega;x)=\frac{C_\Delta}{(2\sigma)^\Delta (u+n\cdot x-i\sigma\epsilon)^\Delta}\label{externalline}
\eea where 
$\sigma$ is to take into account the incoming and outgoing states.
Its integral representation would be 
\bea 
D(u,\Omega;x)=D_\Delta\int_0^\infty d\omega \omega^{\Delta-1}e^{-i\sigma\omega (u+n\cdot x-i\sigma\epsilon)}
\eea where
\be 
D_{\Delta}=\left(\frac{i}{2}\right)^\Delta\frac{ C_\Delta}{\Gamma(\Delta)}.
\ee 

We should also design the vertices for the field $\Phi$ in the bulk. As massless $\Phi^4$ theory,
we assume that there is only one type of vertex in the bulk which corresponds to the interaction of four fields $\Phi$. For each vertex located at $x$, we should integrate it with 
\be -i\lambda \int d^4x.\label{verticesPhi4}
\ee Therefore, the generalized $\Phi^4$ theory is described by the propagator \eqref{feynmanprop}, the external line \eqref{externalline} and the vertex \eqref{verticesPhi4} as well as the symmetry factor.
\subsection{Tree level}
As in the previous sections, the tree-level four-point Carrollian amplitude in generalized $\Phi^4$ theory is 
\bea 
C_{(4)}^{\text{tree}}&\equiv& \langle\prod_{j=1}^4 V(u_j,\Omega_j)\rangle\nn\\&=&-i\lambda \int d^4x \prod_{j=1}^4 D(u_j,\Omega_j;x)\nn\\&=&-i\lambda D_\Delta^4\int d^4x\prod_{j=1}^4 \int_0^\infty d\omega_j\omega_j^{\Delta-1}e^{-i\sigma_j\omega_j(u_j+n_j\cdot x-i\sigma_j\epsilon)}.
\eea The integration of the bulk point $x$ leads to the conservation of momentum $p_j=\sigma_j\omega_jn_j$ and we still fix $z_1=0,z_3=1,z_4=\infty$ using Lorentz transformation, then 
\bea 
C_{(4)}^{\text{tree}}&=&{ -i\lambda D_\Delta^4}\prod_{j=1}^4 \int_0^\infty d\omega_j\omega_j^{\Delta-1}e^{-i\sigma_j\omega_ju_j}(2\pi)^4\delta^{(4)}(\sum_{j=1}^4 p_j)\nn\\&=&-\lambda F_\Delta(z)\int_0^\infty d\omega \omega^{4\Delta-5}e^{-i\omega\chi}
\eea where $\chi$ is exactly the same as \eqref{chicomplete} and the generalized function $F_\Delta(z)$ is 
\bea 
F_{\Delta}(z)=iD_\Delta^4 (2\pi)^4\frac{(1+z^2)^\Delta}{2^{2-\Delta}}z^{2-2\Delta}(z-1)^{2-2\Delta}\Theta(z-1)\delta(\bar z-z).
\eea As before, we already set $\sigma_1=\sigma_2=-1,\sigma_3=\sigma_4=1$ to simplify notation. For $\Delta=1$, the tree-level result is the same as the massless $\Phi^4$ theory. For $\Delta>1$, the integral is convergent 
\bea 
C_{(4)}^{\text{tree}}=-\lambda F_\Delta(z)\Gamma (4 \Delta -4) (i \chi)^{4-4 \Delta }.\label{singularity}
\eea The amplitude is fine for general choices of $u_i$ except for a hyperplane in the  $(u_1,u_2,u_3,u_4)$ space defined by the equation 
\be 
\chi=0.
\ee  Note that the singularity of \eqref{singularity} becomes a branch point in massless $\Phi^4$ theory. The physical meaning of this singular plane is unclear at this moment.
\subsection{Loop corrections}
Now we may evaluate the loop corrections to ensure that the generalized $\Phi^4$ theory is well defined. In the $s$ channel, the 1-loop correction is 
 \bea 
C_{(4)}^{\text{1-loop,$s$ channel}}&=&\frac{(-i\lambda)^2}{2}\int d^4x \int d^4 y D(u_1,\Omega_1;x)D(u_2,\Omega_2;x)G^2_F(x,y)D(u_3,\Omega_3;y)D(u_4,\Omega_4;y)\nn\\&=&-\frac{\lambda^2}{2}D_\Delta^4 \prod_{j=1}^4 \int_0^\infty d\omega_j \omega_j^{\Delta-1}e^{-i\sigma_j\omega_j u_j}\int \frac{d^4q}{(2\pi)^4}G_F(p_1+p_2+q)G_F(q)(2\pi)^4\delta^{(4)}(\sum_{j=1}^4 p_j)\nn\\&=&\frac{(-1)^{-2\Delta} \lambda^2}{2(4\pi)^2}\frac{\Gamma(2-2\Delta)\Gamma^2(\Delta)}{\Gamma(2\Delta)\Gamma^2 (2-\Delta)}F_\Delta(z)\left(-\frac{4}{z-1}\right)^{2\Delta-2}\Gamma (8 \Delta -8) (i \chi)^{8-8 \Delta },\quad\Delta>1.\nn\\
\eea

 The integral of the momentum $q$ is divergent superficially and can be obtained by analytic continuation with dimensional regularization \cite{tHooft:1972tcz,Cicuta:1972jf}. 
At the last step, we have set $\Delta>1$. We will discuss a bit more on the coefficient  
\be 
e^{\text{1-loop}}_\Delta=\frac{\Gamma (2-2 \Delta ) \Gamma^2 (\Delta ) \Gamma (8 \Delta -8)}{\Gamma (2 \Delta ) \Gamma^2 (2-\Delta )}.
\ee For general $\Delta>1$, it is well defined except for $\Delta=\frac{3}{2},\frac{5}{2},\cdots$, where the coefficient $e_{\Delta}$ is always divergent. Another notable feature is that $e_\Delta$ is always zero for integer $\Delta=2,3,4,\cdots$.

Now we can also include the $t,u$ channel, then the 1-loop correction for the generalized $\Phi^4$ theory is 
\bea 
C_{(4)}^{\text{1-loop}}=\frac{(-1)^{-2\Delta}\lambda^2}{2(4\pi)^2}F_{\Delta}(z)e^\text{1-loop}_\Delta \left\{\left(-\frac{4}{z-1}\right)^{2\Delta-2}+\left(\frac{4}{z(z-1)}\right)^{2\Delta-2}+\left(\frac{4}{z}\right)^{2\Delta-2}\right\}(i\chi)^{8-8\Delta}.\nn\\
\eea We conclude that the 1-loop correction is free from divergences for $\Delta>1$ and $\Delta\not=\frac{3}{2},\frac{5}{2},\cdots$. For 2-loop corrections,  the $\mathcal{M}$ matrix of the first diagram of figure \ref{2loop} is proportional to the square of the one-loop correction
\bea 
C_{(4)}^{\text{2-loop},1}=-\frac{(-1)^{-4\Delta}\lambda^3}{(4\pi)^4} F_\Delta(z)e_{\Delta}^{\text{2-loop},1}\left\{\left(-\frac{4}{z-1}\right)^{4\Delta-4}+\left(\frac{4}{z(z-1)}\right)^{4\Delta-4}+\left(\frac{4}{z}\right)^{4\Delta-4}\right\}(i\chi)^{12-12\Delta}\nn\\
\eea with 
\bea 
e_\Delta^{\text{2-loop},1}=\frac{ \Gamma (2-2 \Delta )^2 \Gamma (\Delta )^4 \Gamma(12\Delta-12)}{4\Gamma (2-\Delta )^4 \Gamma \left(2\Delta \right)^2}.
\eea The second diagram vanishes and the contribution of the  third diagram
is 
\bea 
C_{(4)}^{\text{2-loop},3}=-\frac{(-1)^{-4\Delta}\lambda^3}{(4\pi)^4} F_\Delta(z)e_{\Delta}^{\text{2-loop},3}\left\{\left(-\frac{4}{z-1}\right)^{4\Delta-4}+\left(\frac{4}{z(z-1)}\right)^{4\Delta-4}+\left(\frac{4}{z}\right)^{4\Delta-4}\right\}(i\chi)^{12-12\Delta}\nn\\
\eea with 
\bea 
e_\Delta^{\text{2-loop},3}=\frac{ \Gamma (4-4 \Delta ) \Gamma (2-2 \Delta ) \Gamma (\Delta )^2 \Gamma (3 \Delta -2)^2\Gamma(12\Delta-12)}{\Gamma (2-\Delta )^4 \Gamma \left(2\Delta \right) \Gamma (4 \Delta -2)}.
\eea
There is a subtlety in the calculation. The Feynman parameterization leads to an integral \bea 
\int_0^1 dx\int_0^{1-x}dy\  x^{3\Delta+d-7}(1-x-y)^{3\Delta+d-7}y^{3-2\Delta-d/2}
\eea which is divergent for $\Delta>1$ and $d=4$. We regularize this integral by restricting $d$ to the range $6-3\Delta<d<8-4\Delta$ and then continue it to $d=4$. 
 The rest of Feynman diagrams are trivial, since the 1PI bubbles on external legs evaluate zero. Therefore, the nonvanishing two-loop graphs are simply those without 1PI subgraphs on external legs. 
Sum up all Feynman diagrams contributing to 2-loop corrections and then we can derive the 2-loop correction for the generalized $ \Phi^4 $ theory:
\bea 
C_{(4)}^{\text{2-loop}}=-\frac{(-1)^{-4\Delta}\lambda^3}{(4\pi)^4} F_\Delta(z)e_{\Delta}^{\text{2-loop}}\left\{\left(-\frac{4}{z-1}\right)^{4\Delta-4}+\left(\frac{4}{z(z-1)}\right)^{4\Delta-4}+\left(\frac{4}{z}\right)^{4\Delta-4}\right\}(i\chi)^{12-12\Delta}\nn\\
\eea with

\bea 
e_{\Delta}^{\text{2-loop}}=\frac{\Gamma(2-2\Delta)\Gamma(\Delta)^2\Gamma(12\Delta-12)}{\Gamma(2-\Delta)^4\Gamma(2\Delta)}\left(\frac{\Gamma(4-4\Delta)\Gamma(3\Delta-2)^2}{\Gamma(4\Delta-2)}+\frac{\Gamma(2-2\Delta)\Gamma(\Delta)^2}{4\Gamma(2\Delta)}\right)
\eea 

The conclusion is that the four-point Carrollian amplitude of the  generalized $\Phi^4$ theory is free from UV and IR divergences up to two-loop level.
It seems that the generalized $\Phi^4$ theory is always finite and the $\beta$ function vanishes, although this conjecture should be checked at higher loops.

\section{Conclusion and outlook}\label{conc}
In this paper, we work out the details on the  Carrollian amplitude in the framework  of bulk reduction. Based upon the correspondence between the asymptotic states and the fundamental fields at future/past null infinity, we build the connection between scattering amplitude and Carrollian amplitude. The Carrollian amplitude could be regarded as the correlator with the operators inserted at $\mathcal{I}^+/\mathcal{I}^-$. We derive the Feynman rules to calculate the Carrollian amplitude in Carrollian space. The Feynman rules include the boundary-to-boundary propagator which connects the $\mathcal{I}^-$ to $\mathcal{I}^+$ and the external line which links a bulk field and a boundary field. We also need the standard Feynman rules in the bulk. In this representation, no Fourier transform is needed. Taking advantage of these preparations, we delve into the four-point Carrollian amplitude for massless $\Phi^4$ theory. As a first step, we make use of the Lorentz transformation to fix three operators to $z=0,1,\infty$. The resulting amplitude only depends on the cross ratio of  the celestial sphere as well as $\chi$, a linear superposition of the inserting time which itself is fixed by translation invariance. Then we compute the four-point Carrollian amplitude up to two loops. At the tree level, the  Carrollian amplitude is a linear function of $I_0$, whose coefficient is an analytic function of the cross ratio $z$. Interestingly, the function $I_0$ also appears in the two-point Carrollian amplitude, although the argument is slightly different. Our results match with recent literature after a suitable Lorentz transformation. At the one-loop level, the Carrollian amplitude is a quadratic polynomial of $I_0$, whose coefficients are also functions of the cross ratio. The branch points of the coefficients are related to the unitarity of the scattering amplitude. We find a discontinuity for the Carrollian amplitude when crossing the branch cuts, similar to the Optical theorem of the momentum space scattering amplitude. At the two-loop level, the Carrollian amplitude is a cubic polynomial of $I_0$, whose coefficients are much more involved functions of the cross ratio. However, the branch points are still located at $0,1,\infty$. 
Note that what we calculated is the renormalized Carrollian amplitude instead of the bare Carrollian amplitude. The coupling constant in the expressions depends on the observational energy scale $M$. We have checked the consistency of the renormalized Carrollian amplitude under renormalization group flow, i.e., Callan-Symanzik equation of the Carrollian amplitude is satisfied. One should also notice that the infrared divergences are avoided by carefully discarding the IR modes of one of the boundary fields.
Based on these results, it is natural to conjecture that the $n$-loop results for the perturbative Carrollian amplitude is a polynomial of $I_0$ with degree $n+1$
\bea 
C^{\text{$n$-loop}}_{(n)}=F(z)\sum_{j=0}^{n+1} a_j I_0^j
\eea where $a_j$ are functions of the cross ratio and free from UV divergences. They depend on the renormalized coupling constant and the energy scale as well as the IR cutoff $\omega_0$.  There are various open questions that deserve further study. 
\begin{itemize}
    \item We only compute the four-point Carrollian amplitude for  massless scalar theory up to 2-loop level in this work. It is nice to extend it to higher loops and a resummation of the perturbative corrections is also interesting. One can also extend the results to higher point  Carrollian amplitudes. The massless $
\Phi^4$ theory suffers dynamical symmetry breaking due to Coleman-Weinberg mechanism \cite{Coleman:1973jx} where the effective potential receives quantum corrections such that the vacuum expectation value of $\Phi$ becomes nontrivial. As a consequence, the massless scalar requires a nonvanishing mass. Still, the Carrollian amplitude for massless $\Phi^4$ theory deserves study since all the method in this work can be extended to nontrivial theories with nonvanishing helicities. 
    \item Our results show that Carrollian amplitude is a natural quantity in Carrollian space for massless theory, even at the loop level. All the nice properties in momentum space scattering amplitude may reflect in Carrollian amplitude. In this work, the four-point Carrollian amplitude only depends on two variables $z$ and $\chi$ as a consequence of Poincar\'e symmetry. The two variables factorize in the Carrollian amplitude. The Carrollian amplitude, as a function of complex $z$, has branch cuts  at the loop level, which reflects the unitarity of the $S$ matrix. On the other hand, as a function of $\chi$, it also has  a branch point or isolated singularity at $\chi=0$ which is a hyperplane in the parametric space of the retarded time. In the program of $S$ matrix bootstrap \cite{Chew:1961}, the analytic properties, 
   such as poles and branch cuts are rather important to determine the complete $S$ matrix. In the Carrollian amplitude approach, the poles and branch cuts are  transferred to the analytic properties of the four-point amplitude, which is the function of two complex variables.
    There may be a ``Carrollian amplitude bootstrap'' by imposing nice properties on the Carrollian amplitude and determining it completely. This program deserves further study in the future.    
    
    \item Feynman rule for Carrollian amplitude from boundary field theory. In our work, the bulk theory is the  well-known massless scalar theory and the boundary theory is found by bulk reduction. Given a boundary theory without the knowledge of the bulk theory, it would be fine to find out the corresponding Feynman rules to derive the Carrollian amplitude. Based on this observation, we try to design the Feynman rules in a toy model, i.e., generalized $\Phi^4$ theory to construct the Carrollian amplitude.  
   The spectral density in our toy model takes the form of the unparticle of scale dimension $\Delta$ whose Lagrangian is unusual. Using Feynman rules instead of bulk scattering amplitudes, we finally arrive at results of Carrollian amplitudes free from UV and IR divergences at two loops, which suggests that the model may possess better UV and IR properties. If we further relax the constraint on $\Delta$, we notice that the propagator here may be related to the higher derivative scalar field theories which may be traced back to \cite{1950PhRv...79..145P}.
    Higher derivative theories usually have better  ultraviolet behavior than the familiar second derivative theories. However, generally, they may suffer Ostrogradsky instability \cite{Ost:1850} and in the quantum version, there would be ghosts with negative norms which make the theory non-unitary \cite{Stelle:1976gc}. There are also discussions on the higher derivative theories motivated by AdS/CFT \cite{Brust:2016gjy} and it is shown that the interacting theory may capture the critical behaviour around the Wilson-Fisher fixed point\cite{Gracey:2017erc}. Interestingly, it has been shown that higher derivative theories can be free from negative modes in some special cases \cite{Mannheim:2004qz}. It would be rather interesting to explore various aspects  of these theories in the context of Carrollian amplitude. 
  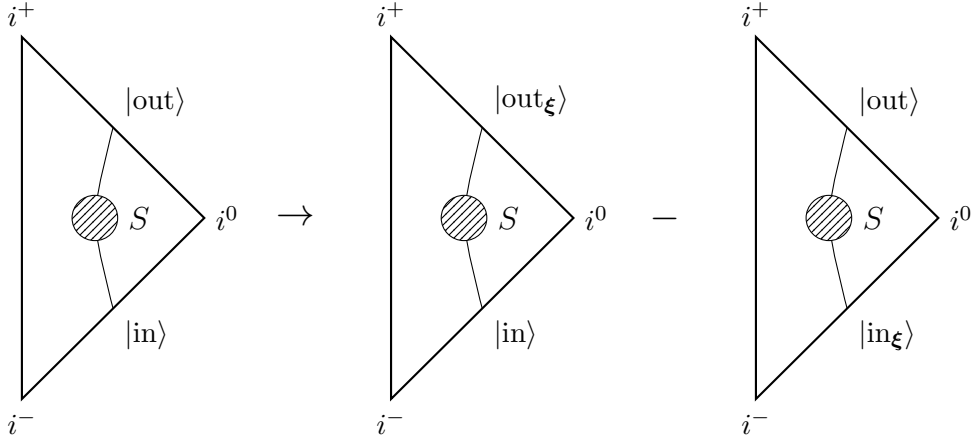
\begin{figure}
    \centering
    \begin{tikzpicture}[scale=0.6]
    \draw[thick] (0,4) node[above]{\small $i^+$} -- (2,2) node[above right]{\small $\ket{\text{out}}$} -- (4,0) node[right]{\small $i^0$}  -- (2,-2) node[below right]{\small $\ket{\text{in}}$} -- (0,-4) node[below]{\small $i^-$} -- cycle;
    \draw plot[smooth,tension=0.44] coordinates{(2,2) (1.6,0) (2,-2)};
    \fill[white](1.6,0) circle (0.5);
    \filldraw[pattern=north east lines] (1.6,0) circle (0.5);
    \node[right] at (2.1,0) {$S$};
    \node at (6,0) {\large $\rightarrow$};
  \end{tikzpicture}\hspace{0.4cm}
  \begin{tikzpicture}[scale=0.6]
    \draw[thick] (0,4) node[above]{\small $i^+$} -- (2,2) node[above right]{\small $\ket{\text{out}_{\bm\xi}}$} -- (4,0) node[right]{\small $i^0$}  -- (2,-2) node[below right]{\small $\ket{\text{in}}$} -- (0,-4) node[below]{\small $i^-$} -- cycle;
    \draw plot[smooth,tension=0.44] coordinates{(2,2) (1.6,0) (2,-2)};
    \fill[white](1.6,0) circle (0.5);
    \filldraw[pattern=north east lines] (1.6,0) circle (0.5);
    \node[right] at (2.1,0) {${S}$};
     \node at (6,0) {\large $-$};
  \end{tikzpicture}\hspace{0.4cm}
  \begin{tikzpicture}[scale=0.6]
    \draw[thick] (0,4) node[above]{\small $i^+$} -- (2,2) node[above right]{\small $\ket{\text{out}}$} -- (4,0) node[right]{\small $i^0$}  -- (2,-2) node[below right]{\small $\ket{\text{in}_{\bm\xi}}$} -- (0,-4) node[below]{\small $i^-$} -- cycle;
    \draw plot[smooth,tension=0.44] coordinates{(2,2) (1.6,0) (2,-2)};
    \fill[white](1.6,0) circle (0.5);
    \filldraw[pattern=north east lines] (1.6,0) circle (0.5);
    \node[right] at (2.1,0) {${S}$};
  \end{tikzpicture}
    \caption{Variation of $S$ matrix under infinitesimal Carrollian diffeomorphism. The asymptotic states are denoted as $|\text{in}\rangle$ and $|\text{out}\rangle$. They are transformed to the states $|\text{in}_{\bm\xi}\rangle$ and $|\text{out}_{\bm\xi}\rangle$, respectively. }
    \label{cardiff}
\end{figure}
    \item Carrollian diffeomorphism. In this work, we derive the Carrollian amplitude without touching the Carrollian diffeomorphism, which is shown to transform the boundary fields in a nontrivial way in a series of papers\cite{Liu:2022mne,Liu:2023qtr,Liu:2023gwa,Li:2023xrr,Liu:2023jnc,Liu:2024nkc}. For example, for a general supertranslation which is generated by $\bm\xi_f$, we have
    \bea 
    \delta_{\bm f}\Sigma(u,\Omega)=-f(u,\Omega)\dot\Sigma(u,\Omega).
    \eea  In the language of boundary state $|\Sigma(u,\Omega)\rangle$, the infinitesimal transformation of the boundary state would be 
    \bea 
   |\Sigma'(u,\Omega)\rangle=|\Sigma(u,\Omega)\rangle-f(u,\Omega)|\dot\Sigma(u,\Omega)\rangle.
    \eea In terms of the flux operators $Q_{\bm\xi}$ defined at the null boundary, the transformation may be written as 
    \bea 
    |\Sigma'(u,\Omega)\rangle=|\Sigma(u,\Omega)\rangle+Q_{\bm\xi}|\Sigma(u,\Omega)\rangle.
    \eea This shows that the asymptotic state is transformed to another state which is physically distinguishable and the Fock space of the incoming and outgoing states should be organized by the Carrollian diffeomorphism.
    As a consequence, the $S$ matrix 
    \be 
    {}_{\text{out}}\langle \prod_{k=m+1}^{m+n}\Sigma(u_k,\Omega_k)|\prod_{k=1}^m \Xi(v_k,\Omega_k^P)\rangle_{\text{in}}
   \ee should be transformed to 
   \bea 
   &&{}_{\text{out}}\langle \prod_{k=m+1}^{m+n}\Sigma'(u_k,\Omega_k)|\prod_{k=1}^m \Xi'(v_k,\Omega_k^P)\rangle_{\text{in}}\\&=&{}_{\text{out}}\langle \prod_{k=m+1}^{m+n}\Sigma(u_k,\Omega_k)|\prod_{k=1}^m \Xi(v_k,\Omega_k^P)\rangle_{\text{in}}+{}_{\text{out}}\langle \prod_{k=m+1}^{m+n}\Sigma(u_k,\Omega_k)|Q_{\bm\xi}|\prod_{k=1}^m \Xi(v_k,\Omega_k^P)\rangle_{\text{in}}.\nn
   \eea The flux operator $Q_{\bm\xi}$ is  $Q^{(+/-)}_{\bm\xi}$ at $\mathcal{I}^{+/-}$ and therefore the last term of the above equation would be 
   \bea 
 \delta_{\bm\xi} S(m\to n)= \langle \prod_{k=m+1}^{m+n}\Sigma(u_k,\Omega_k)|Q^{(+)}_{\bm\xi}S-SQ^{(-)}_{\bm\xi}|\prod_{k=1}^m \Xi(v_k,\Omega_k^P)\rangle,\label{cd}
   \eea where the minus sign before $Q_{\bm\xi}^{(-)}$ is from sign flip by the  definition of the flux operator at $\mathcal{I}^-$. The notation $ \delta_{\bm\xi} S(m\to n)$ is the transformation of $m\to n$ scattering matrix under (infinitesimal) Carrollian diffeomorphism.  The variation of $S$ matrix under infinitesimal Carrollian diffeomorphism is shown in figure \ref{cardiff}. Usually, for a Poincar\'e transformation, the $S$ matrix should be invariant and the previous equation becomes zero, leading to the  Ward identity. However, for a general Carrollian diffeomorphism, the equation \eqref{cd} is already nonvanishing at the classical level\cite{Li:2023xrr}
   \bea 
   Q_{\bm\xi}^{(+)}-Q_{\bm\xi}^{(-)}=\frac{1}{2}\int_{\text{bulk}} d^4x T^{\mu\nu}\delta_{\bm\xi}g_{\mu\nu}.\label{cdclass}
   \eea When there is no anomaly, the above classical equation may be turned into the quantum version \cite{Liu:2024nkc}
   \bea 
  \delta_{\bm\xi}S(m\to n)=\frac{1}{2}\int_{\text{bulk}}d^4x\  {}_{\text{out}}\langle \prod_{k=m+1}^{m+n}\Sigma(u_k,\Omega_k)| T^{\mu\nu}\delta_{\bm\xi}g_{\mu\nu}|\prod_{k=1}^m \Xi(v_k,\Omega_k^P)\rangle_{\text{in}}.\label{stressins}
   \eea  Note that the flux operator is constructed from the hard part of the flux. For BMS transformation, the right-hand side of \eqref{stressins} should be equal to the contribution from one additional soft mode insertion, as has been found by \cite{Strominger:2013jfa}
   \bea 
   \delta_{\bm\xi}S(m\to n)={}_{\text{out}}\langle \prod_{k=m+1}^{m+n}\Sigma(u_k,\Omega_k)|\  \text{soft graviton}\ |\prod_{k=1}^m \Xi(v_k,\Omega_k^P)\rangle_{\text{in}}.
   \eea 
   For Carrollian diffeomorphism, the right-hand side may be modified to the insertion of a graviton into the scattering amplitude which is induced by Carrollian diffeomorphism (CD) in the bulk\bea 
    \delta_{\bm\xi}S(m\to n)={}_{\text{out}}\langle \prod_{k=m+1}^{m+n}\Sigma(u_k,\Omega_k)|\  \text{graviton induced by CD}\ |\prod_{k=1}^m \Xi(v_k,\Omega_k^P)\rangle_{\text{in}}.\label{grav}
   \eea The equivalence between \eqref{stressins} and \eqref{grav} should be checked in the context of flat holography.
   The essential part of the right-hand side of \eqref{stressins} is the insertion of the stress tensor in the Carrollian amplitude
   \bea 
   {}_{\text{out}}\langle \prod_{k=m+1}^{m+n}\Sigma(u_k,\Omega_k)| T^{\mu\nu}(x)|\prod_{k=1}^m \Xi(v_k,\Omega_k^P)\rangle_{\text{in}}.
   \eea  In the momentum space, the energy-momentum tensor matrix becomes  
   \bea 
 {}_{\text{out}}\langle \bm p_{\bm m+1}\cdots \bm p_{m+n}|T^{\mu\nu}(q)|\bm p_1\cdots \bm p_{m}\rangle_{\text{in}}
   \eea and reduces to the gravitational form factors \cite{1963JETP...16.1343K,Pagels:1966zza,Jaffe:1989jz} in the $1\to 1$ scattering process. It would be very interesting to explore this issue in the future. 
\end{itemize} 
\vspace{10pt}
{\noindent \bf Acknowledgments.} 
The work of J.L. was supported by NSFC Grant No. 12005069.

\appendix
\section{Doppler effect}\label{doppler}
We assume that the light is emitted by an inertial observer $A$. In the reference frame $\Sigma_A$ of the inertial observer, the light propagates in the direction of $\bm n$ and its frequency is $\omega_{A}=\omega$. Then in the frame $\Sigma_A$ the four-momentum of the light is 
\bea 
p^\mu=\omega(1,\bm n).
\eea The light is detected by another observer $B$ with constant velocity $\bm\beta$ with respect to $A$, four-velocity of $B$ in the frame $\Sigma_A$ is 
\be 
u_B^\mu=\gamma(1,\bm\beta),\qquad \gamma=(1-\beta^2)^{-1/2}.
\ee Therefore, the frequency observed by $B$ is 
\bea 
\omega_{B}=-u^\mu_B p_\mu=\gamma\omega(1-\bm\beta\cdot\bm n).
\eea There is a redshift for the frequency of the light which is exactly the $\Gamma$ defined in the context
\bea 
\frac{\omega_B}{\omega_A}=\gamma(1-\bm\beta\cdot\bm n)=\Gamma.
\eea 

\section{Split representation of the Feynman propagator}\label{iep}
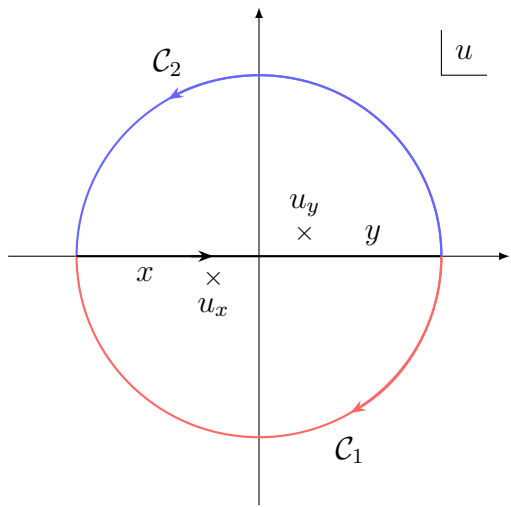
\begin{figure}
    \centering
    \begin{tikzpicture}[scale=1.5]
    \draw [-latex] (-2.2, 0) -- (2.2, 0);
    \draw [-latex] (0, -2.2) -- (0, 2.2);
    \draw (2,1.6)--(1.6,1.6)--(1.6,2);
    \node at(1.8,1.8){$u$};
    \node [below] at (-1, 0) {$x$};
    \node [above] at (1, 0) {$y$};
    \node  at (0.4, 0.2) {\footnotesize $\times$};
    \node  at (-0.4, -0.2) {\footnotesize $\times$};
    \node  at (-0.4, -0.45) {$u_x$};
    \node  at (0.4, 0.45) {$u_y$};
     \draw[thick, blue!60] (1.6,0) arc (0:180:1.6);
    \draw[thick, red!60] (-1.6,0) arc (180:360:1.6);
    \draw[-{Stealth[length=2mm,width=1.5mm]},thick] (-0.6,0) -- (0.-0.4,0);
    \draw [thick] (-1.6, 0)-- (1.6, 0);
    \draw[-{Stealth[length=2mm,width=1.5mm]},thick, blue!60] (1.6,0) arc (0:120:1.6);
    \draw[-{Stealth[length=2mm,width=1.5mm]},thick,red!60] (1.6,0) arc (360:300:1.6);
    \node[below] at (0.8,-1.5) {$\mathcal{C}_1$};
    \node[above] at (-0.8,1.5) {$\mathcal{C}_2$};
  \end{tikzpicture}
    \caption{The integrand is singular at $u=u_x$ and $u_y$ in the complex $u$ plane. We also show the contours $\mathcal{C}_1$ (the red curve) and $\mathcal{C}_2$ (the blue curve) to evaluate the integral.  }
    \label{contours}
\end{figure}

In this section, we will derive the split representation of the Feynman propagator \eqref{split}. We will prove it for $x^0>y^0$. The right-hand side of \eqref{split} is 
\bea 
\text{RHS}&=&2i\int du d\Omega D^*(u,\Omega;x)\partial_u D(u,\Omega;y)\nn\\&=&-\frac{i}{32\pi^4}\int du d\Omega \frac{1}{(u+n\cdot x+i\epsilon)(u+n\cdot y-i\epsilon)^2}.
\eea The integrand has two poles in the complex $u$ plane 
\bea 
u_x=-n\cdot x-i\epsilon,\quad u_y=-n\cdot y+i\epsilon
\eea where $u_x$ locates in the lower  plane and $u_y$ locates in the upper plane. We may use residue theorem to evaluate the integration. In the first method, we choose the contour $\mathcal{C}_1$ in the lower complex $u$ plane and only the residue at $u=u_x$ contributes 
\bea 
\text{RHS}&=&-2\pi i \times \left(-\frac{i}{32\pi^4}\right)\int d\Omega\ \text{Res}_{u=u_x} \frac{1}{(u+n\cdot x+i\epsilon)(u+n\cdot y-i\epsilon)^2}\nn\\&=&-\frac{1}{16\pi^3}\int d\Omega \frac{1}{\left(n\cdot(x-y)+i\epsilon\right)^2}.\label{firstmeth}
\eea 
We may also choose the contour $\mathcal{C}_2$ in the upper complex $u$ plane and then we can pick up the residue at $u=u_y$
\bea 
\text{RHS}&=&2\pi i \times \left(-\frac{i}{32\pi^4}\right)\int d\Omega\ \text{Res}_{u=u_y} \frac{1}{(u+n\cdot x+i\epsilon)(u+n\cdot y-i\epsilon)^2}\nn\\&=&-\frac{1}{16\pi^3}\int d\Omega \frac{1}{\left(n\cdot(x-y)+i\epsilon\right)^2}. \label{secondmeth}
\eea The result matches with the one using the first method. The contour $\mathcal{C}_1$ and $\mathcal{C}_2$ as well as the poles $u_x,u_y$ are shown in figure \ref{contours}. Now the integration becomes 
\bea 
\text{RHS}&=&-\frac{1}{16\pi^3}\int d\Omega \frac{1}{\left(-(x^0-y^0-i\epsilon)+\bm n\cdot (\bm x-\bm y)\right)^2}\nn\\&=&-\frac{1}{16\pi^3}\int \sin\gamma d\gamma d\phi \frac{1}{\left(-(x^0-y^0-i\epsilon)+ |\bm x-\bm y|\sin\gamma\right)^2}\nn\\&=&\frac{1}{4\pi^2}\frac{1}{|\bm x-\bm y|^2-(x^0-y^0-i\epsilon)^2}.
\eea Since $x^0>y^0$, the result can be rewritten as 
\bea 
\text{RHS}=\frac{1}{4\pi^2\left((x-y)^2+i\epsilon\right)}=G_F(x,y).
\eea 

\section{The cross ratio and  $s,t,u$ channels}\label{zvalue}
For brevity, we define 
\bea 
\Theta_{\sigma_1\sigma_2\sigma_3\sigma_4}= \Theta(-\frac{\sigma_4}{z\sigma_1})\Theta(\frac{1+z^2}{z(1-z)}\frac{\sigma_4}{\sigma_2})\Theta(-\frac{2}{1-z}\frac{\sigma_4}{\sigma_3}).
\eea In the context, we find that $\Theta_{--++}$ is nonvanishing only for $z>1$
\bea 
\Theta_{--++}=\Theta(z-1).
\eea In the following table \ref{physicalconst}, we find the range of $z$ with nonvanishing $\Theta_{\sigma_1\sigma_2\sigma_3\sigma_4}$ for all permutations of $\sigma_1\sigma_2\sigma_3\sigma_4$.
\begin{table}
\begin{center}
\renewcommand\arraystretch{1.5}
    \begin{tabular}{|c|c|c|c||c|}\hline
 $\sigma_1$&$\sigma_2$& $\sigma_3$&$\sigma_4$&$z$\\\hline\hline
$-$&$-$&+&+&$(1,\infty)$\\\hline
 $-$&+&$-$&+&$(0,1)$\\\hline
  $-$&+&+&$-$&$(-\infty,0)$\\\hline
    +&$-$&$-$&+&$(-\infty,0)$\\\hline
      +&$-$&+&$-$&$(0,1)$\\\hline
        +&+&$-$&$-$&$(1,\infty)$\\\hline
\end{tabular}
\caption{\centering{The value of $z$ with nonvanishing $\Theta_{\sigma_1\sigma_2\sigma_3\sigma_4}$}}\label{physicalconst}
\end{center}
\end{table} It shows that the value of $z$ should be in the range $z>1$ for $s$ channel, $0<z<1$ for $t$ channel and $z<0$ for $u$ channel.

\section{Incomplete Gamma function and related Integrals}\label{gamma}
In the context, we have defined the following integral 
\bea 
J(q,\chi,\omega_0)=\int_0^\infty \frac{d\omega}{\omega^{1-q}}e^{-i\omega \chi}\Theta(\omega-\omega_0)
\eea 
which can be used to generate the following classes of integrals
\bea 
J_n(\chi,\omega_0)=\int_0^\infty \frac{d\omega}{\omega}\log^n\omega e^{-i\omega \chi}\Theta(\omega-\omega_0)=\lim_{q\to 0}\frac{d^n}{dq^n}J(q,\chi,\omega_0).\label{Indef}
\eea 
By inserting a $-i\epsilon$ into $\chi$ factor, the integral can be found as 
\bea 
J(q,\chi,\omega_0)=(i \chi )^{-q} \Gamma (q,i\omega_0 \chi)\label{Jq}
\eea where $\chi$ should be understood as $\chi-i\epsilon$. The incomplete Gamma function is defined as \cite{1972hmfw.book.....A}
\bea 
\Gamma(q,x)=\int_x^{\infty }dt \ t^{q-1} e^{-t}.
\eea 
It is the Gamma function for $x=0$
\be 
\Gamma(q,x=0)=\Gamma(q).
\ee 
The Digamma function is the derivative of the logarithmic of the Gamma function 
\bea 
\psi(q)=\frac{d}{dq}\log\Gamma(q)=\frac{\Gamma'(q)}{\Gamma(q)}.\label{Digamma}
\eea The Digamma function is analytic on the complex plane except for the points $q=0,-1,-2,-3,\cdots$ where the function is singular. The Digamma function satisfies the Recurrence formula
\be 
\psi(q+1)=\psi(q)+\frac{1}{q}.\label{rec}
\ee 
The Euler constant $\gamma_E$ is related to the Digamma function by 
\be 
\psi(1)=-\gamma_E.\label{Eulergamma}
\ee Therefore, we may use the relation \eqref{rec} to find the exact value of $\psi(n),\ n=1,2,3,\cdots$. The Duplication formula for the Digamma function is 
\be 
\Psi(2q)=\frac{1}{2}\psi(q)+\frac{1}{2}\psi(q+\frac{1}{2})+\log 2.
\ee Setting $q=\frac{1}{2}$ and combining  with \eqref{Eulergamma}, we find 
\bea 
\psi(\frac{1}{2})=-\gamma_E-2\log 2.
\eea It follows that 
\be 
\psi(\frac{3}{2})=2-\gamma_E-2\log 2.
\ee 
The $n$-th derivative of the Digamma function is defined as 
\bea 
\psi^{(n)}(q)=\frac{d^n}{dq^n}\psi(q).
\eea 
In the IR region $\omega_0\to 0$ and near $q=0$, we may write the result \eqref{Jq} as 
\bea 
J(q,\chi,\omega_0)=\Gamma(q)(i\chi)^{-q}-\frac{\omega_0^q}{q}+\mathcal{O}(\omega_0).
\eea 
By taking the limit $q\to 0$, we find 
\bea 
\lim_{q\to 0}\frac{d^n}{dq^n}J(q,\chi,\omega_0)=\lim_{q\to 0}\frac{d^n}{dq^n}[\Gamma(q)(i\chi)^{-q}-\frac{\omega_0^q}{q}+\mathcal{O}(\omega_0)]
\eea  which follows that 
\bs\begin{align}
   \lim_{q\to 0}J(q,\chi,\omega_0)&=-I_0,\\
   \lim_{q\to 0}\frac{d}{dq}J(q,\chi,\omega_0)&=\frac{I_0^2}{2}-I_0\log\omega_0+\frac{\pi^2}{12},\label{c10b}\\
   \lim_{q\to 0}\frac{d^2}{dq^2}J(q,\chi,\omega_0)&=-\frac{I_0^3}{3}+I_0^2\log\omega_0-I_0(\log^2\omega_0+\frac{\pi^2}{6})+\frac{1}{6}[\pi^2\log\omega_0+2\psi^{(2)}(1)],\\
   \lim_{q\to 0}\frac{d^3}{dq^3}J(q,\chi,\omega_0)&=\frac{I_0^4}{4}-I_0^3\log\omega_0+I_0^2\frac{1}{4}(\pi^2+6\log^2\omega_0)-I_0[\log\omega_0^3+\frac{\pi^2}{2}\log\omega_0+\psi^{(2)}(1)\nn\\&\quad +\frac{1}{4} \pi ^2 \log ^2\omega_0+\psi ^{(2)}(1) \log\omega_0+\frac{3 \pi ^4}{80}]
\end{align}\es where we have defined
\bea 
I_0=\gamma_E+\log i\omega_0\chi.
\eea The integrals $ J_n(\chi,\omega_0)$ are always finite for any $n=0,1,2,3,\cdots$.

\section{Callan-Symanzik equation for two-point Carrollian amplitude}\label{cseqn}
The Callan-Symanzik equation for two-point Carrollian amplitude is 
\bea 
M\frac{\partial}{\partial M}C_{(2)}+\beta\frac{\partial}{\partial\lambda}C_{(2)}-2\gamma C_{(2)}=0.
\eea Since $C_{(2)}$ has been fixed to \eqref{functiong} by Poincar\'e symmetry up to a normalization factor $N$, the equation becomes a constraint for $N$
\bea 
M\frac{\partial}{\partial M}N+\beta\frac{\partial}{\partial\lambda}N-2\gamma N=0.\label{2ptcs}
\eea Since $\gamma\not=0$, the normalization $N$ should depend on the energy scale $M$ and the coupling constant $\lambda$. Then it is evolved into the scale $\mu$ through the solution of \eqref{2ptcs}  
\bea 
N(\bar{\lambda}(\mu))=\exp[2\int_{M}^\mu \frac{d\mu'}{\mu'}\gamma(\bar{\lambda}(\mu'))]N(\lambda).
\eea 
Now we will calculate the normalization factor $N$ up to 2-loop. The relevant diagram is the sunset diagram
\bea 
i\mathcal{M}_{\text{sunset}}^{\text{2-loop}}(p)&=&\frac{(-i\lambda)^2}{6}\bar{M}^{2\varepsilon}\int\frac{d^dp_1}{(2\pi)^d}\int\frac{d^dp_2}{(2\pi)^d}\frac{-i}{p_1^2}\frac{-i}{p_2^2}\frac{-i}{(p-p_1-p_2)^2}\nn\\&=&i\frac{\lambda^2}{6\times(4\pi)^d}\bar{M}^{2\varepsilon}\Gamma(3-d)B(\frac{d}{2}-1,\frac{d}{2}-1)B(\frac{d}{2}-1,d-2)(p^2)^{d-3}\nn\\&=&i\frac{\lambda^2}{12\times(4\pi)^4} p^2(-\frac{1}{\varepsilon}+\log\frac{p^2}{M^2}-\log 4-\frac{5}{4}),
\eea where the Beta function $B(p,q)$ is defined as 
\be 
B(p,q)=\frac{\Gamma(p)\Gamma(q)}{\Gamma(p+q)}.
\ee 
The divergence ${1}/{\varepsilon}$ is canceled by a counter diagram and then the 2-loop correction to the two-point amplitude is \be 
i\mathcal{M}^{\text{2-loop}}(p)=i\frac{\lambda^2}{12\times(4\pi)^4}p^2(\log\frac{p^2}{M^2}-\log 4-\frac{5}{4}).
\ee 
The result is the same as the Problem 10.3 of \cite{1995iqft.book.....P}.
It seems that the amplitude is 0 on shell since the particle is massless and then the two-point Carrollian amplitude receives no contribution at 2-loop level. However, we will work out the details to transform it to the Carrollian amplitude 
\bea 
C_{(2)}^{\text{2-loop}}&=&
\left(\frac{1}{8\pi^2 i}\right)^2\int_0^\infty d\omega_1\int_0^\infty d\omega_2e^{i\omega_1u_1-i\omega_2u_2}(2\pi)^4\delta^{(4)}(p_1-p_2)i p_1^2 U(p_1^2)\label{2pt2loop}
\eea where we have defined a function $U(p^2)$ through 
\bea 
U(p^2)=\frac{\lambda^2}{12\times(4\pi)^4}(\log\frac{p^2}{M^2}-\log 4-\frac{5}{4})
\eea which is divergent in the limit $p^2\to 0$.
Note that we have flipped the sign of the incoming momentum in the context and parameterized the momenta $p_1$ and $p_2$ as 
\bea 
p_1=\omega_1n_1,\quad p_2=\omega_2 n_2.
\eea 
The Dirac delta function is 
\bea 
(2\pi)^4 \delta^{(4)}(p_1-p_2)=(2\pi)^4 \delta^{(3)}(\bm p_1-\bm p_2)\delta(\omega_1-\omega_2).
\eea The last term is divergent as $\delta(\omega_1-\omega_2)=\delta(0)$ on-shell since $\bm p_1=\bm p_2$. However, we also notice that $p_1^2=0$ and we will argue that the product 
$
(2\pi)^4\delta^{(4)}(p_1-p_2)p_1^2
$ could be replaced by 
\bea 
(2\pi)^4\delta^{(4)}(p_1-p_2)p_1^2\to  \text{const.}\times \omega_1 (2\pi)^3\delta^{(3)}(\bm p_1-\bm p_2)\label{rep}
\eea in the integrand. The factor $\omega_1$ is fixed by dimensional analysis and the constant may be fixed by comparing it with the tree-level two-point Carrollian amplitude which is a Fourier transform of the two-point amplitude
\bea 
C_{(2)}^{\text{tree}}=\left(\frac{1}{8\pi^2 i}\right)^2\int_0^\infty d\omega_1\int_0^\infty d\omega_2e^{i\omega_1u_1-i\omega_2u_2}(2\pi)^4\delta^{(4)}(p_1-p_2) (ip_1^2).\label{c2pt}
\eea  The replacement rule \eqref{rep} follows from comparing \eqref{c2pt} with \eqref{bbprp} and \eqref{delta3} in which $\text{const.}=-2 i$. Then the 2-loop correction has the same form as the tree level, except a $\mathcal{O}(\lambda^2)$ correction from $U(p^2_1)$. The logarithmic IR divergence of $U(p_1^2)$ may be cured by setting $p_1^2$ to a spacelike momentum $-\mu^2$. As will be shown, the IR divergence doesn't affect the discussion. We will not focus on it in this work. Therefore, we find the following correction for the normalization factor $N$
\bea 
N^{\text{2-loop}}=N^{\text{tree}}(1+U(-\mu^2)).
\eea Note that \bea 
M\frac{\partial}{\partial M}N^{\text{2-loop}}=-2\times\frac{\lambda^2}{12\times (4\pi)^4}N^{\text{tree}}=2\gamma^{(2)}\lambda^2N^{\text{tree}}
\eea and $\beta=\mathcal{O}(\lambda^2)$, the Callan-Symanzik equation \eqref{2ptcs} is satisfied.

\bibliography{refs}
\end{document}